%% file: ssmpaper.tex
\documentclass[a4paper,11pt]{article}
\pdfoutput=1 

\usepackage{jcappub} 

\usepackage[T1]{fontenc} 

\usepackage{graphicx}
\usepackage{dcolumn}
\usepackage{amssymb}
\usepackage{amsmath}
\usepackage{color}
\usepackage{verbatim}
\usepackage{multirow}
\usepackage{soul}

\usepackage{tabularx}

\usepackage{subfigure}

\graphicspath{{figures/}}

\def\al{\alpha} 
\def\be{\beta} 
\def\ga{\gamma}
\def\de{\delta}
\def\ep{\epsilon}

\def\th{\theta}

\def\ka{\kappa}
\def\la{\lambda}

\def\ta{\tau}

\def\ph{\phi}

\def\om{\omega}
\def\De{\Delta}
\def\Ga{\Gamma}

\def\Om{\Omega}

\def\pa{\partial}

\newcommand{\ben}{\begin{equation}}
\newcommand{\een}{\end{equation}}
\newcommand{\bea}{\begin{eqnarray}}
\newcommand{\eea}{\end{eqnarray}}
\newcommand{\ba}{\begin{array}}
\newcommand{\ea}{\end{array}}
\newcommand{\bal}{\begin{align}}
\newcommand{\eal}{\end{align}}
\newcommand{\bit}{\begin{itemize}}
\newcommand{\eit}{\end{itemize}}

\newcommand{\vev}[1]{\left\langle#1\right\rangle}

\newcommand{\bk}{\textbf{k}}
\newcommand{\bq}{\textbf{q}}
\newcommand{\bv}{\textbf{v}}
\newcommand{\bx}{\textbf{x}}

\newcommand{\half}{\frac{1}{2}}

\newcommand{\dbar}[2]{\frac{d^{#1}{#2}}{(2\pi)^{#1}}}
\newcommand{\debar}[2]{{(2\pi)^{#1}}\de({#2})}

\newcommand{\mcA}{\mathcal{A}}

\newcommand{\AdInd}{\Gamma} 
\newcommand{\aln}{\al_\text{n}}

\newcommand{\BrePosRat}{r_\text{b}}
\newcommand{\bqtil}{\tilde{\bq}}

\newcommand{\cs}{c_\text{s}} 
\newcommand{\cwgfitfun}{C}

\newcommand{\Devw}{\De_\text{w}}

\newcommand{\enDen}{e}
\newcommand{\en}{\rho}
\newcommand{\enAv}{\bar{\en}}

\newcommand{\fluidV}{\overline{U}_\text{f}}  
\newcommand{\fluidL}{L_\text{f}} 

\newcommand{\gaWF}{\tilde{\ga}} 

\newcommand{\HN}{H_\text{n}} 
\newcommand{\Hn}{H_\text{n}}
\newcommand{\hdot}{\dot{h}}

\newcommand{\lafft}{\tilde{\la}}

\newcommand{\nb}{n_\text{b}}
\newcommand{\Nb}{N_\text{b}}

\newcommand{\OmGW}{\Omega_\text{gw}}
\newcommand{\OmGWscaled}{\tilde\Omega_\text{gw}}
\newcommand{\omtil}{\tilde{\omega}}

\newcommand{\pf}{p_\text{f}}
\newcommand{\Pspec}[1]{{\mathcal P}_{#1}}
\newcommand{\PspecGW}{\Pspec{\text{gw}}}

\newcommand{\qtil}{\tilde{q}}

\newcommand{\Rbc}{R_*} 

\newcommand{\rGW}{\enDen_\text{gw}}

\newcommand{\SpecDen}[1]{P_{#1}}
\newcommand{\SpecDenGW}{\tilde P_{\text{gw}}}

\newcommand{\strPar}{\alpha}
\newcommand{\ssmfitfun}{M}

\newcommand{\tNL}{\ta_\text{nl}}
\newcommand{\tf}{t_\text{f}}
\newcommand{\Tstar}{T_*}

\newcommand{\tCoh}{\tau_\text{c}}

\newcommand{\tLife}{\tau_\text{v}}
\newcommand{\TN}{T_\text{n}} 
\newcommand{\tN}{t_\text{n}} 
\newcommand{\tn}{t_\text{n}} 
\newcommand{\Tc}{T_\text{c}} 
\newcommand{\tc}{t_{\text{c}}} 
\newcommand{\tInit}{t_\text{i}}
\newcommand{\TInit}{T_\text{i}}
\newcommand{\Ttil}{\tilde{T}}

\newcommand{\uetcTen}{U_\Pi}

\newcommand{\vw}{v_\text{w}} 
\newcommand{\vsh}{v_\text{sh}} 
\newcommand{\vCJ}{v_\text{CJ}} 
\newcommand{\Vu}{V}
\newcommand{\Vb}{V_\text{b}}
\newcommand{\Vtot}{V_\text{tot}}
\newcommand{\vWF}{\tilde{v}} 
\newcommand{\vSF}{\tilde{v}^\text{sh}}
\newcommand{\vfft}{\tilde{v}} 
\newcommand{\vq}{\vel{\bq}} 

\newcommand\vel[1]{v_{#1}}

\newcommand{\Vol}{{\mathcal V}}
\newcommand{\vip}{v_\text{ip}}

\newcommand{\wAv}{\bar{w}}

\newcommand{\xiw}{\xi_\text{w}}
\newcommand{\xish}{\xi_\text{sh}}

\definecolor{newgreen}{RGB}{10,100,20}

\makeatletter
\def\Let@{\def\\{\notag\math@cr}}
\makeatother

\subheader{NORDITA-2019-083, HIP-2019-29/TH}

\title{Gravitational waves from first order cosmological phase transitions in the Sound Shell Model}

\newcommand{\addressSussex}{Department of Physics \& Astronomy, University of Sussex, \\Brighton, BN1 9QH, United Kingdom}
\newcommand{\HIPetc}{
		Department of Physics and Helsinki Institute of Physics,
		\\PL 64,  
		FI-00014 University of Helsinki,
		Finland
	}

\author[a,b]{Mark Hindmarsh} 
\emailAdd{mark.hindmarsh@helsinki.fi}
\affiliation[a]{\HIPetc}
\affiliation[b]{\addressSussex}

\author[c]{and Mulham Hijazi}
\affiliation[c]{School of Physics and Astronomy, 
University of Manchester\\
Oxford Road, Manchester, 
M13 9PL, 
United Kingdom}
\emailAdd{mulhamhijazi@gmail.com}

\abstract{We calculate gravitational wave power spectra 
from first order early Universe phase transitions using the Sound Shell Model.
The model predicts that the power spectrum depends on the mean bubble separation, 
the phase transition strength, the phase boundary speed, with the overall frequency scale set by the nucleation temperature. 
There is also a dependence on the time evolution of the bubble nucleation rate. 
The gravitational wave peak power and frequency are in good agreement with 
published numerical simulations, where bubbles are nucleated simultaneously. 
Agreement is particularly good for detonations, but the total power for deflagrations is predicted higher than numerical simulations show, indicating refinement of the model of the transfer of energy to the fluid is needed for accurate computations.
We show how the time-dependence of the bubble nucleation rate affects the shape of the power spectrum: an exponentially rising 
nucleation rate produces higher amplitude gravitational waves at a longer wavelength than simultaneous nucleation.  
We present an improved fit for the predicted gravitational wave power spectrum in the form of a double broken power law, where the two breaks in the slope happen at wavenumber corresponding to the mean bubble separation and the thickness of the fluid shell surrounding the expanding bubbles, which in turn is related to the difference of the phase boundary speed from the speed of sound. 
}


\begin{document}
\maketitle


\section{Introduction}
\label{sec:intro}

With the direct detection of  gravitational waves from the merger of binary black holes \cite{Abbott:2016blz} 
and binary neutron stars \cite{TheLIGOScientific:2017qsa}, 
the approval of the Laser Interferometer Space Antenna (LISA) \cite{Audley:2017drz},
and the outstanding performance of LISA Pathfinder \cite{Armano:2018kix},  
there has been a growing interest in the prospects for gravitational wave observatories. 
LISA will enable us to probe a period around 10 ps after the Big Bang, 
which is of particular interest for a first order electroweak phase transition.

In theories beyond the Standard Model, the Higgs and any other scalar fields involved 
in electroweak symmetry-breaking can become trapped in 
a metastable state as the Universe expands and cools \cite{Kirzhnits:1976ts}. 
Thermal or quantum fluctuations drive the field over or through an activation barrier,
in small bubbles of the stable phase, which expand and fill up the entire space \cite{Coleman:1977py,Linde:1981zj}. 

Some of the potential energy in the scalar fields is converted into the kinetic energy of the cosmological fluid surrounding the bubbles \cite{Steinhardt:1981ct}, which is a source of shear stress, and therefore gravitational radiation \cite{Witten:1984rs}. 
The power spectrum of the gravitational waves contains information about the thermodynamic parameters of the transition, and therefore about the underlying theory.  An important programme is make this connection as precise as possible, 
in order that future gravitational wave observations at the Laser Interferometer Space Antenna (LISA) \cite{Audley:2017drz} 
can be used to probe new physics.

It has become clear through numerical simulations that  
the sound waves produced by expanding bubbles are a very important 
source of gravitational waves \cite{Hindmarsh:2013xza,Hindmarsh:2015qta,Hindmarsh:2017gnf}.  
A model for these acoustically-produced gravitational waves was been put forward \cite{Hindmarsh:2016lnk}, 
which can make precise predictions for the amplitude and shape of the power spectrum as a function of parameters of the phase transition.  

The model exploits the fact that colliding sound waves generate gravitational waves, whose power spectrum can be computed if the velocity power spectrum of the fluid is known, and is Gaussian. 
The model proposes that the velocity power spectrum is determined by the sound shells --
compression waves -- surrounding the expanding bubbles of the stable phase, whose shapes are easily computable in 
relativistic hydrodynamics. 
Simple predictions for power laws have been made \cite{Hindmarsh:2016lnk} and shown to work quite well \cite{Hindmarsh:2017gnf}. 
An important feature of the model is that there are two length scales in the power spectrum: the mean bubble separation and the sound shell thickness, which decreases as the wall speed approaches the speed of sound. 
This has also been qualitatively supported by the numerical simulations.

In this work we pursue the detailed predictions of the sound shell model, showing the velocity and gravitational wave power spectra 
for various wall speeds and transition strengths, and making a comparison with numerical simulations.
The power laws observed	 in \cite{Hindmarsh:2017gnf} are confirmed, 
and the shape around the 
peak of the power spectra reproduced for a range of wall speeds. 
We correct an error made in \cite{Hindmarsh:2016lnk}, leading to revised predictions for the power laws at low wavenumber.

The model also agrees well with the peak power for detonations, but less well for deflagrations.  The velocity and 
gravitational wave power spectra observed in numerical simulations of deflagrations are lower than predicted.  
Recent simulations \cite{Cutting:2019zws} have shown that there  
is a suppression of kinetic energy, by an amount which grows with the strength of the transition. 
This suppression needs to be incorporated in future versions of the model. 

Furthermore, the numerical simulations of the coupled scalar field hydrodynamic system were performed with 
bubbles nucleated simultaneously, rather than at the more realistic exponentially rising rate.  
Numerical simulations in a simplified model indicate that with exponential nucleation, 
the peak of the power spectra moves to lower values of $k\Rbc$,
and that the gravitational waves amplitude increases significantly \cite{Weir:2016tov}. 
Our predictions accommodate both nucleation histories, and reproduce this behaviour.
Further work is needed here too, in the form of large scalar-hydrodynamic simulations with exponential nucleation.

Other models of gravitational wave generation from sound waves in the fluid following a first order phase transition have also been put forward, focusing on the dynamics of the expanding compression waves shells in real space \cite{Jinno:2016vai,Konstandin:2017sat}. 
While not as successful in accounting for the shape of the power spectrum around the peak, they suggest that there are small 
signals at very low frequencies compared to the peak not included in the Sound Shell Model.

The Sound Shell Model in the form presented here 
does not include effects from the non-linear evolution of the fluid perturbations, which 
are already important for strong transitions during the collision phase \cite{Cutting:2019zws}, 
and may lead to quite different behaviour in transitions with extreme supercooling where 
very thin ultra relativistic fluid shells are 
propagating in a near-vacuum \cite{Jinno:2019jhi}.

Non-linearity is also to effects on timescales of order $\tNL = \Rbc/\fluidV$, where $\Rbc$ is the mean bubble separation and $\fluidV$ the 
enthalpy-weighted RMS fluid velocity.. These non-linear effects include the generation of shocks, turbulence and the damping of the fluid flow.  
Turbulence can also be generated in the collision phase by the interaction between fluid shells, and by the bubble wall dynamics \cite{Cutting:2019zws}.  

Observable gravitational waves are likely to have large enough RMS fluid velocities that  $\tNL \lesssim \HN^{-1}$ 
\cite{Hindmarsh:2017gnf,Ellis:2018mja,Ellis:2019oqb}, 
showing up as changes in the shape of the gravitational power spectrum. For example, 
the gravitational wave power spectrum from turbulent flows has been modelled 
\cite{Gogoberidze:2007an,Caprini:2009yp,Niksa:2018ofa} 
and while the predictions of the asymptotic power laws depend on assumptions about the velocity correlations, 
they are quite different  from purely acoustic production.  
Recent numerical simulations of magnetohydrodynamic turbulence indicate another characteristic power law 
to the low-frequency side of the peak \cite{Pol:2019yex}. 

The true power spectrum of a strong phase transition is likely to be a complicated mixture of effects, 
especially if magnetic fields are involved, and 
the Sound Shell Model can be viewed as a first step in the understanding of the gravitational wave power spectrum from 
first order phase transition in the early Universe.  

\section{Cosmological first order phase transitions}

A first order thermal phase transition proceeds by the nucleation, growth and merger of bubbles of the stable phase in the supercooled metastable phase of the cosmological fluid \cite{Steinhardt:1981ct,Gyulassy:1983rq,Hogan:1984hx,DeGrand:1984uq,Kajantie:1986hq,Enqvist:1991xw,Csernai:1992as,Turner:1992tz,Espinosa:2010hh}.  
The transition is signalled by the scalar order parameter $\phi$ gaining a large expectation value inside the bubble, which normally spontaneously breaks a symmetry of the theory. 

The dynamics of the transition are controlled by the bulk free energy density, or equivalently the pressure, which is a function of both temperature and $\ph$. 
At the critical temperature $\Tc$ the pressure is the same in the two phases. Below $\Tc$ 
the pressure inside the bubble is higher than outside (as the free energy density is lower), the bubbles expand, collide, and the stable phase eventually fills space.  Generally, friction between the plasma and the phase boundary ensures that the bubble wall expands at a constant speed $\vw$, which is determined by the pressure difference and the coupling between the order parameter and the fluid \cite{Moore:1995si,Huber:2011aa,Huber:2011aa}.
However, if this coupling is sufficiently weak, or the supercooling sufficiently large, the friction may be insufficient to prevent the wall from continuing to accelerate in a so-called run-away \cite{Bodeker:2009qy,Bodeker:2017cim,Ellis:2019oqb}. 
The sound shell model does not apply in the run-away scenario; the so-called envelope approximation 
\cite{Kosowsky:1991ua,Huber:2008hg} gives an order-of-magnitude estimate of the gravitational wave power. 
A more accurate form for the power spectrum of colliding vacuum bubbles has been found numerically in Ref.~\cite{Cutting:2018tjt}.  

Assuming that the friction with the fluid is sufficiently important that the bubble wall speed asymptotes to a constant $\vw$, 
the way the Universe changes from the symmetric to the broken phase can be calculated from the Euclidean action for nucleating a critical bubble of the new phase, $S(t)$. This decreases from infinity at the time $\tc$ at which the Universe passes through the critical temperature $\Tc$. The nucleation rate per unit volume is
\ben
p(t) = p_0 e^{-S(t)},
\label{e:NucRat}
\een
where the time dependence in $p_0$ can be ignored as a first approximation, and $p_0 \sim \Tc^4$.
At the same level of approximation, the phase transition occurs when the nucleation rate per unit volume reaches one bubble per Hubble volume per Hubble time, $p \sim H^4$.  
We assume this occurs while the Universe is still radiation dominated, rather than becoming vacuum energy dominated. In the latter case the analysis goes differently \cite{Guth:1981uk}.

If the nucleation rate is an increasing function of time, 
the fraction of the Universe in the symmetric phase $h(t)$ behaves to a very good approximation as \cite{Enqvist:1991xw} 
\ben
h(t) = \exp\left( - e^{\be(t - \tf)} \right), 
\label{e:hExp}
\een
where $\be = d\ln p/dt \simeq 
-S'(\tf)$ is the transition rate parameter, and $\tf$ is the time at which the fraction of the Universe in the symmetric phase $h(t)$ is reduced to a fraction $1/e$ \cite{Enqvist:1991xw}. This is determined implicitly by the equation
\ben
\label{e:tfEq}
p_0 e^{-S(\tf)} = \frac{(S'(\tf))^4}{8\pi\vw^3}.
\een
This is also the time at which the volume-averaged bubble nucleation rate $\Ga(t) = h(t)p(t)$ reaches its maximum, which we can call the nucleation time $\tN$.
The final bubble density, obtained by integrating $\Ga(t)$ over all time, is 
\ben
\nb = 8\pi \frac{\be^3}{\vw^3}
\een
We term this scenario exponential nucleation \cite{Cutting:2018tjt}.

In the case that $S(t)$ has a minimum, at say $t_0$, bubble nucleation is concentrated at around $t_0$, and the fraction in the symmetric phase behaves at late times as
\ben
h(t) = \exp\left( - \frac{4\pi}{3} \nb \vw^3 (t - t_0)^3 \right),
\label{e:hSim}
\een
where $\nb$ is the asymptotic bubble density.  Hence the Universe behaves as if all bubbles were nucleated at $t = t_0$.
If the action is expanded around its minimum as $S(t) = S_0 + \half \be_2^2(t - t_0)^2$, the asymptotic bubble density is 
\ben
\nb = \frac{\sqrt{2\pi}}{\beta_2} p_0 e^{-S(t_0)} 
\een
We term this scenario simultaneous nucleation  \cite{Cutting:2018tjt}.

The mean bubble separation, which sets the length scale for the fluid perturbations, is in both cases defined as 
\ben
\Rbc = \nb^{-\frac13},
\een
and the transition completes in a time of order $\Rbc/\vw$. 
In the exponential case, the transition time can be taken to be $\beta^{-1}$. 
In the simultaneous case, it is convenient to define an effective transition rate parameter from the bubble density, 
$\be_\text{eff} = \vw/(8\pi)^\frac{1}{3}\Rbc$, 
so that 
\ben
h_\text{sim}(t) = \exp\left( - \frac{1}{6} \beta_\text{eff}^3(t - t_0)^3 \right) .
\een
The nucleation scenarios are reviewed in more detail in Appendix \ref{a:BubNuc}.

Relativistic hydrodynamics dictates how the fluid responds around the expanding bubbles.  The fleld-fluid system has energy-momentum tensor
\begin{align}
T^{\mu\nu} &= (\enDen+p)u^\mu u^\nu  + g^{\mu\nu}\left( p - \half (\pa \phi)^2 \right) +  \pa^\mu \phi \pa^\nu \phi  ,
\end{align}
with $\enDen$ the energy density, $p$ the pressure,  $u^\mu$ the fluid 4-velocity, and $g^{\mu\nu}$ the space-time metric.
Conservation of energy-momentum across the bubble wall is the basis of the computation of the fluid velocity and enthalpy density $w = \enDen + p$.

If the bubbles expand at a constant speed, the fluid velocity and enthalpy density settle down to a self-similar radially-symmetric profile \cite{KurkiSuonio:1984ba,Kamionkowski:1993fg,KurkiSuonio:1995pp,KurkiSuonio:1995vy}. 
The fluid responds in one of the following ways: a compression wave with a leading shock ahead of the subsonic wall in a deflagration;
a compression wave behind the supersonic wall in a detonation;
or both in the case of a supersonic deflagration (hybrid).

The fluid profiles are entirely determined by 
the wall speed $\vw$, the sound speed $\cs(T)$, and the transition strength parameter $\strPar(T)$ evaluated at the nucleation temperature $\TN$. 
The transition strength parameter is defined 
from the enthalpy density and the trace anomaly difference between the symmetric and broken phases, where the trace anomaly is  
\ben
\th = \frac{1}{4} (\enDen - 3p).
\een
This is the precise definition of what is meant by the potential energy of the scalar field in this context. 
The transition strength parameter is then
\ben
\label{e:StrParDef}
\aln = \frac{4}{3}\frac{\th_\text{s}(\TN) - \th_\text{b}(\TN)}{w(\TN)},
\een
where the subscripts s and b denote the symmetric (metastable) and broken (stable) phases. 

Note the distinction  between the trace anomaly difference and the latent heat density
$L = w_\text{s}(\Tc) - w_\text{b}(\Tc)$.  
Using $\th = w/4 - p$, and the definition of the critical temperature, one immediately sees that
\ben
\th_\text{s}(\Tc) - \th_\text{b}(\Tc) = L/4.
\een
In the bag model of the equation of state near a phase transition, 
the trace anomaly difference is a temperature-independent constant, 
equal to the vacuum energy difference between the two phases.

As the bubbles collide and merge, their surrounding compression waves become propagating sound waves, 
with a characteristic length scale $\fluidL$ set by the mean bubble separation. 

Non-linearities in the fluid operate on a  timescale $\tNL = \fluidL/\fluidV$, 
where $\fluidV$ is the RMS fluid velocity,  
and can generate further shocks and turbulence, and 
leads to the eventual dissipation of the fluid perturbations on the same timescale.
The fluid perturbations are a source of gravitational waves throughout the collision, acoustic and non-linear phases of the transition. 
If $\tNL$ is much greater than the Hubble time, the acoustic phase is the dominant source.  
This is the case if $\tNL \Hn = (\Hn \fluidL)/\fluidV \gg 1$. 
The mean bubble separation must be less than the Hubble distance in order for the phase transition to complete.

Acoustic production has been extensively studied by 3-dimensional numerical simulations of the coupled fluid-field system 
\cite{Hindmarsh:2013xza,Hindmarsh:2015qta,Hindmarsh:2017gnf}. 
The simulations revealed a power spectrum peaked at a wavelength around the average bubble separation $\Rbc$, with a power-law $k^{-p}$ at wavenumber $k \gg \Rbc^{-1}$. Where the power law is clear, the index was somewhere in the range $-3 \lesssim p \lesssim -4$.  There was also evidence for some structure in the peak: where the bubble wall speed $\vw$ was closer to the speed of sound, the peak was broader. 

In a previous work \cite{Hindmarsh:2016lnk}, one of us outlined a model for the acoustic gravitational wave power spectrum, called the Sound Shell Model.
It is based on the observation that the sound waves set up by the compression shells around the expanding bubble of the stable phase continue to propagate after the phase boundaries driving them have disappeared. As the bubbles expand, 
the radial velocity field around the bubble $v(r,t)$ takes a self-similar invariant profile $\vip(\xi)$, with $\xi = r/t$. 
When the bubbles collide, the invariant profile is assumed to become the initial condition for a sound wave.
Recent simulations \cite{Cutting:2019zws} have shown that there is significant interaction in the collision phase of deflagrations in strong transitions, 
which suppress the fluid kinetic energy and generate turbulence.
 
The subsequent local fluid velocity is the superposition of the waves from many sound shells, and can be treated as a Gaussian random field, as the velocity field at any point is the resultant of the sound shells from a very large (and increasing) number of bubbles. The power spectrum of the velocity field is computable from the velocity profile $\vip(\xi)$, and the gravitational wave power spectrum can then be computed from the velocity field by a simple convolution of the power spectrum \cite{Hindmarsh:2015qta}. 
The model is distinguished from earlier work \cite{Kosowsky:2001xp,Caprini:2006jb,Caprini:2007xq,Caprini:2009fx} by the recognition that long-lasting sound waves are the main source of gravitational radiation, and by the computation of their power spectrum from the hydrodynamic solution. 

In \cite{Hindmarsh:2016lnk}, it was shown that the model makes clear predictions for the general shape of the velocity and gravitational wave power spectra: they are double broken power laws, with the breaks in slope at wave-numbers $k$ dictated by the mean bubble separation $\Rbc$, and the sound shell thickness  $\De\Rbc = \Rbc\Devw$, where $\Devw = |\vw - \cs|/\vw$ for 
wall speeds which are not much less than $c$.  
For large $k\De\Rbc$, the gravitational wave power law index is $-3$.  
For wall speeds near the speed of sound $\cs$, the sound shell is thin, and there is a characteristic $k^{1}$ power law in the range  $\Rbc^{-1} \lesssim k \lesssim \De\Rbc^{-1}$. 

Denoting the velocity field wave number by $q$, the velocity power spectrum index for large $q\De\Rbc$ is $-1$. 
For wall speeds near the speed of sound $\cs$ there is a  $q^{1}$ power law between
the two breaks.

These predictions have been studied in numerical simulations \cite{Hindmarsh:2017gnf}.
The gravitational wave power spectrum shows a peak at around $k\Rbc = 10$, with 
a clear $k^{-3}$ behaviour at high $k$ for detonations,
although the power law seems slightly steeper for the one thick-shell deflagration studied. 
There is a slowly rising plateau consistent with $k^1$ 
for transitions with a very thin sound shell. 
The velocity power spectra show similar agreement. 
The simulations are not large or long enough to 
determine the low wavenumber behaviour of the gravitational waves.

In any case, an error was made in \cite{Hindmarsh:2016lnk} which affects the prediction for low wavenumber spectra.  In setting the initial conditions for the sound 
wave from the self-similar sound shell around the bubble, insufficient attention was paid to the need to respect energy conservation, and the resulting velocity power spectrum did not obey the causality conditions derived in Ref.~\cite{Durrer:2003ja}.  Here, we find that the corrected causal velocity power spectrum goes as $q^5$ (instead of $q^3$) at low $q$, and that the resulting gravitational wave power spectrum goes as $k^9$ (instead of $k^5$) at low $k$.  

In the following we distinguish between the spectral density of a field $f(\bx)$ with Fourier coefficients $f_\bk$,
$\SpecDen{f} = |f_\bk|^2$, and the power spectrum  $\Pspec{f} = {k^3}|f_\bk|^2/{2\pi^2}$.

\section{Gravitational wave power spectrum from phase transitions}

\subsection{Gravitational wave power spectrum from shear stress correlator}

We assume that the phase transition completes in much less than a Hubble time, and so that we can neglect the expansion of the Universe.\footnote{In fact, the fluid equations have a scale symmetry which means that the expansion of the Universe can be scaled out of the equations \cite{Brandenburg:1996fc,Hindmarsh:2015qta}.}  
The fluid and the scalar field are a source of metric perturbations, which 
in the synchronous gauge produce a change in the space-time interval
\[
ds^2 = -dt^2 + (\de_{ij} + h_{ij} ) dx^i dx^j
\]
The metric perturbations ${h}_{ij}$ are sourced by shear stress, the transverse-traceless (`tensor') part of EM tensor $\Pi_{ij}$, 
\ben
\ddot{h}_{ij} - \nabla^2 h_{ij} = 16 \pi G \Pi_{ij}
\een
The energy-momentum tensor 
contains contributions from both the fluid and the scalar field, 
\begin{align*}
T^\text{f}_{ij} &= (e+p)\ga^2v_iv_j + p\de_{ij} \\
T^{\phi}_{ij} &= \pa_i\phi\pa_j\phi -\half(\pa\phi)^2 \de_{ij} .
\end{align*}
It is conventional to include the effective potential for the scalar field $V_T(\phi)$ in the fluid pressure $p$.

A particular solution for the gravitational wave equation in terms of $h_{ij}$ is given by:
\begin{align}
h_{ij}(\textbf{k},t)=(16\pi G)\Lambda_{ij,kl}(\textbf{k})\int_{0}^{t}dt'\frac{\sin[k(t-t')]}{k}T_{kl}(\textbf{k},t')
\end{align}
where $\Lambda_{ij,kl}(\textbf{k})=P_{ik}(\textbf{k})P_{jl}(\textbf{k})-\frac{1}{2}P_{ij}(\textbf{k})P_{kl}(\textbf{k})$ , and $P_{ij}(\textbf{k})=\delta_{ij}-\hat{k}_{i}\hat{k}_{j}.$
We will assume that the energy-momentum is the result of an isotropic random process.

The gravitational wave energy density is 
\ben
\rGW = \frac{1}{32\pi G} \overline{\dot{h}_{ij} (x)\dot{h}_{ij} (x)}
\een
where the line over the expression denotes an average over many  wavelengths and periods. 
We define $P_{\dot{h}}(\textbf{k},t)$, the spectral density of the time derivative of the perturbations in the metric, such that:
\begin{align}
\left\langle 
\dot{h}_{\textbf{k}}^{ij}(t)\dot{h}_{\textbf{k}}^{ij}(t) 
\right\rangle 
= P_{\dot{h}}(\textbf{k},t)(2\pi)^{3}\delta(\textbf{k}+\textbf{k}'),
\end{align}
where the angle brackets denote an average over the random process generating the gravitational waves.
In terms of the spectral density, the gravitational wave energy density is 
\ben
\rGW = \frac{1}{32\pi G}  \int \frac{dk k^2}{2\pi^2} \SpecDen{\hdot}(k)
\een
It is often convenient to use the power spectrum of $\dot{h}$, which we define as 
\( \Pspec{\hdot}\). 
The gravitational wave power spectrum is defined as the contribution to the density fraction in gravitational waves per logarithmic wavenumber interval, or 
\ben
\PspecGW(k) \equiv \frac{d \OmGW}{d \ln(k)} = \frac{1}{\enAv}  \frac{1}{32\pi G} \Pspec{\hdot}(k) = \frac{1}{12 H^2} \Pspec{\hdot}(k) .
\een
It suffices to consider the tensor
\ben
\tau_{ij}=\gamma^{2} w v_{i}v_{j} + \pa_i\phi \pa_j\phi
\een
as the source of the shear stress, as the diagonal pressure term is not traceless.
Hence
\begin{align}
\left\langle 
\dot{h}_{\textbf{k}_{1}}^{ij}(t)\dot{h}_{\textbf{k}_{2}}^{ij}(t)
\right\rangle 
&=(16\pi G)^{2}\int_{0}^{t}dt_{1}dt_{2}\cos[k_{1}(t-t_{1}]\cos[k_{2}(t-t_{2})]\\
&\times\Lambda_{ij,kl}(\textbf{k})
\left\langle 
\ta_{\textbf{k}_{1}}^{ij}(t_{1})\ta_{\textbf{k}_{2}}^{kl}(t_{2})
\right\rangle 
\end{align}
By defining the unequal time correlator (UETC) of the fluid shear stress $\uetcTen$ from
\begin{align}
\Lambda_{ij,kl}(\textbf{k})
\left\langle 
\ta_{\textbf{k}_{1}}^{ij}(t_{1})\ta_{\textbf{k}_{2}}^{kl}(t_{2})
\right\rangle 
=\uetcTen(k_{1},t_{1},t_{2})(2\pi)^{3}\delta(\textbf{k}_{1}+\textbf{k}_{2})
\end{align}
we can easily, by inspection, obtain an expression for the spectral density as:
\begin{align}
P_{\dot{h}}(k,t)=(16\pi G)^{2}\int_{0}^{t}dt_{1}\int_{0}^{t}dt_{2}\cos[k(t-t_{1})]\cos[k(t-t_{2})]\uetcTen(k,t_{1},t_{2}).
\end{align}
Averaging over a number of oscillations at wavenumber $k$, we have
\begin{align}
\label{e:GWPSfromUETC}
P_{\dot{h}}(k,t)=(16\pi G)^{2} \half \int_{0}^{t}dt_{1}\int_{0}^{t}dt_{2}\cos[k(t_1 - t_{2})]\uetcTen(k,t_{1},t_{2}).
\end{align}
Now we are motivated to compute the fluid shear stress UETC $\uetcTen(k_{1},t_{1},t_{2})$.

\subsection{Shear stress correlator from sound waves}

We assume that the dominant source of shear stress is the fluid,\footnote{This is a very accurate assumption unless the bubble walls are sufficiently weakly coupled that they run away, that is, accelerate until collision \cite{Bodeker:2009qy}. 
The sound shell model does not apply in this case.} and 
that the fluid velocities are non-relativistic, so that 
\ben
\tau_{ij} \simeq \wAv  v_{i}v_{j}.
\een
Fluctuations around the mean enthalphy $\wAv$ are higher order in the perturbations and will be neglected. 

We denote the Fourier transform of the velocity field as 
\ben
\vfft^i_\bq(t) = \int d^3 x v^i(\bx,t)e^{-i\bq\cdot \bx}. 
\een
Based on the results of numerical simulations \cite{Hindmarsh:2013xza,Hindmarsh:2015qta,Hindmarsh:2017gnf}, it is justifiable to assume that the velocity field is irrotational and statistically homogeneous, in which case the the two-point function can be written 
\ben
\label{e:LonVelUETC}
\vev{\vfft^i_{\bq_1}(t_1)\vfft^{*j}_{\bq_2}(t_2)} 
= 
\hat{q}_{1}^{i}\hat{q}_{1}^{j} G(q_1,t_1,t_2) 
\debar3{\bq_1 - \bq_2}.
\een
The leading term in the shear stress UETC is then \cite{Kosowsky:2001xp,Caprini:2006jb,Caprini:2007xq}
\begin{align}
\Lambda_{ij,kl}(\textbf{k})
\left\langle 
\ta_{\textbf{k}_{1}}^{ij}(t_{1}) \ta_{\textbf{k}_{2}}^{kl}(t_{2})
\right\rangle 
&=
\wAv^2
\int \dbar3{q_1} \dbar3{q_2}
\left[
\vev{\vfft^i_{\bq_1}(t_1)\vfft^{k}_{\bq_2}(t_2) }  \vev{\vfft^{*j}_{\bqtil_1}(t_1)\vfft^{*l}_{\bqtil_2}(t_2) } \right.\\ 
&+ 
\left. \vev{\vfft^i_{\bq_1}(t_1)\vfft^{*l}_{\bqtil_2}(t_2) }  \vev{\vfft^{*j}_{\bqtil_1}(t_1)\vfft^{k}_{\bq_2}(t_2) } 
\right]
\Lambda_{ij,kl}(\textbf{k}_1)
\end{align}
where $\bqtil_{1,2} = \bq_{1,2} - \bk$, and the third term in the Wick expansion has been dropped, as it is removed by the transverse-traceless projector. 

It then follows that 
\begin{align}
\uetcTen(k;t_1,t_2) 
&=  \wAv^2  \int \dbar3{q} \frac{q^2}{\qtil^2} (1-\mu^2)^2
G(q,t_1,t_2)G(\qtil,t_1,t_2),
\label{e:SSuetcInt}
\end{align}
where $\bqtil = \bq - \bk$ and $\mu = \hat{\bq}\cdot\hat{\bk}$, and we have used 
\begin{align}
\Lambda_{ij.kl}(\textbf{k})\hat{q}^{i}\hat{\tilde{q}}^j\hat{\tilde{q}}^k\hat{q}^{l}=\frac{1}{2}(1-\mu^{2})^{2}\frac{q^{2}}{\tilde{q}^{2}}.
\end{align}
Any non-Gaussianity in the velocity field will lead to an extra contribution from the connected four-point correlator; 
we will assume that this is negligible. 

When the velocity field is caused by sound waves, and velocities are non-relativistic, the fluid variables obey a linearised wave equation 
\bea
\frac{\dot \enDen}{w} + \pa_j v^j &=& 0, \\
\dot v^i + \frac{\pa^i p}{w} &=& 0 .
\eea
Defining an energy fluctuation variable
\ben
\label{e:EneFluVar}
\la(x) = \frac{\enDen(x) - \bar{\enDen}}{\bar{w}} ,
\een
where $\bar{w}$ and $\bar\enDen$ are the mean enthalpy and energy densities,
we can write in Fourier space
\bea
\label{e:LinEOMa}
\dot \lafft_\bq + iq_j \vfft_\bq^j &=& 0, \\
\label{e:LinEOMb}
\dot \vfft_\bq^i + \cs^2 iq^i \lafft_\bq &=& 0 . 
\eea
The general solution for the velocity consists of a superposition of plane waves:
\begin{align}
v^{i}(\textbf{x},\ t)=\int\frac{d^{3}q}{(2\pi)^{3}}(v_{\textbf{q}}^{i}e^{-i\omega t+i\textbf{q}\cdot \textbf{x}}+v_{\textbf{q}}^{*i}e^{i\om t-i\textbf{q}\cdot \textbf{x}})
\label{e:vPlaWav}
\end{align}
where $\omega = \cs q$.
Note the distinction between the plane wave amplitudes $\vq^i$ and the 
Fourier transform of the velocity field  $\vfft^i_\bq(t)$. 
Similarly, $\la(\bx,t)$, with Fourier transform $\lafft_\bq(t)$, can also be expanded in plane wave coefficients $\la_\bq$.

Being longitudinal, the plane wave coefficients for the velocity can be written
\ben
v_{\textbf{q}}^{i} = \hat{q}^i v_{\textbf{q}},
\een
where $\hat{q}^i = q^i/ q$ and $q = |\bq|$.
By taking the Fourier transform of (\ref{e:vPlaWav}) 
we find that at some initial time $\tInit$ the plane wave amplitude is 
\ben
\label{e:PlaWavCoeCor}
\vel{\bq} = \half \left( \hat{q}^i\tilde{v}_\bq^i(\tInit) - \cs \lafft_\bq(\tInit) \right) e^{i \om \tInit},
\een
This expression replaces Eq.~(13) in Ref.~\cite{Hindmarsh:2016lnk},  where instead of 
$-\cs \lafft_\bq(\tInit)$, the expression $ + \dot\vfft_\bq(\tInit)/i\om$ was used. 
The reason for preferring (\ref{e:PlaWavCoeCor}) is discussed in Appendix \ref{s:AppIniCon}.

The plane wave coefficients $\vel{\bq_1}$ and $\vel{\bq_2}^*$ are not independent, 
and we write
\ben
\vel{\bq} = \vel{\bq}^{*} e^{2 i \theta_q(\tInit)},
\een
where $\th_{q} = \om\tInit + \varphi_v$, and $\varphi_v = \arg(\vel{\bq})$.
Then 
one can show that 
\begin{align}
\label{e:VuetcFull}
G(q,t_1,t_2) &=   
2 \SpecDen{v}(q)  \cos[\om(t_1-t_2)]  
+
2 Q_v(q)  \cos[\om(t_1 + t_2 ) - 2\bar{\th}_{q}] 
\end{align}
where 
$\SpecDen{v}$ and $Q_v$ 
are spectral densities of the plane wave amplitudes, defined from 
\bea
\vev{\vel{\bq_1} \vel{\bq_2}^{*} } 
&=& \SpecDen{v}(q_1)\debar3{\bq_1 - \bq_2}, \\
\vev{\vel{\bq_1} \vel{-\bq_2} } 
&=& Q_v(q_1) e^{2 i \bar{\th}_{q_1} } \debar3{\bq_1 - \bq_2}. 
\eea
The equal time correlator $G(q,t,t)$ is non-negative, and so $|Q_v(q)| \le P_v(q)$.

The random process which generates the sound waves is the collision of bubbles, resulting in the removal of the forcing term which maintains the self-similar velocity field around the bubble. Bubble collisions occur at a range of times, and so the ensemble average includes an average over 
initial times $\tInit$ at which the velocity field becomes a freely propagating sound wave. Without being specific about the distribution, one can see from (\ref{e:PlaWavCoeCor}) that in constructing $Q_v(q)$ an average over $e^{2 i\om \tInit}$ is taken.  This factor is not present in the construction of $P_v(q)$.  One therefore expects $Q_v(q)$ to be less than $P_v(q)$, and significantly less for frequencies $\om \gg 1/\De \tInit$, where $\De \tInit$ is the width of the probability distribution for $\tInit$, the duration of the phase transition.
We will therefore neglect $Q_v$ in comparison to $P_v$, and write 
\ben
\label{e:VelUETC}
G(q,t_1,t_2) = 2P_v(q) D(\om,t_1,t_2),
\een
which defines the decoherence function $D$, approximately given by 
\ben
D(\om,t_1,t_2) \simeq \cos[\om(t_1 - t_2)]. 
\een
Hence the velocity correlator is stationary (depends only on $t_1 - t_2$) at high frequencies. 
The argument that the velocity correlator is stationary breaks down for low frequencies, $\om \ll 1/\De \tInit$, and we should expect to see some dependence on $t_1+t_2$ there as well.  However, we will see that these terms contribute subdominantly to the gravitational wave power spectrum.

Substituting the expression for the velocity UETC (\ref{e:VelUETC}) into the shear stress UETC, we find 
\begin{align}
\uetcTen(k,\ t_{1},\ t_{2})= 4 \wAv^{2}\int\frac{d^{3}q}{(2\pi)^{3}}\frac{q^{2}}{\tilde{q}^{2}}(1-\mu^{2})^{2}P_{v}(q)P_{v}(\tilde{q})
\cos(\om t_-) \cos(\omtil t_-),
\end{align}
where $t_- = t_1 - t_2$.
We can perform an angular integration, and change the integration over $\mu$ to an integration over $\qtil$, using 
\ben
\mu = \frac{q^2 + k^2 - \qtil^2}{2kq}.
\een
The result is
\begin{align}
\label{e:SSMuetc}
\uetcTen(k,\ t_{1},\ t_{2})= \frac{4 \wAv^{2}}{4\pi^2 k}
\int_0^\infty dq \int_{|q-k|}^{q+k} d\qtil q \qtil \frac{q^2}{\qtil^2} (1 - \mu^2)^2  
P_{v}(q)P_{v}(\tilde{q})
\cos(\om t_-) \cos(\omtil t_-). 
\end{align}
In view of the discussion around the non-stationary behaviour of the velocity correlations at low wavelengths, we should also expect to see a non-stationary behaviour in the shear stress UETC for $k \De \tInit \ll 1$.

\subsection{Gravitational wave power spectrum from sound waves}

By substituting the UETC from a random field of sound waves (\ref{e:SSMuetc}) into the general expression 
for the spectral density of $\dot{h}$ (\ref{e:GWPSfromUETC}),
we get 
\begin{align}
P_{\dot{h}}(k,t)&=(16\pi G)^2 \frac{4 \wAv^{2}}{4\pi^2 k}
\int_0^\infty dq \int_{|q-k|}^{q+k} d\qtil q \qtil \frac{q^2}{\qtil^2} (1 - \mu^2)^2  
P_{v}(q)P_{v}(\tilde{q})\Delta(t,k,q,\tilde{q}), 
\end{align}
where
\begin{align}
\Delta(t,k,q,\tilde{q}) &\simeq \half \int_{0}^t dt_1 \int_{0}^t dt_2 \cos(kt_-)\cos(\omega t_-)\cos(\tilde{\omega}t_-).
\end{align}
Defining $t_+ = (t_1+t_2)/2$, we have
\begin{align}
\Delta(t,k,q,\tilde{q}) &= \half \int_{0}^t dt_+ \int_{-2t_+}^{2t_+} dt_- \cos(kt_-)\cos(\omega t_-)\cos(\tilde{\omega}t_-).
\end{align}
Hence $\De(t,k,q,\tilde{q})$ grows at a rate 
\begin{align}
\dot\Delta(t,k,q,\tilde{q}) &= \half \int_{-2t}^{2t} dt_- \cos(kt_-)\cos(\omega t_-)\cos(\tilde{\omega}t_-), 
\end{align}
which asymptotes at large times to a $\de$-function, 
\ben
\dot\Delta(t,k,q,\tilde{q})  \to \frac{\pi}{8} \sum_{\pm\pm\pm} \de( \pm k \pm \om \pm \omtil).
\een
Only $k - \om - \omtil$ can vanish, and so the growth rate of the spectral density asymptotes to 
\begin{align}
\label{e:SpeDenHdoAsy}
\dot P_{\dot{h}}(k,t)\to(16\pi G)^2 \frac{4 \wAv^{2}}{4\pi^2 k}
\int_0^\infty dq \int_{|q-k|}^{q+k} d\qtil q \qtil \frac{q^2}{\qtil^2} (1 - \mu^2)^2  
P_{v}(q)P_{v}(\tilde{q})  \frac{\pi}{4} \delta(k-\omega-\tilde{\omega})
\end{align}
It is straightforward to see that only the stationary terms in the shear stress UETC contribute to this secular growth;
the non-stationary contributions inherited from the neglected term $2Q_v(q)\cos(2\om t_+ - 2\bar{\th}_\bq)$ in the velocity UETC (\ref{e:VuetcFull}) 
will produce only oscillatory contributions to the gravitational wave spectral density $\dot P_{\dot{h}}(k,t)$, and can be dropped.

We can now perform the integral over $\qtil$ in (\ref{e:SpeDenHdoAsy}), 
ending up with an expression for the asymptotic growth rate of the spectral density, which we indicate by dropping the indication of time dependence, 
\begin{align}
\dot P_{\dot{h}}(k) = (16 \pi G)^2   \frac{\wAv ^{2}}{4\pi k\cs} \int^{q_+}_{q_-} dq\   \bigg(\frac{q^{3}}{\tilde{q}}\bigg)(1-\mu^{2})^{2}P_{v}(q)P_{v}(\tilde{q}). 
\label{e:SpeDenHdo}
\end{align}
Here, 
\begin{align}
\tilde{q}=k/c_s-q,\ \ \ q_\pm=\frac{k(1\pm c_s)}{2c_s},\ \ \ \mu=\frac{2qc_s-k(1-c_s^2)}{2qc_s^2} .
\end{align}
Now we introduce a scaled velocity spectral density $\tilde{P}_v$ from 
\ben
P_v(q)=L_\text{f}^3 \fluidV^2 \tilde{P}_v(qL_\text{f}), 
\een
where $L_\text{f}$ is a length scale in the velocity field, 
and $\fluidV$ is the RMS fluid velocity.\footnote{In the numerical simulations \cite{Hindmarsh:2015qta,Hindmarsh:2017gnf} $\fluidV$ denotes the enthalpy-weighted RMS fluid velocity (see Appendix \ref{s:HydExpBub}). For the non-relativistic flows we consider, the difference is negligible.}  
We further define $y=k L_\text{f}$ and $z=qL_\text{f}$, so that the asymptotic growth rate of the spectral density is 
\begin{align}
\dot P_{\dot{h}}(y)= \left (16 \pi G \wAv \fluidV^2 \right)^2\frac{L_\text{f}^4}{4\pi y c_s}\int_{z_-}^{z_+}dz\bigg(\frac{z^3}{y/c_s-z}\bigg)(1-\mu^2)^2\tilde{P}_v(z)\tilde{P}_v(y/c_s-z)
\end{align}
where $z_\pm=y(1\pm c_s)/{2c_s}$. Now since the gravitational wave power spectrum is 
\begin{align}
\PspecGW  = \frac{1}{12 H^2}\frac{k^3}{2\pi^2}P_{\dot{h}}, 
\end{align}
its growth rate relative to the Hubble rate is 
\begin{align}
\label{e:GWGroRat}
\PspecGW' = 
\frac{1}{H}\frac{d}{dt} \PspecGW
&= 3 \left(\AdInd \fluidV^2 \right)^2 \left( HL_\text{f} \right) \frac{(k\fluidL)^3}{2\pi^2}\SpecDenGW(kL_\text{f})
\end{align}
with a dimensionless spectral density function  
\begin{align}
\SpecDenGW(y)=\frac{1}{4 \pi y \cs}\left(\frac{1-c_s^2}{c^2_s}\right)^2\int_{z_-}^{z_+}\frac{dz}{z}\frac{(z-z_+)^2(z-z_-)^2}{(z_+ + z_- - z)}\bar{P}_v(z)\bar{P}_v(z_+ + z_- - z). 
\end{align}
Hence, a stationary velocity power spectrum with a lifetime $\tLife$ generates a gravitational wave power spectrum 
\ben
\PspecGW(k) = 3 \left(\AdInd \fluidV^2 \right)^2 (H \tLife) \left( HL_\text{f} \right) \frac{(k\fluidL)^3}{2\pi^2}\tilde{P}_{GW}(kL_\text{f})
\een
Here we see the two characteristic time scales involved in gravitational wave generation, the lifetime of the source, and a coherence time $\tCoh \sim\fluidL$, which is set by the characteristic length scale of the sound waves.

The two important scales relevant for estimating the lifetime of the velocity field are the Hubble time $H^{-1}$ and the timescale for non-linear behaviour, $\tNL = \fluidL/\fluidV$.  If $H \tNL\gg 1$, it can be shown that $H \tLife \to1$ \cite{Hindmarsh:2015qta}, 
i.e. that the effective source lifetime is precisely the Hubble time.
A better estimate of $\tLife$ for higher speed flows, which may be the most relevant ones \cite{Ellis:2018mja}, is clearly vital.

Now we turn to finding an expression for the velocity power spectrum, which is developed from the acoustic sound shell model examined in the next section.

\subsection{Length scales and power laws}
\label{ss:PowLaw}

The characteristic length scale $\fluidL$ of the sound wave spectrum is set by the mean bubble separation $\Rbc$.  There is also another scale in the sound waves, which is the width of the fluid shells surrounding the expanding bubbles, and a factor $|\cs - \vw|$ smaller than $\Rbc$.  We will see that both scales are visible in the gravitational wave power spectrum.

The power laws in the velocity power spectrum can be translated into power laws for the gravitational wave power spectrum.
If the velocity power spectrum goes as $k^n$, the velocity spectral density $P_v(k)$ goes as $k^{n-3}$. 
Consequently the shear stress equal time correlator will go as $k^{2n - 3}$, as it is the integral of a product of spectral densities. 
The dimensionless gravitational wave spectral density $\SpecDenGW$ will go as $k^{2n-4}$, due to the presence of $k^{-1}$ in the expression for the spectral density of $\dot{h}$ (\ref{e:SpeDenHdo}).  
Hence the gravitational wave power spectrum goes as $\mathcal{P}_\text{gw}(k)\propto k^{2n-1}$. 

It was incorrectly\footnote{See Appendix \ref{s:AppIniCon}.} 
argued in \cite{Hindmarsh:2016lnk} that $P_v(k)$ should go as $k^0$ for $k\fluidL \ll 1$, as the velocity field is uncorrelated at large distances. That would imply at low $k\fluidL$ we will have that $\mathcal{P}_v(k)\propto k^{3}$  
and $\PspecGW(k)\propto k^5$.

However, it was pointed out in \cite{Durrer:2003ja} that analyticity constrains power spectra of causal fields at large scales, where causal means that two-point correlations should vanish outside the cosmological horizon. A causal power spectrum is analytic in $q^i$, and so the power spectrum of a vector field with the structure $\hat{q}^{i}\hat{q}^{j} G(q)$ should be bounded by a quantity proportional to $q^2$ at low $q$.
The analyticity constraint therefore changes the low wavenumber power law index to $n=5$. We should therefore expect $\mathcal{P}_v(k)\propto k^{5}$  
and $\PspecGW(k)\propto k^9$.

We will carry out a detailed comparison with the power spectra in the numerical simulations elsewhere; here we note that the 
low-wavenumber velocity power spectra in the numerical simulations do 
appear to be steeper than $n=3$  (see Fig.~3c of Ref.~\cite{Hindmarsh:2017gnf}). The gravitational wave power spectra at low $k$ are dominated by long-wavelength modes generated during the bubble collision phase, where different power laws are found \cite{Huber:2008hg,Cutting:2018tjt}, and 
so the gravitational waves produced during the acoustic phase are obscured.

\section{Velocity power spectrum from colliding bubbles}
\begin{figure}[hb]
   \centering
   \includegraphics[width=0.49\textwidth]{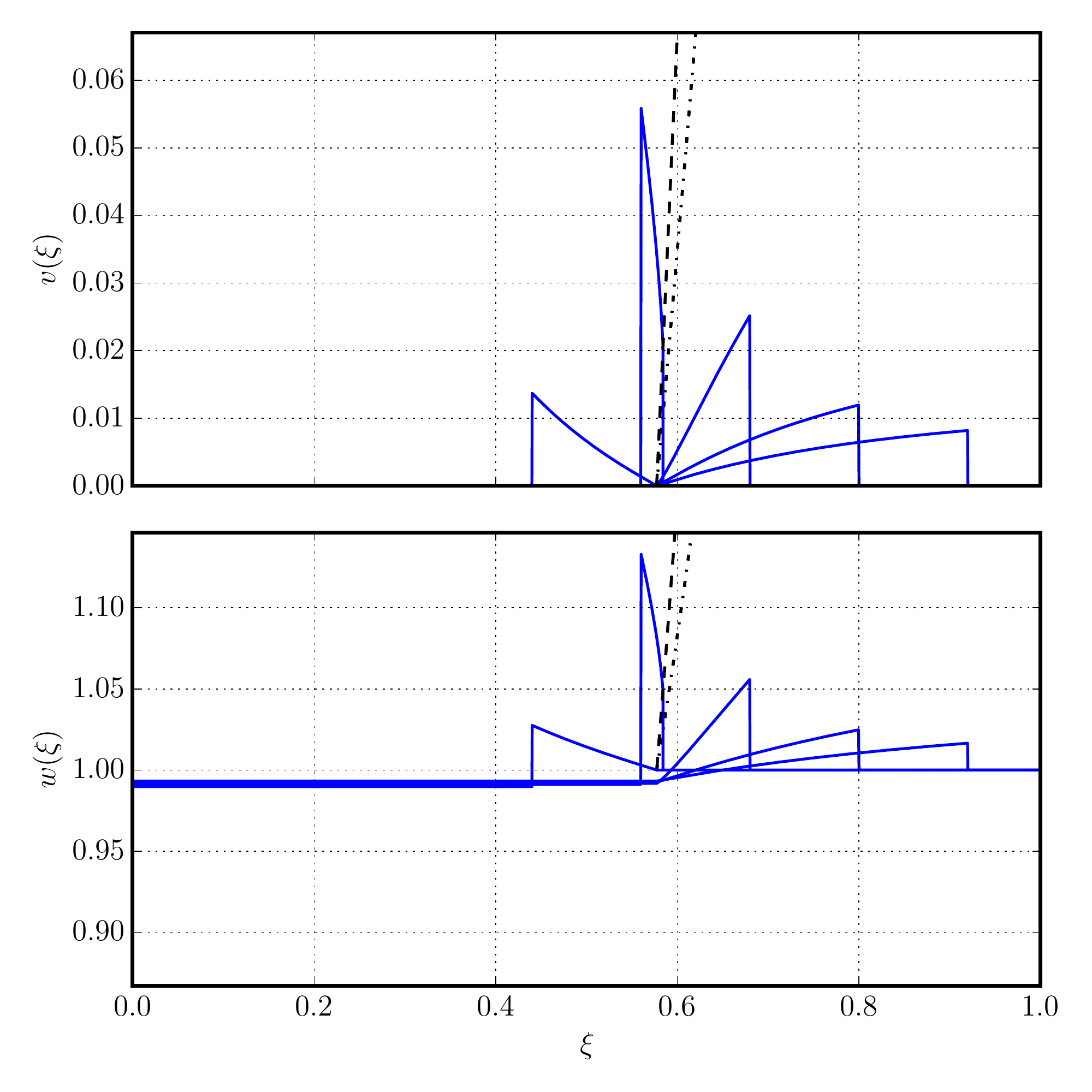} 
   \includegraphics[width=0.49\textwidth]{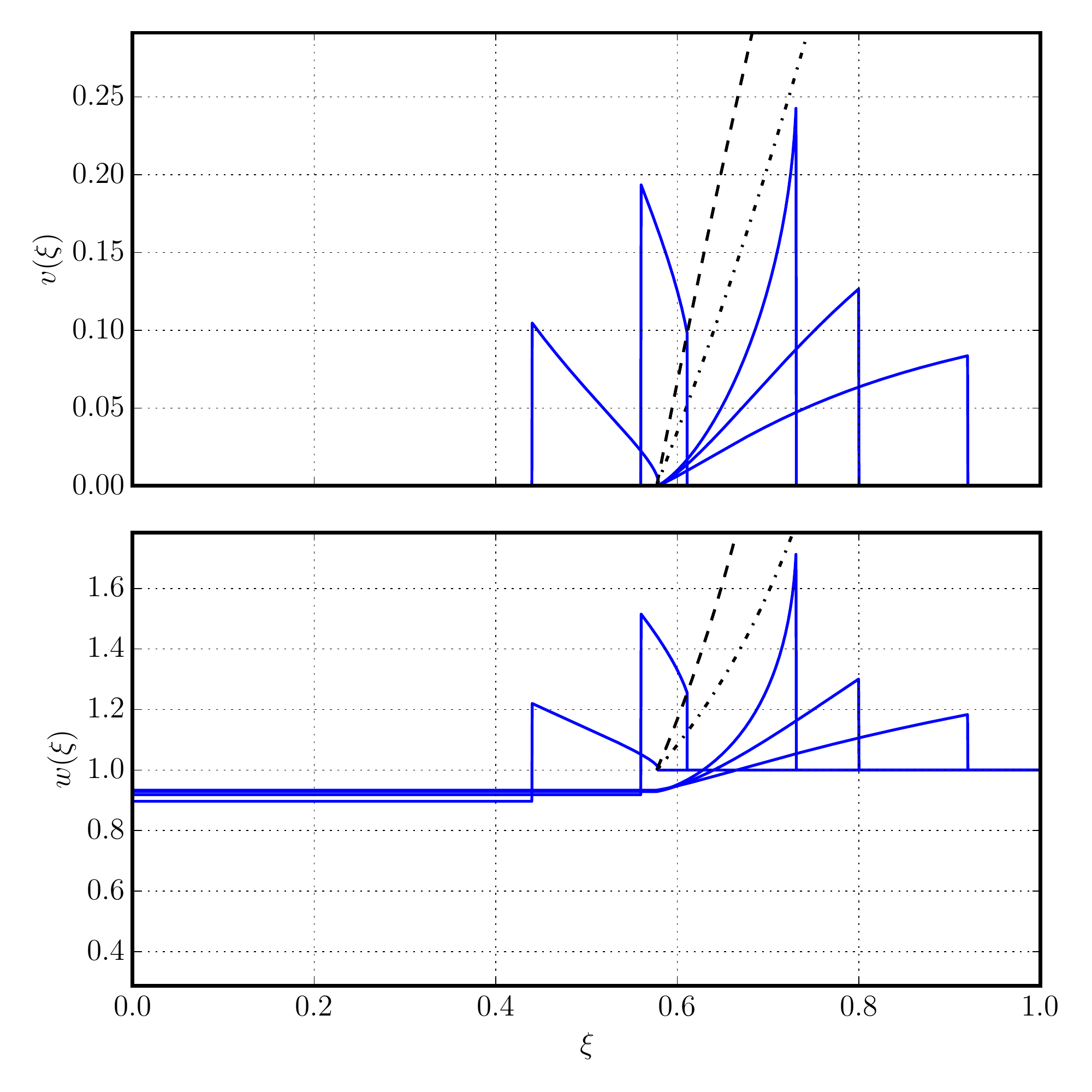} 
   \caption{Self-similar radial velocity $v$ and enthalpy density $w$ profiles as functions of the scaled radius $\xi = R/T$, 
   where $R$ is the distance from the bubble centre and $T$ is the time since nucleation, 
   for wall speeds $\vw = [0.44,0.56,0.68,0.80,0.92]$ 
   and phase transition strengths $\alpha_n = 0.0046$ (left) and $\alpha_n = 0.05$ (right, where the third wall speed 
   is adjusted to $\vw = 0.731$ to better match the simulations in \cite{Hindmarsh:2017gnf}).
   The black dashed line is the curve in $(\xi,v)$ and $(\xi,w)$ planes where shocks must occur.
   The dash-dot line indicates the maximum possible fluid velocity behind a wall in all cases, 
   and the maximum possible enthalpy behind a detonation. 
   See Appendix \ref{s:HydExpBub} for more details.
   }
   \label{f:BubPro}
\end{figure}

In the Sound Shell Model, the velocity field is supposed to be a superposition of the self-similar velocity and enthalpy density profiles generated by randomly placed expanding bubbles which have nucleated at different times $t^{(n)}$. 
Examples of these profiles are shown in Fig.~\ref{f:BubPro}; self-similarity means that they are functions of $\xi = R/T$, where 
$R$ is the distance from the centre, and $T$ is the time since nucleation.

As they collide, the forcing by the scalar field which produced the self-similar profile is removed and the self-similar profile becomes the initial condition for freely propagating sound waves. The collision dynamics is rather complicated; as a first approximation we will suppose that the bubble completely disappears when half of it has merged with the advancing region of the stable phase.  

The resulting velocity field is then a convolution of the single-bubble velocity field with the distribution of bubble lifetimes.
We will see that this crude model produces a velocity power spectrum which works quite well for non-relativistic velocity fields.

\subsection{Velocity field from superposition of single-bubble fluid shells}
\label{ss:VelFieOneBub}

\begin{figure}[t]
   \centering
   \includegraphics[width=0.49\textwidth]{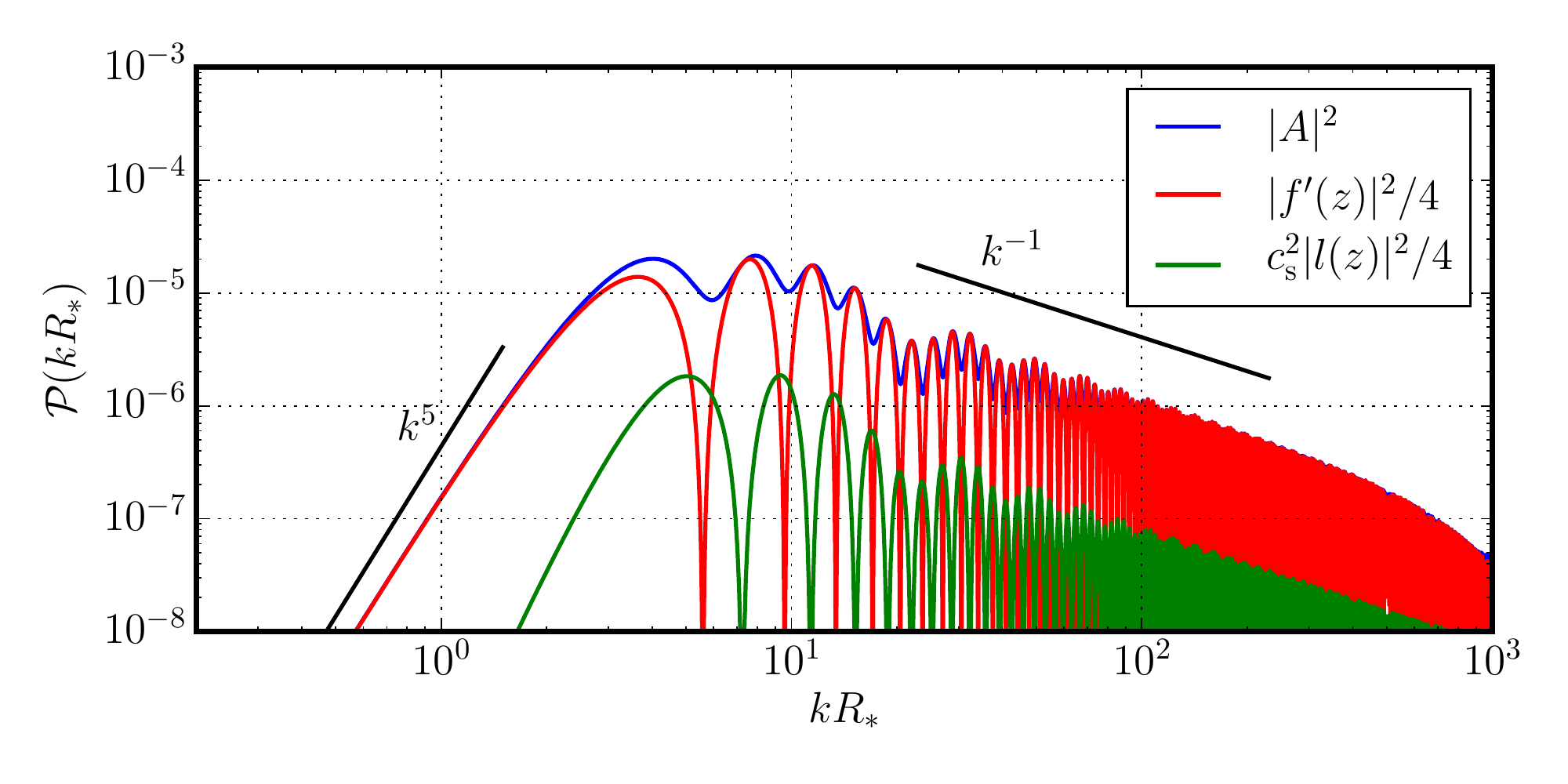} 
   \includegraphics[width=0.49\textwidth]{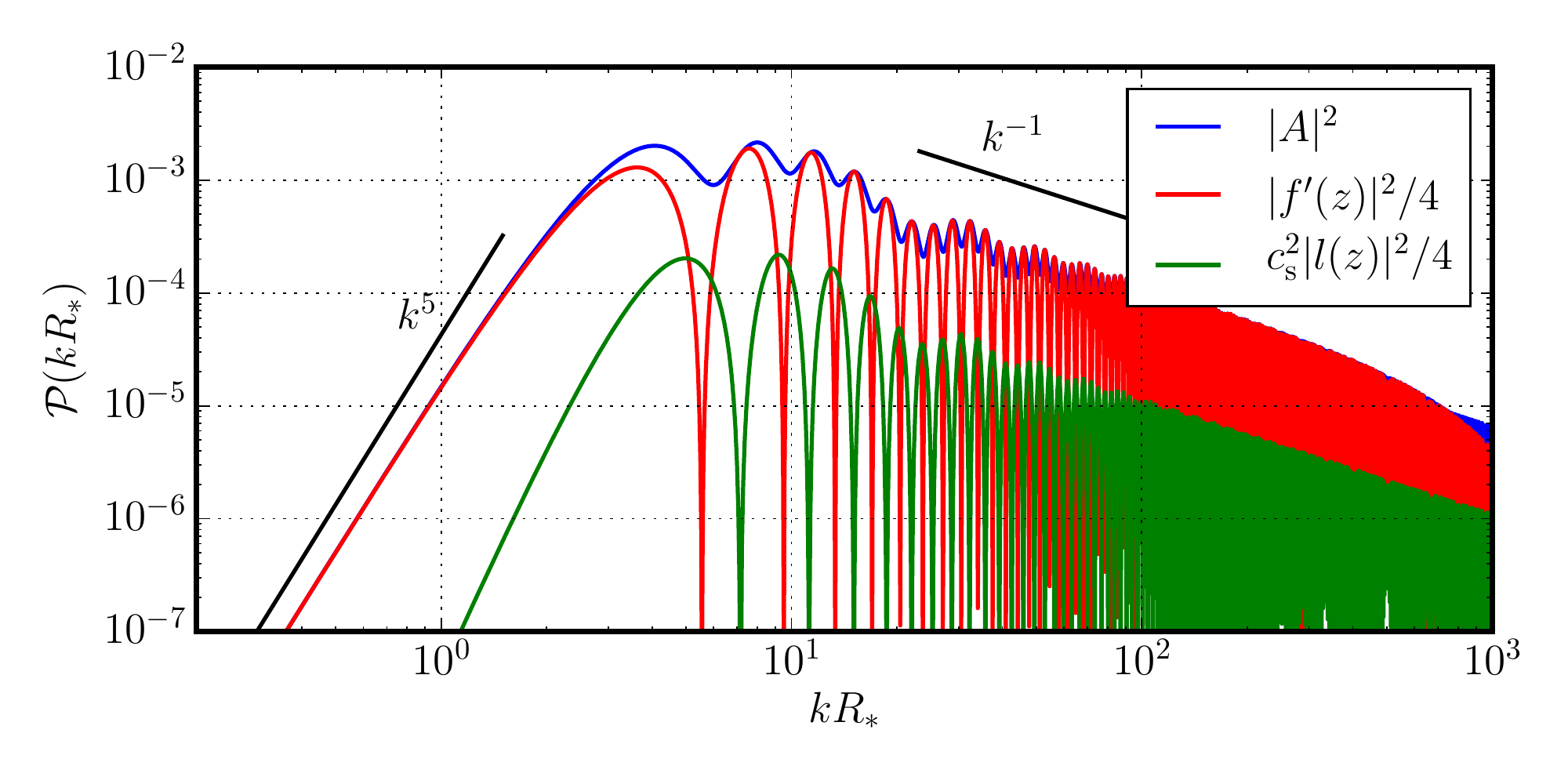} \\
   \includegraphics[width=0.49\textwidth]{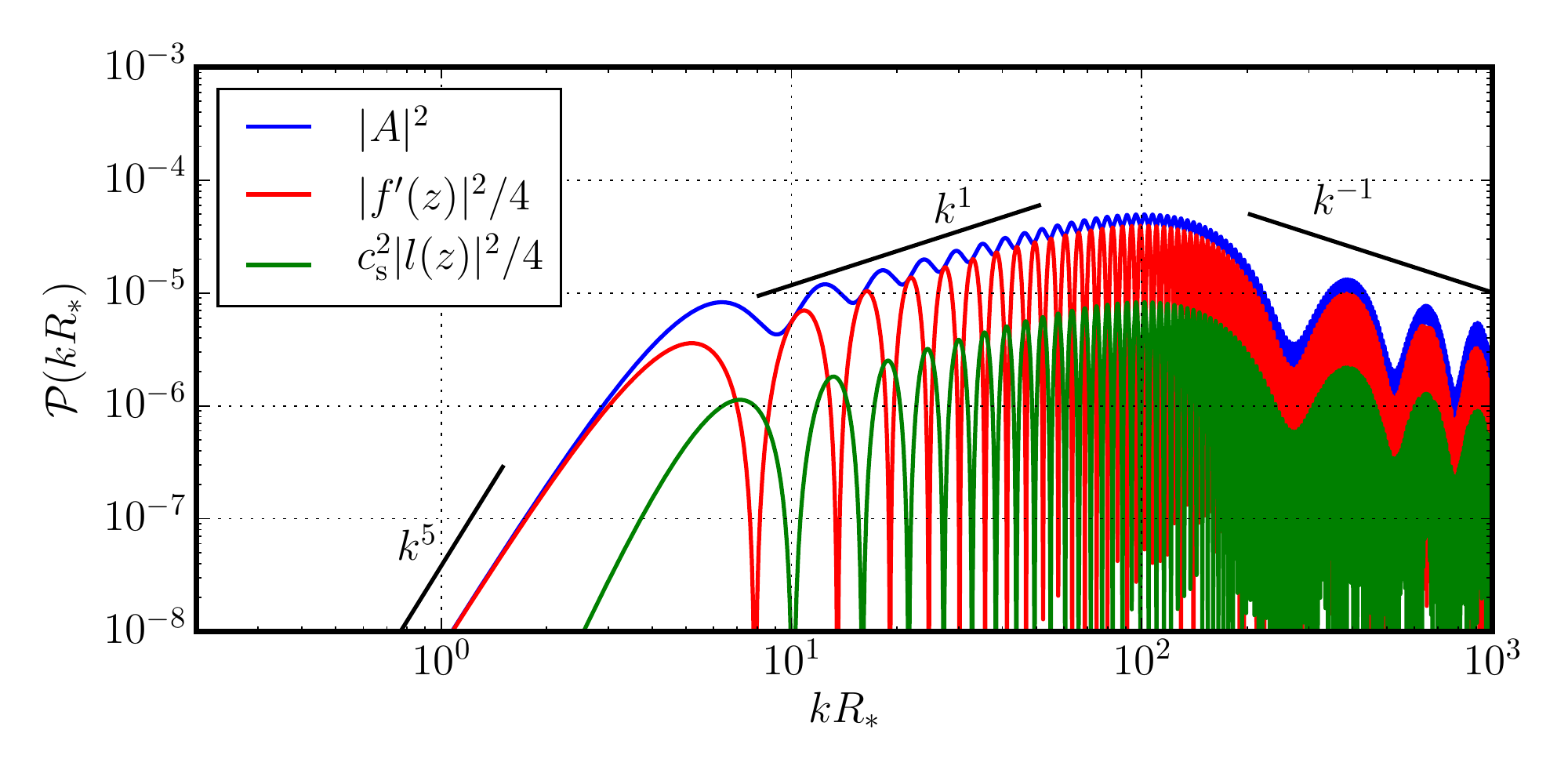} 
   \includegraphics[width=0.49\textwidth]{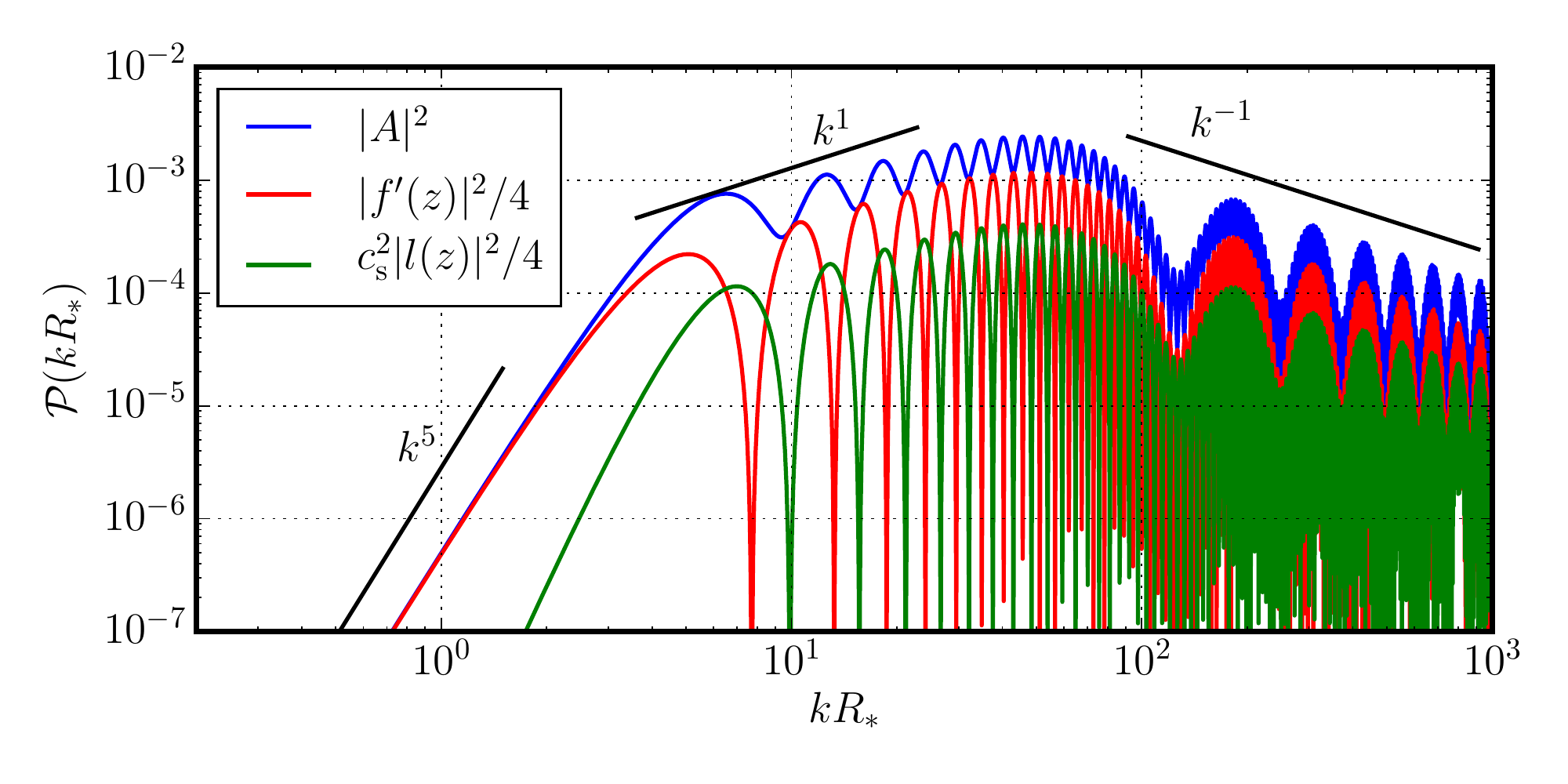} \\
   \includegraphics[width=0.49\textwidth]{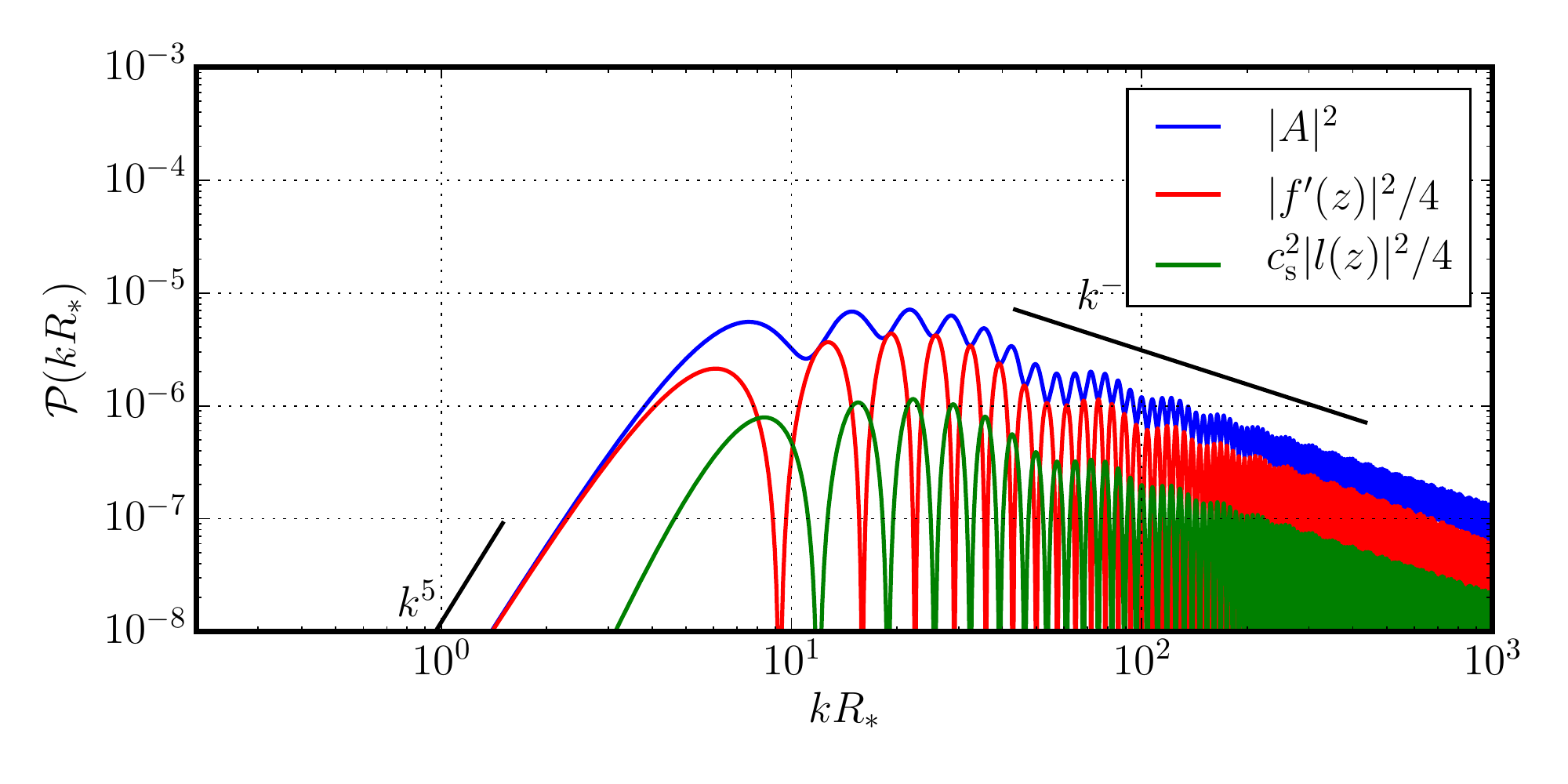} 
   \includegraphics[width=0.49\textwidth]{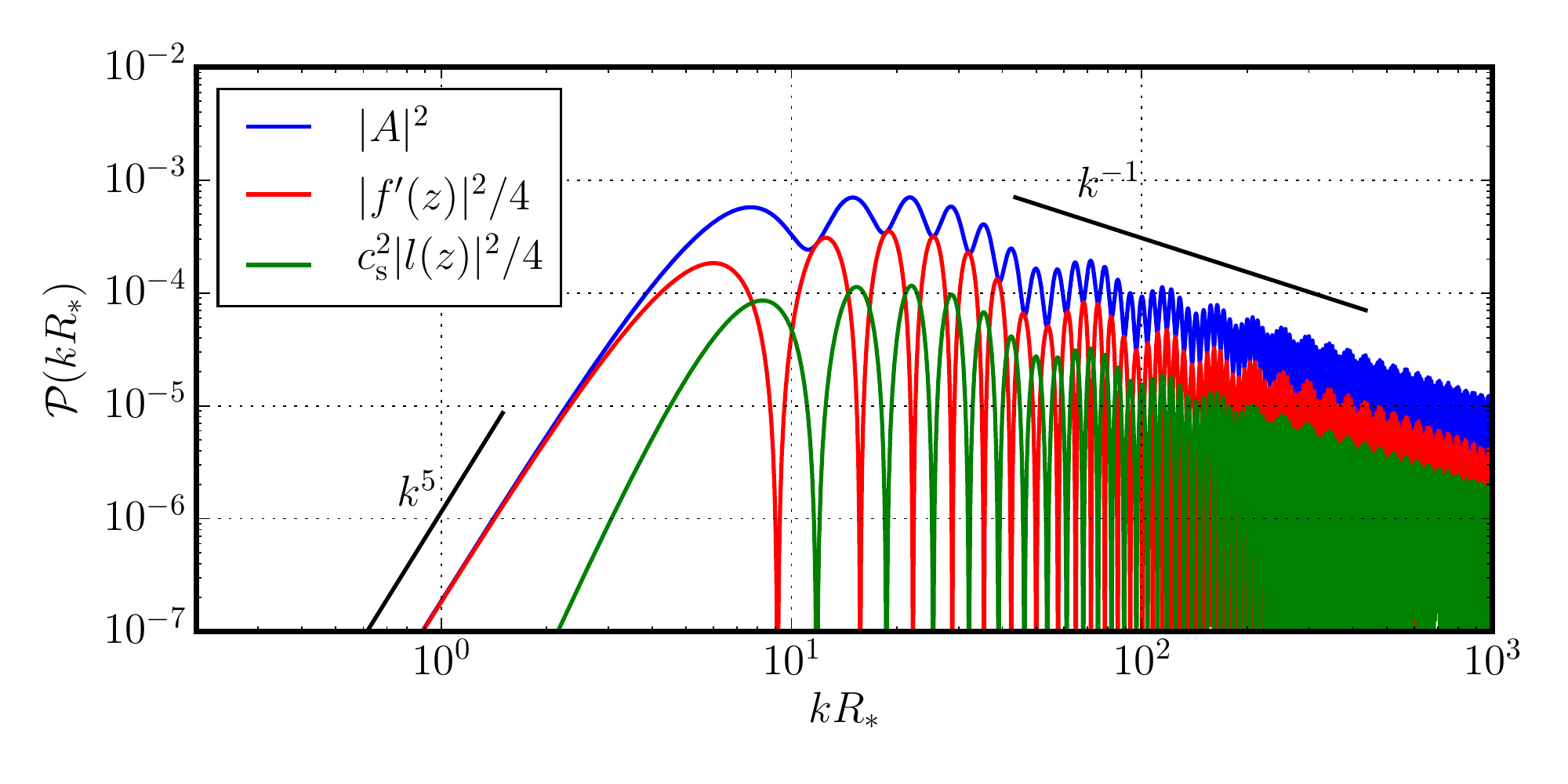} 
   \caption{Single-bubble plane wave power spectra.   Left are weak strength phase transitions, with (top to bottom) $\vw = 0.92$, $0.56$
  and $0.44$.  Right are intermediate phase transitions, with $\vw = 0.92$, $0.56$ and $\vw = 0.44$.  
  See Eqs.~(\ref{e:fDef}), (\ref{e:LamDef}) and  (\ref{e:ADef}) for definitions of the quantities plotted.
  }
   \label{f:OneBubPS}
\end{figure}

We write the velocity field in a large volume $\Vol$ as the sum of velocity fields produced by $\Nb$ individual bubbles,  
\begin{align}
v_i(\textbf{x},t)=\sum\limits_{n=1}^{\Nb} v_i^{(n)}(\textbf{x},t).
\end{align}
Since the velocity field of the $n${th} bubble centered at $\textbf{x}^{(n)}$ is radial, we write:
\begin{align}
v_i^{(n)}(\textbf{x},t) = \frac{R_i^{(n)}}{R^{(n)}}v_{\text{ip}}(\xi)
\end{align}
where $R_i^{(n)}=x_i-x_i^{(n)}$ is the displacement of the $n$th bubble surface from its centre, 
$T^{(n)}=t-t^{(n)}$ is the time since the nucleation of the $n$th bubble, and $\xi=R^{(n)}/T^{(n)}$.
The Fourier transform of the velocity field is 
\begin{align}
\tilde{v}_i^{(n)}(\textbf{q},t)=\int d^3x v_i^{(n)}(\textbf{x},t)e^{-i\textbf{q}\cdot \textbf{x}}=e^{-i\textbf{q}\cdot \textbf{x}^{(n)}}\int d^3R^{(n)}\frac{R^{(n)}_i}{R^{(n)}}v_{\text{ip}}(\xi)e^{-i\textbf{q}\cdot \textbf{R}^{(n)}} .
\end{align}
By defining  $z^{i}=q^{i}T^{(n)}$, and changing the integration variable from $R^{(n)}$ to $\xi$, we rewrite the previous equation as 
\begin{align}
\tilde{v}_{i}^{(n)}(\textbf{q},t)=e^{-i\textbf{q}\cdot \textbf{x}^{(n)}}i(T^{(n)})^{3}\frac{\partial}{\partial z_{i}}\bigg(\int d^3\xi\frac{1}{\xi}v_{\text{ip}}(\xi)e^{-iz^i\xi^i}\bigg)
\end{align}
Now we perform the angular integration and define the function $f(z)$, 
\begin{align}
\label{e:fDef}
f(z)&= \int d^{3}\xi\frac{1}{\xi}v_{\text{ip}}(\xi)e^{-iz^{i}\xi^{i}}=\frac{4\pi}{z}\int_{0}^{\infty}d\xi v_{\text{ip}}(\xi)\sin(z\xi)
\end{align}
in terms of which the Fourier-transformed velocity profile is 
\begin{align}
\label{e:vFFT}
\tilde{v}_{i}^{(n)}(\textbf{q},t)=e^{-i\textbf{q}\cdot \textbf{x}^{(n)}}i(T^{(n)})^{3}\hat{z}^i f'(z) .
\end{align}
In order to complete the computation of the initial conditions for the sound waves we also need $\lafft(\textbf{q},t)$, the Fourier-transformed energy perturbation variable (\ref{e:EneFluVar}).
Performing similar steps, we find
\ben
\lafft^{(n)}(\textbf{q},t) = e^{-i\textbf{q}\cdot \textbf{x}^{(n)}}(T^{(n)})^{3} l(z), 
\een
with 
\ben
\label{e:LamDef}
l(z) = \frac{4\pi}{z} \int_0^\infty d\xi \la_\text{ip}(\xi) \xi \sin(z\xi). 
\een
We suppose that the entire fluid perturbation around the $n$th bubble becomes the initial condition for a sound wave at a time
$\tInit^{(n)}$.
It follows that the contribution of this bubble to the plane wave amplitude is 
\begin{align}
v_{\textbf{q},i}^{(n)}=i(\TInit^{(n)})^3 \hat{z}_ie^{i\omega \tInit-i\textbf{q}\cdot \textbf{x}^{(n)}}A(z), 
\end{align}
where $\TInit^{(n)} = \tInit^{(n)} - t^{(n)}$, the lifetime of the bubble, and 
\begin{align}
\label{e:ADef}
A(z)=\frac{1}{2}\left[f'(z)+ i \cs l(z)\right].
\end{align}
Note that $f'$ and $\la$ are real, so that
\ben
\label{e:A2Def}
|A(z)|^2 = \frac{1}{4}\left[(f'(z))^2+ (\cs l(z))^2\right].
\een
The function $|A(z)|^2$ contains information about the shape of the fluid shells.  The main features are
the wall speed, the fluid shell thickness, and the peak amplitude \cite{Hindmarsh:2016lnk}.
Because of the finite support between the sound speed, the wall speed and the shock speed, 
the functions are oscillatory, giving rise to the ``ringing'' observed in the fluid power spectra when bubbles are 
nucleated simultaneously \cite{Hindmarsh:2015qta,Hindmarsh:2017gnf}.
The functions $|A(z)|^2$, $|f'(z)|^2/4$ and $\cs^2|l(z)|^2/4$ are plotted for selected values of wall speed and transition strength 
in Fig.~\ref{f:OneBubPS}.
One can see how the spatial frequency of the ringing is set by the bubble size, modulated by the spatial frequency of 
corresponding to the shell thickness.  One can also see how the envelopes satisfy the power laws 
described in subsection \ref{ss:PowLaw}.

We can now compute the plane wave amplitude correlation function for $\Nb$ randomly-placed bubbles in a volume $\Vol$, 
\begin{align}
\left\langle v_{\textbf{q}_1}^{i}v_{\textbf{q}_2}^{*j}\right\rangle
&=\sum\limits_{m=1}^{\Nb}\sum\limits_{n=1}^{\Nb}  \left\langle (\TInit^{(m)})^3(\TInit^{(n)})^3\hat{z}^i\hat{z}'^{j}A(z)A^*(z') \right.\\
&\times \left.  e^{-i\textbf{q}_1\cdot\textbf{x}^{(m)}+i\textbf{q}_2\cdot\textbf{x}^{(n)}} e^{i(\om_1-\om_2)\tInit} \right\rangle ,
\end{align}
where the average is over the ensemble of bubble locations $\textbf{x}^{(n)}$, nucleation times $t^{(n)}$, and collision times $\tInit^{(n)}$.
Let us first average over locations of bubbles nucleated between $t'$ and $t' + dt'$, and colliding between $\tInit$ and $\tInit + d\tInit$, for which 
\begin{align}
 \sum\limits_{m=1}^N\sum\limits_{n=1}^N 
\left\langle e^{-i\textbf{q}_{1}\cdot \textbf{x}^{(m)}+i\textbf{q}_{2}\cdot \textbf{x}^{(n)}}\right\rangle
=
d^{2} P \frac{\Nb}{\Vol}  (2\pi)^{3}\delta(\textbf{q}_{1}-\textbf{q}_{2}),
\end{align}
where $d^2 P(t',\tInit)$ is the joint probability for nucleating and colliding in the given time ranges. The $\de$-function in the wave number forces $\om_1 = \om_2$, and therefore removes the dependence on the absolute collision time $\tInit$, leaving an average over bubble lifetimes $\TInit$. Denoting the probability density distribution of lifetimes $n(\TInit) = (\Nb/\Vol) dP(\TInit)/d\TInit$, we have 
\begin{align}
\left\langle v_{\textbf{q}_1}^{i}v_{\textbf{q}_2}^{*j}\right\rangle
&= \int d\TInit n(\TInit) \TInit^{6}
\hat{z}^{i}\hat{z}^{j}|A(z)|^{2}(2\pi)^{3}\delta(\textbf{q}_{1}-\textbf{q}_{2}).
\end{align}
We define the mean bubble separation $\Rbc$ by $\Rbc^3 = \Nb/\Vol$ in
the infinite volume limit. 
Hence, $\int n(\TInit)d\TInit=1/R_*^3$, 
and we can write 
\begin{align}
n(\TInit)d\TInit=\frac{\beta}{R_{*}^{3}}\nu(\beta \TInit)d\TInit
\end{align}
where $\nu(\beta T)$ is the bubble lifetime distribution function, 
and we have introduced a rate $\beta$, so that $\nu$ is dimensionless, and normalised so that $\int \nu (x) dx = 1$. 
As the notation suggests, it will be convenient to take this rate to be the nucleation rate parameter, defined after Eq.~\ref{e:hExp}, so that 
\ben
\beta = (8\pi)^{\frac13} \frac{\vw}{\Rbc}.
\een
This can be considered the definition of $\be$ in the case of simultaneous nucleation. 
Hence the spectral density of the plane wave components of the velocity field is 
\begin{align}
P_{v}(q)= 
\frac{1}{\be^6\Rbc^3}
\int d\tilde{T}\nu(\tilde{T})\tilde{T}^{6}|A(\tilde{T}q/\beta)|^{2}
\end{align}
where $\tilde{T}=\beta \TInit$. 
The velocity power spectrum is then 
\begin{align}
\label{e:VelPowSpe}
\Pspec{\vfft}(q) &= 2 \frac{q^3}{2\pi^2}P_v(q) = 
\frac{2}{(\be\Rbc)^3} \frac{1}{2\pi^2} \left( \frac{q}{\beta} \right)^3 
\int d\tilde{T}\nu(\tilde{T})\tilde{T}^{6}|A(\tilde{T}q/\beta)|^{2}.
\end{align}
We remind the reader that the factor two arises from the relationship between the plane wave amplitudes and the 
velocity Fourier transform (\ref{e:VelUETC}).

\subsection{Collision time distribution}

The last link in the chain of argument is the dimensionless collision time distribution, $\nu(\tilde{T})$. 
The whole notion of a single collision time for a bubble is clearly an over-simplification, as mentioned at the start of the last section.  
Different parts of the bubble wall collide at different times, and the collision surface forms a complicated weighted Voronoi tessellation of space.  Even before collision, in the case of deflagrations 
fluid shells will interact and change the pressure difference driving the wall, and hence the wall propagation speed \cite{Cutting:2019zws}.  
As the bubbles collide, the phase boundary shape becomes rather complicated, and contains regions of high curvature, which move faster than the walls of the uncollided bubbles. 

The velocity field generated by the complex collision dynamics will therefore not always resemble 
a superposition of colliding fragments of spherical shells. 
However, we will continue with this simplest idea and examine the results.

\begin{figure}[t]
 \centering
   \includegraphics[width=0.7\textwidth]{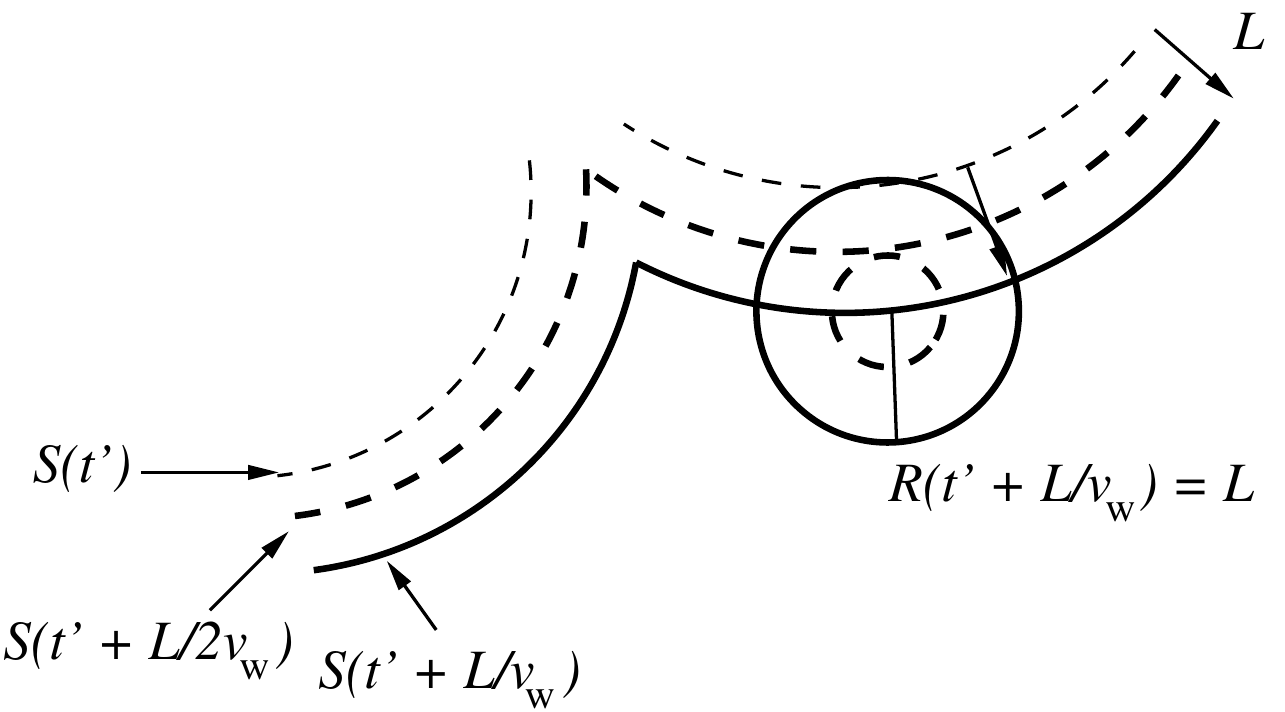} 
   \caption{Bubble collision.  A bubble nucleated at time $t'$ at distance $L$ from the advancing phase boundary $S(t')$ (thin dashed line). 
   The bubble wall and the phase boundary both move at speed $\vw$. 
   At time $t' + L/2\vw$ the bubble makes contact with the boundary (thick dashed lines).  
   At time $t' + L/\vw$ the boundary would have reached the nucleation site, had the nucleation not taken place,  
and approximately half the bubble has been destroyed (solid lines).  We take this time to mark the end of the bubble, 
so that its lifetime is $T = L/\vw$, when its radius is $R = L$.  
   }
   \label{f:BubCol}
\end{figure}

The key elements in the computation are 
the bubble nucleation rate per unit volume $p(t)$, 
the fractional volume remaining in the metastable phase $h(t)$, 
the area per unit volume of the phase boundary $\mcA(t)$,
and the speed at which the phase boundary advances $\vw$.

Bubbles nucleating within a distance $[L,L+dL]$ of the phase boundary, between times $[t',t'+dt']$, 
will have a first collision with the boundary between times $[t_1,t_1+dt'/2]$, where $t_1 = t' + L/2\vw$, and with radii in the range $[L/2,(L+dL)/2]$ (see Fig.~\ref{f:BubCol}).
The time between nucleation and first collision is $T_1= L/2\vw$. We will call this the first collision time of the bubble.

The phase boundary reaches the nucleation site at a time in the range $[t, t + dt']$, where $t = t' + L/\vw$, when the radius of the remaining part of the bubble $R$ is in the range  $[L,L+dL]$.  The time taken 
is $T= L/\vw$. We will take this to be the lifetime of the bubble.

Hence the density of bubbles which will collide with radii between $[R,R+dR]$ nucleated in the time interval $[t, t + dt']$, is 
the density of bubbles nucleated in the volume $\mcA(t' + L/\vw) dL$ in that time interval.  Given that those bubbles will have radius $R$ when they are ``destroyed'' after time $T = R/\vw$, we can write 
\ben
d^2 n = [ \mcA(t' + R/\vw) dR ] [ p(t') dt'].
\een
Hence the bubble size distribution is 
\ben
\frac{dn}{dR} =  \int_{\tc}^\infty \mcA(t' + R/\vw)   p(t') dt' ,
\een
where $\tc$ is the time the Universe reaches the critical temperature and bubble nucleation begins. 
The area per unit volume of the phase boundary is related to the rate of change of the volume fraction in the metastable phase, 
\ben
\mcA = - \frac{1}{\vw} \frac{dh}{dt}.
\een
We will now evaluate the bubble lifetime distribution in two scenarios for the bubble nucleation history.

\subsubsection{Exponential nucleation}
With exponential nucleation, we have from (\ref{e:hExp})
\ben
\mcA(t) = \frac{\be}{\vw } e^{\beta(t - \tf)} \exp\left[ - e^{\beta(t - \tf)} \right],
\een
where $\tf$ is the time at which $h = 1/e$.
We recall that the nucleation probability per unit volume rises approximately exponentially with time, and can be written 
\ben
p(t) = \pf e^{\beta(t - \tf)}.
\een
Hence
\ben
\frac{dn}{dR} =  \frac{\pf}{\vw} \be \int_{\tc}^\infty dt'  e^{\beta(t' - \tf + R/\vw)} \exp\left[ - e^{\beta(t' - \tf+ R/\vw)} \right]  e^{\beta(t' - \tf)}.
\een
Assuming that $\be\tf \gg 1$ and that $R \ll \tf$ (i.e. that the nucleation rate is much larger than the Hubble rate, and that bubbles are much smaller than the Hubble length), 
\bea
\frac{dn}{dR} &=&  \frac{\pf}{\vw} e^{-\beta R/\vw}
 \be \int_{\tc}^\infty dt'   \left[e^{\beta(t' - \tf + R/\vw)}\right]^2 \exp\left[ - e^{\beta(t' - \tf+ R/\vw)} \right] , 
 \nonumber\\
&=&  \frac{\be \nb}{\vw} e^{-\beta R/\vw},
\eea
where we have used the relation $\nb = \pf/\be$ (\ref{e:NbPf})..
The lifetime of a bubble is $T = R/\vw$.  Thus the fraction of bubbles with lifetimes in the range $[T,T+dT]$ is
\ben
f_\text{exp}(T) =  \vw \frac{dn}{dR} = \frac{1}{\Tstar} e^{-T/\Tstar},
\label{e:fExpNuc}
\een
where $\Tstar = 1/\be$.
Hence
\ben
\nu_\text{exp}(\tilde{T}) =  e^{-\tilde{T}}. 
\label{e:nuExpNuc}
\een

\subsubsection{Simultaneous nucleation}
With simultaneous nucleation, we have (\ref{e:hSim})
\ben
\mcA(t) = \frac{1}{2} \frac{\be_\text{eff}^3}{\vw}  (t - \tn)^2 \exp\left[ - \frac{1}{6} \be_\text{eff}^3  (t - \tn)^3 \right],
\een
and the nucleation rate per unit volume can be taken to be a $\delta$-function, 
\ben
p(t') =  {\nb}  \de(t' - \tn).
\een
Hence
\bea
\frac{dn}{dR} 
&=& 
  \int_{\tc}^\infty dt' \frac{1}{2} \frac{\be_\text{eff}^3}{\vw}  (t' - \tn + R/\vw)^2 \exp\left[ - \frac{1}{6} \be_\text{eff}^3  (t' - \tn + R/\vw)^3 \right] {\nb}  \de(t' - \tn), \nonumber\\
&=&
{\nb}  \frac{1}{2} \frac{\be_\text{eff}^3}{\vw}  (R/\vw)^2 \exp\left[ - \frac{1}{6} \be_\text{eff}^3  (R\vw)^3 \right] .
\eea
Defining a timescale for the simultaneous case $\Tstar = 1/\be_\text{eff}$, 
the fraction of bubbles with first collision in the time interval $[T,dT]$ after nucleation is 
\ben
\label{e:fSimNuc}
f_\text{sim}(T) =   \frac{1}{2\Tstar} \left(\frac{T}{\Tstar}\right)^2 \exp\left[ - \frac{1}{6}\left(\frac{T}{\Tstar}\right)^3 \right] .
\een
Hence
\ben
\label{e:nuSimNuc}
\nu_\text{sim}(\tilde{T}) =   \frac{1}{2} \tilde{T}^2 \exp\left( - \frac{1}{6}\tilde{T}^3 \right) .
\een

\subsection{RMS fluid velocity}
\label{ss:RMSFluVel}

The mean square fluid velocity can be recovered from the velocity power spectrum 
\ben
\fluidV^2 = \int \frac{dq}{q}\mathcal{P}_{\vfft}(q) 
=  \frac{2}{(\be\Rbc)^3}
 \int d\tilde{T}\nu(\tilde{T})\tilde{T}^{3} \int dz \frac{z^2}{2\pi^2}|A(z)|^{2},
\een
where $z = \Ttil q/\be$.
Hence 
the RMS fluid velocity is obtained from the third moment of the bubble lifetime distribution, denoted $\nu_3$. 
For both the simultaneous (\ref{e:nuSimNuc}) and exponential lifetime distributions (\ref{e:nuExpNuc}), $\nu_3 = 6$, and so 
\ben
\label{e:Ubarf2SSM}
\fluidV^2 = \frac{3}{4\pi\vw^3} \int dz \frac{z^2}{2\pi^2}2 |A(z)|^{2}, 
\een
for both nucleation histories. 
Hence comparison of $\fluidV$ computed with the two different nucleation histories is a good test of numerical accuracy.

It is commonly assumed that  the mean square fluid velocity 
is given by the mean square fluid velocity around a single bubble with the invariant profile, 
which can be computed directly from the solution $v(\xi)$, $w(\xi)$
or from the Fourier-transform of the velocity field \ref{e:vFFT}, giving 
\ben
\label{e:Ubari1d}
\bar{U}_{f,3}^{\rm 1d}  = \frac{3}{4\pi\vw^3} \int dz \frac{z^2}{2\pi^2} |f'(z)|^{2}.
\een
It is clear from the difference between (\ref{e:Ubarf2SSM}) and (\ref{e:Ubari1d}) that the 
strict equality does not hold. However, we can observe that (\ref{e:A2Def}) 
$2 |A(z)|^{2} = \frac{1}{2}\left(|f'(z)|^2+ \cs^2 | l(z)|^2 \right)$.
Arguing that $\cs^2 | l(z)|^2 \simeq [f'(z)|^2$ through the equations of motion (\ref{e:LinEOMa},\ref{e:LinEOMb}), 
we see that an approximate equality should hold. 

To test this approximate equality, 
in Table \ref{t:UbarCom} we show $\fluidV$ computed in the Sound Shell Model, with simultaneous and exponential nucleation, and
that computed around a single bubble, for a selected set of phase transition parameters relevant for existing simulations.
We see that even for weak transitions with low fluid velocities, the RMS velocities can differ by up to about 20\%.
We also confirm that simultaneous and exponential nucleation histories give RMS fluid velocities agreeing to the per mille level, 
with one exception, ($\alpha$, $v_{\rm w}$) = (0.05, 0.72). In this exceptional case the invariant profile 
has a very narrow shock front in front of the wall (see Fig.~\ref{f:BubPro}) which makes the calculation 
sensitive to the numerical resolution at high wavenumber.  

The calculations were made with $N_\xi = 5000$ points on the $\xi$ lattice, $N_z = 10000$ wavenumbers 
logarithmically spaced in the range $10^{-1} \le k\Rbc \le  10^3$, and $N_{\tilde{T}} = 1000$ time points logarithmically distributed in the range $0.01 \le \tilde{T} \le 20$.

\begin{table}
\centering
\input{table_1dh_compare}
\caption{ \label{t:UbarCom}
Comparison between the RMS fluid velocity (divided by $10^3$) predicted by the Sound Shell Model 
for simultaneous and exponential nucleation ($\bar{U}_{f,3}^{\rm sim}$, $\bar{U}_{f,3}^{\rm exp}$) and that 
around a single bubble $\bar{U}_{f,3}^{\rm 1d}$, computed by the method described in Appendix \ref{s:HydExpBub} around 
Eq.~\ref{e:OneBubUbar}. 
The Sound Shell Model predictions for $\bar{U}_{f}$ are independent of the nucleation history: 
the minor differences between the RMS velocities are due to numerical error. 
}
\end{table}

\section{Shape of the power spectrum}

In the LISA Cosmology Working Group report \cite{Caprini:2015zlo} the
acoustic gravitational wave power spectrum was modelled using a broken
power law function:
\begin{equation}
 \frac{d \OmGW(k)}{d \ln(k)} 
=  (\HN\Rbc) A \cwgfitfun(s)
\label{e:CWGfit}
\end{equation}
where
\begin{equation}
\label{e:CWGFitFun}
\cwgfitfun(s) = s^3\left(\frac{7}{4+3s^2}\right)^{7/2},
\end{equation}
and 
\begin{equation}
s = \frac{k\Rbc}{(k\Rbc)_{\rm p}}.
\end{equation}
The dimensionless parameters $A$ and $(k\Rbc)_{\rm p}$ determine the
magnitude and the location of the maximum of the power spectrum,
respectively.  The form of the function is motivated by the results
from hydrodynamical gravitational wave production simulations \cite{Hindmarsh:2015qta}.  
The power spectrum of the ansatz at
small $k$ is $\propto k^3$, turning over to $\propto k^{-4}$ at large
$k$.
The peak angular frequency is 
approximately $(k\Rbc)_{\rm p} \simeq 10$, while 
the amplitude factor is related to the kinetic energy fraction through 
\ben
\frac{343}{360}\sqrt{\frac{7}{3}} A = 3 \left(\Gamma \fluidV^2\right)^2 \OmGWscaled,
\een
where
\ben
\label{e:OmGWscaled}
\OmGWscaled =  \frac{1}{2\pi^2}\int dx x^2  \SpecDenGW(x).
\een
Numerical simulations of weak and intermediate strength transitions 
indicate that $\OmGWscaled = \text{O}(10^{-2})$ \cite{Hindmarsh:2017gnf}.

The Sound Shell Model predicts different asymptotic power laws (see subsection \ref{ss:PowLaw}), and a more complicated 
structure around the peak. Nevertheless, it is instructive to fit the form (\ref{e:CWGfit}) to the 
Sound Shell Model prediction in order to compare to the output from numerical simulations, which we shall do in the next section.

In order to better describe the true Sound Shell Model prediction, in the more relevant 
exponential nucleation case, we need a form which can represent the two length scales in the system, 
the mean bubble separation and the fluid shell thickness. This suggests that we should use 
a double broken power law, 
\ben
\frac{d \OmGW(k)}{d \ln(k)} 
=  (\HN\Rbc) A_M \ssmfitfun(s,\BrePosRat)
\label{e:SSMfit}
\een
where
\begin{equation}
\label{e:SSMFitFun}
\ssmfitfun(s,\BrePosRat) = s^9\left(\frac{\BrePosRat^4 + 1}{ \BrePosRat^4 + s^4}\right)^{2} \left(\frac{5}{5 - m + ms^2}\right)^{5/2},
\end{equation}
with $m = (9\BrePosRat^4 + 1)/(\BrePosRat^4 + 1)$.
This introduces another break in the power spectrum at dimensionless wavenumber $(k\Rbc)_\text{b} = \BrePosRat (k\Rbc)_\text{p}$.
Provided $\BrePosRat < 1$, the function peaks at $s=1$ with value $M(1) = 1$. 

The peak power parameter $A_M$ is related to the total power parameter $\OmGWscaled$ through 
\ben
\label{e:SSMPeaPow}
\mu(r_\text{b}) A_M = 3 \left(\Gamma \fluidV^2\right)^2 \OmGWscaled,
\een
with
\ben
\mu(r_\text{b}) = \int_0^\infty \frac{ds}{s} M(s,\BrePosRat).
\een
An approximate expression for $\mu$, accurate to about 10\% over the relevant range $0 < \BrePosRat < 1$, is $\mu(\BrePosRat) = 4.78 - 6.27\BrePosRat + 3.34\BrePosRat^2$.

\section{Power spectra: predictions and comparisons}

We use the Sound Shell Model, as described above, to calculate velocity and gravitational wave power spectra for a range of  
wall velocities $\vw$ and strength parameters $\aln$. 
using a python module PTtools developed for this task.
We chose the values $(\aln,\vw)$ used in the numerical simulations described in 
\cite{Hindmarsh:2017gnf}.  The simulations used simultaneous nucleation; we calculate for both simultaneous and exponential nucleation.

\begin{figure}[t!]
  \subfigure[\ Weak, $\vw = 0.92$]{\includegraphics[width=0.49\textwidth,clip=true]{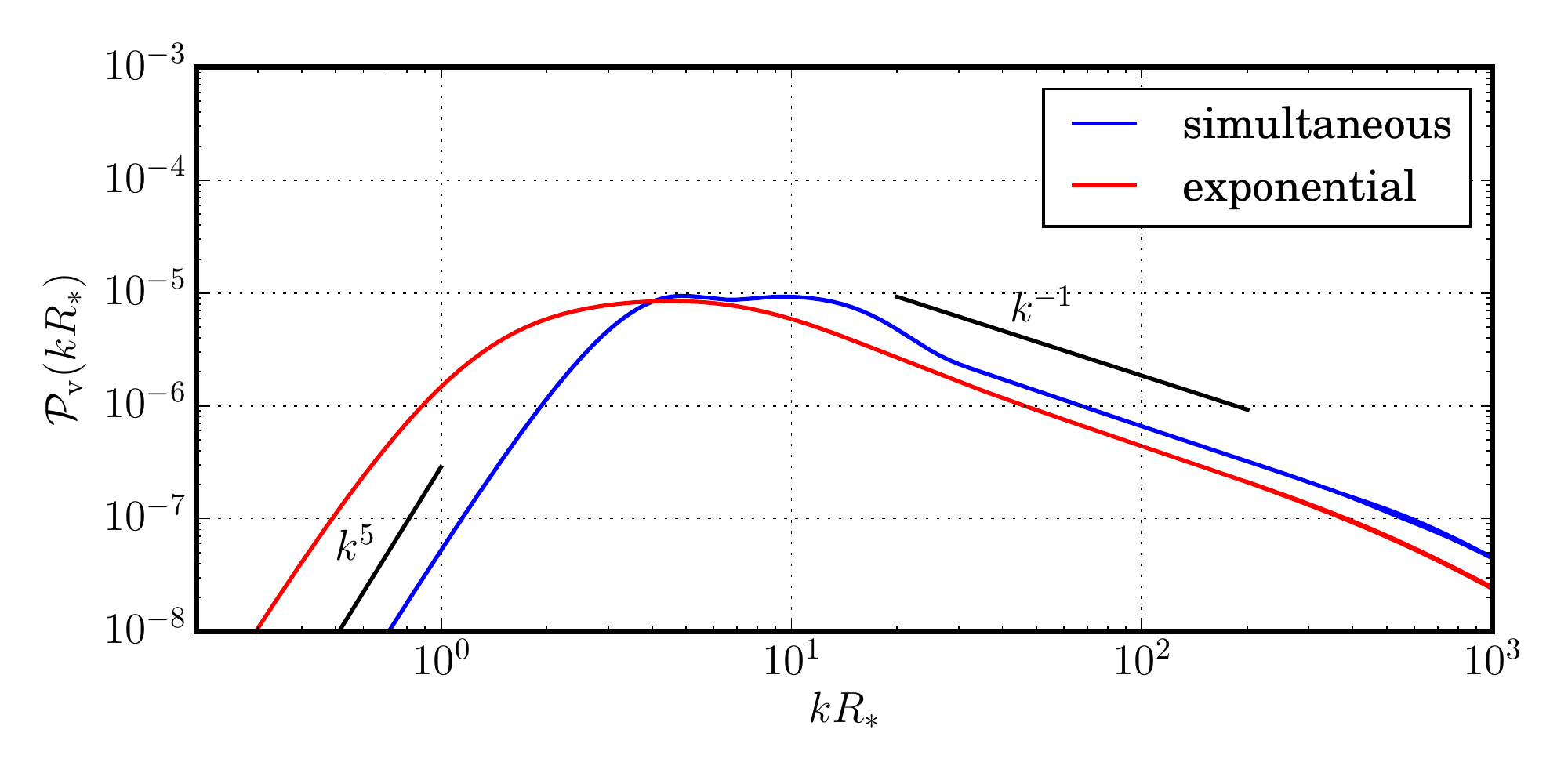}}
  \hfill	
  \subfigure[\ Intermediate, $\vw=0.92$]{\includegraphics[width=0.49\textwidth, clip=true]{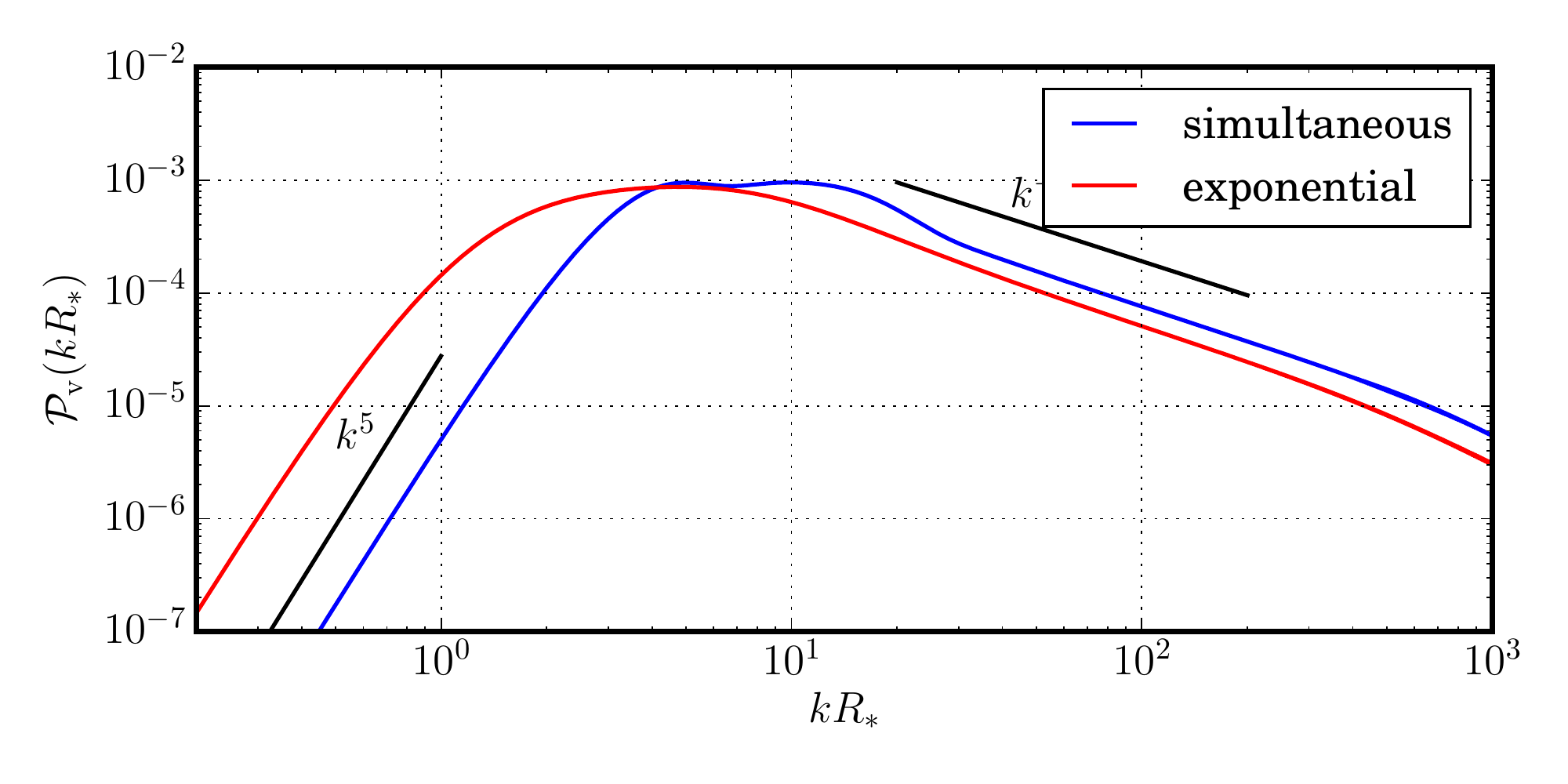}}
  \\
  \subfigure[\ Weak, $\vw=0.80$]{\includegraphics[width=0.49\textwidth,clip=true]{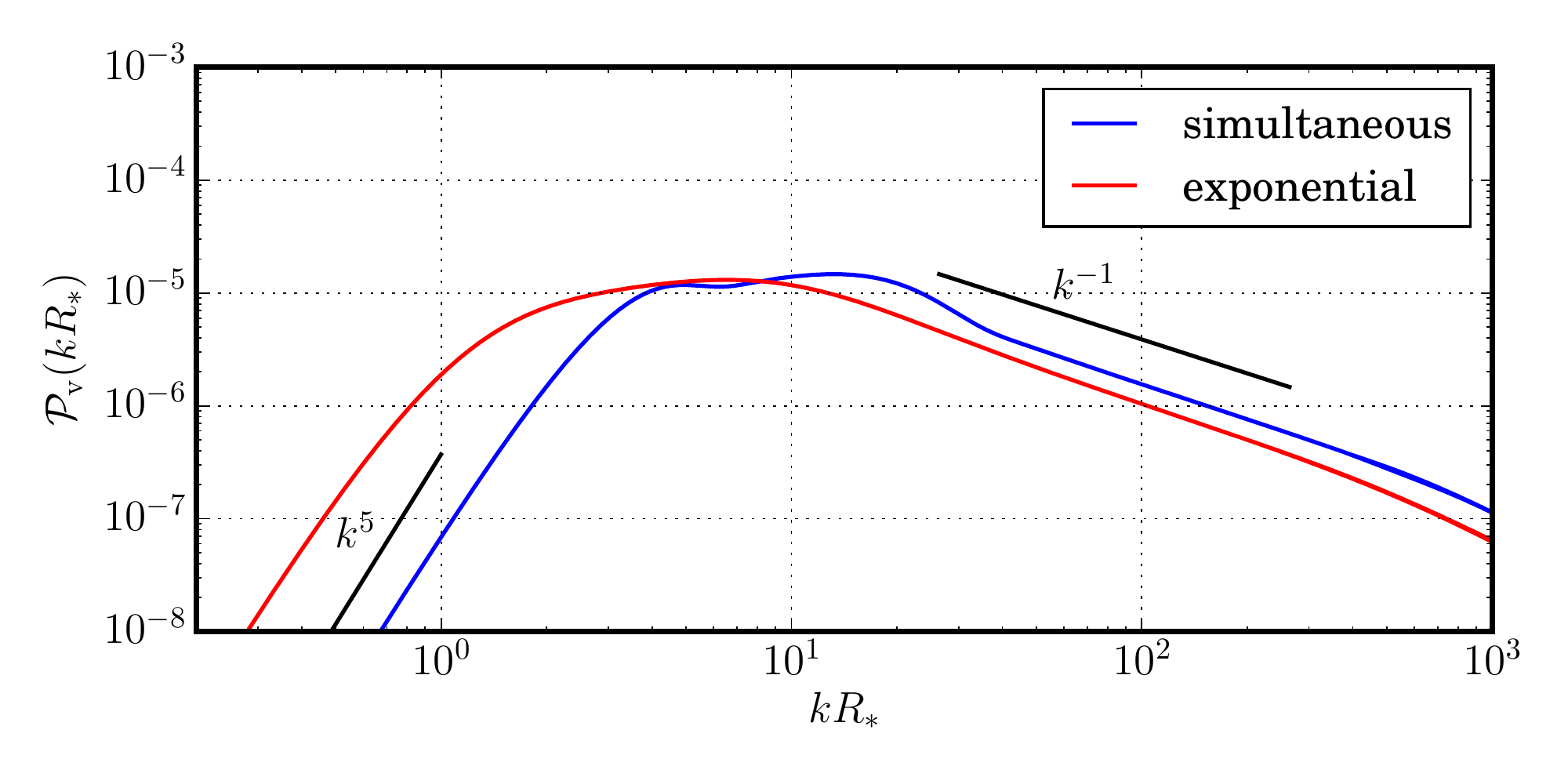}}
  \hfill	
  \subfigure[\ Intermediate, $\vw=0.80$]{\includegraphics[width=0.49\textwidth, clip=true]{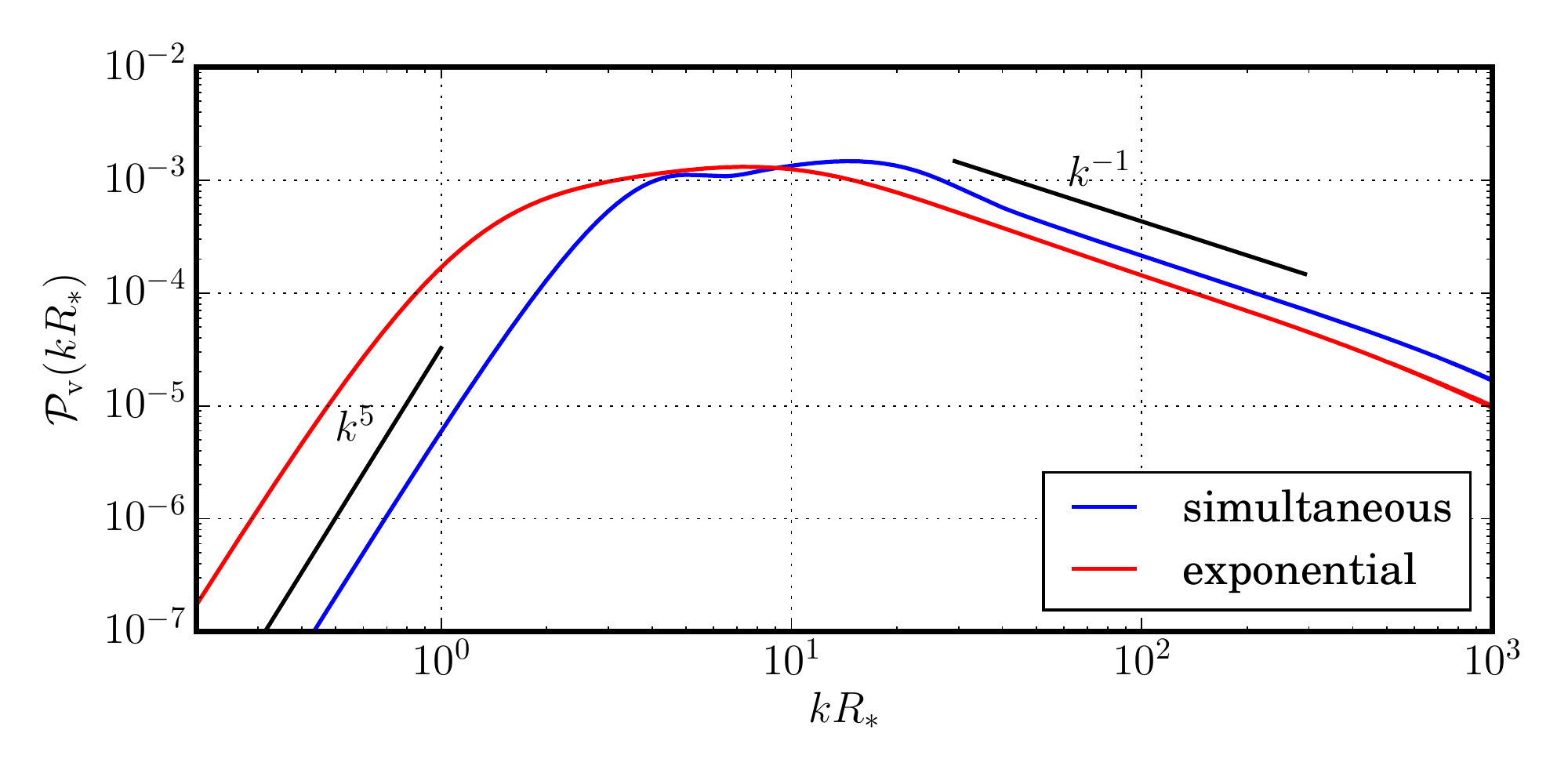}}
  \\
  \subfigure[\ Weak, $\vw=0.68$]{\includegraphics[width=0.49\textwidth,clip=true]{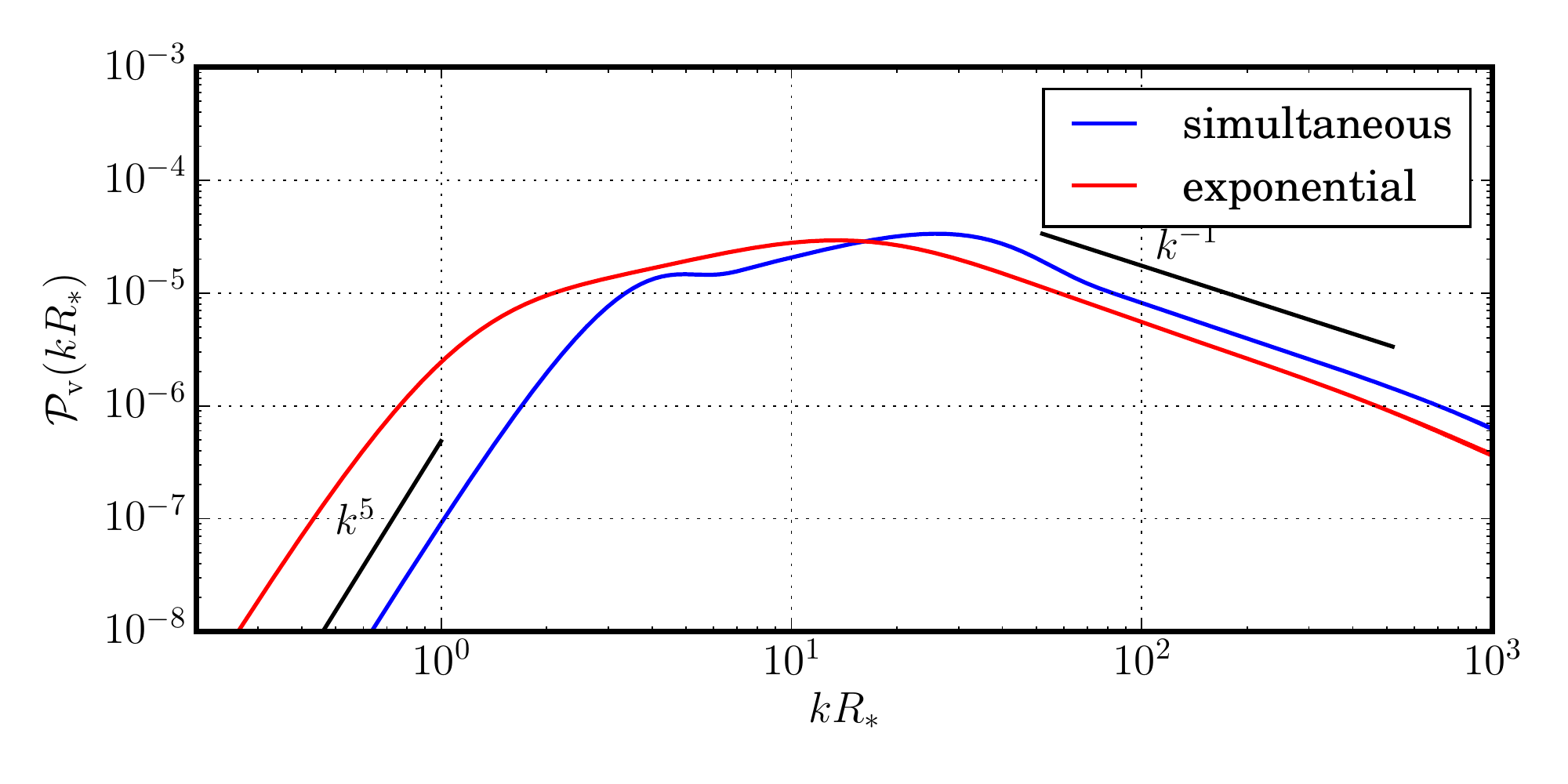}}
  \hfill
  \subfigure[\ Intermediate, $\vw=0.731$]{\includegraphics[width=0.49\textwidth,clip=true]{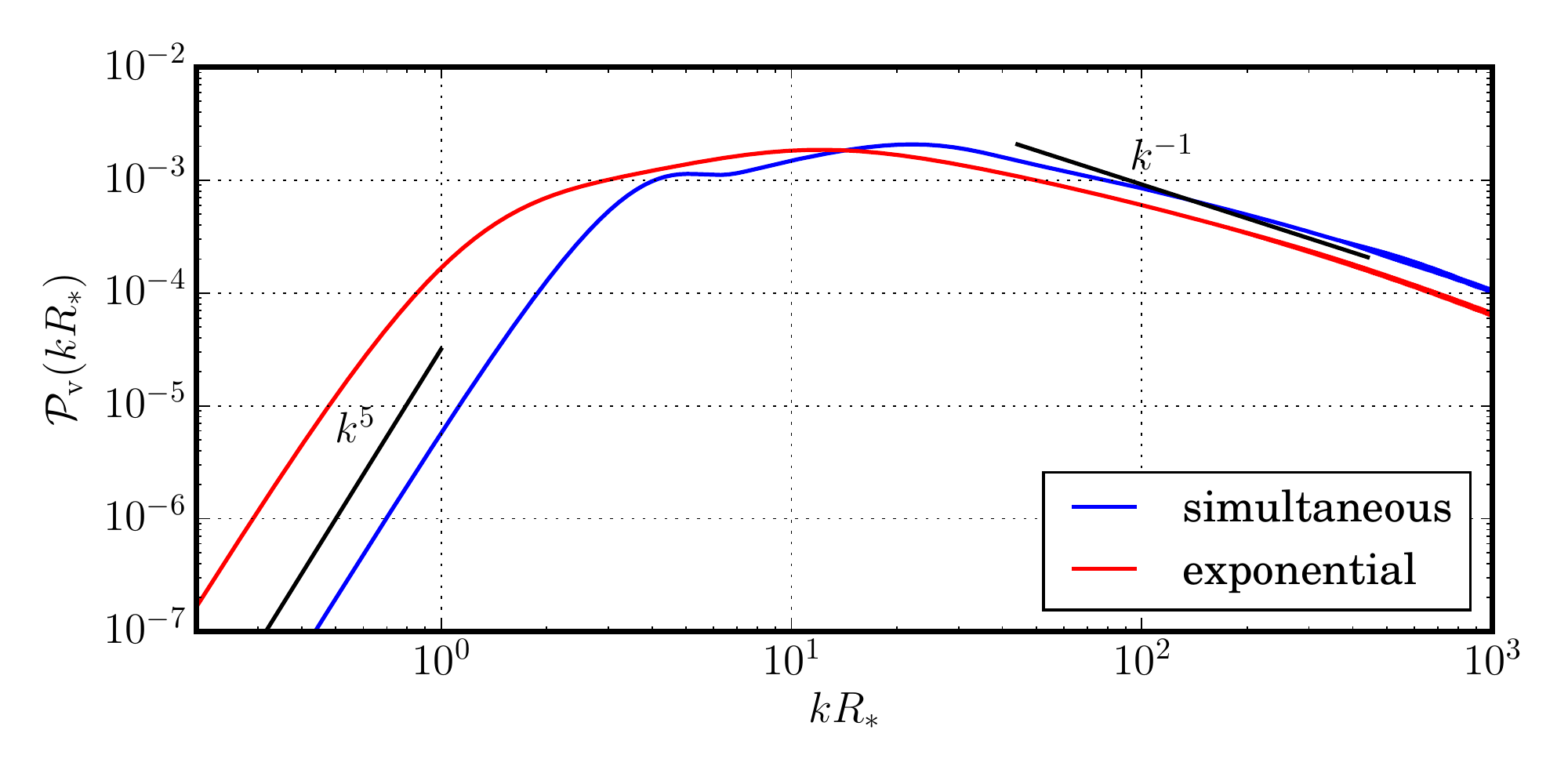}}

\caption{\label{f:VelPSDet} Velocity power spectra for detonations.
Predictions of the sound shell model with simultaneous nucleation are shown in blue, with exponential nucleation in red. 
  Left are weak strength phase transitions, with $\vw = 0.92$, $0.80$
  and $0.68$.  Right are intermediate phase transitions, with $\vw =
  0.92$, $0.80$ and $\vw = 0.731$.  
}
\end{figure}

\begin{figure}[t!]
\subfigure[\  Weak, $\vw=0.56$]{\includegraphics[width=0.49\textwidth,clip=true]{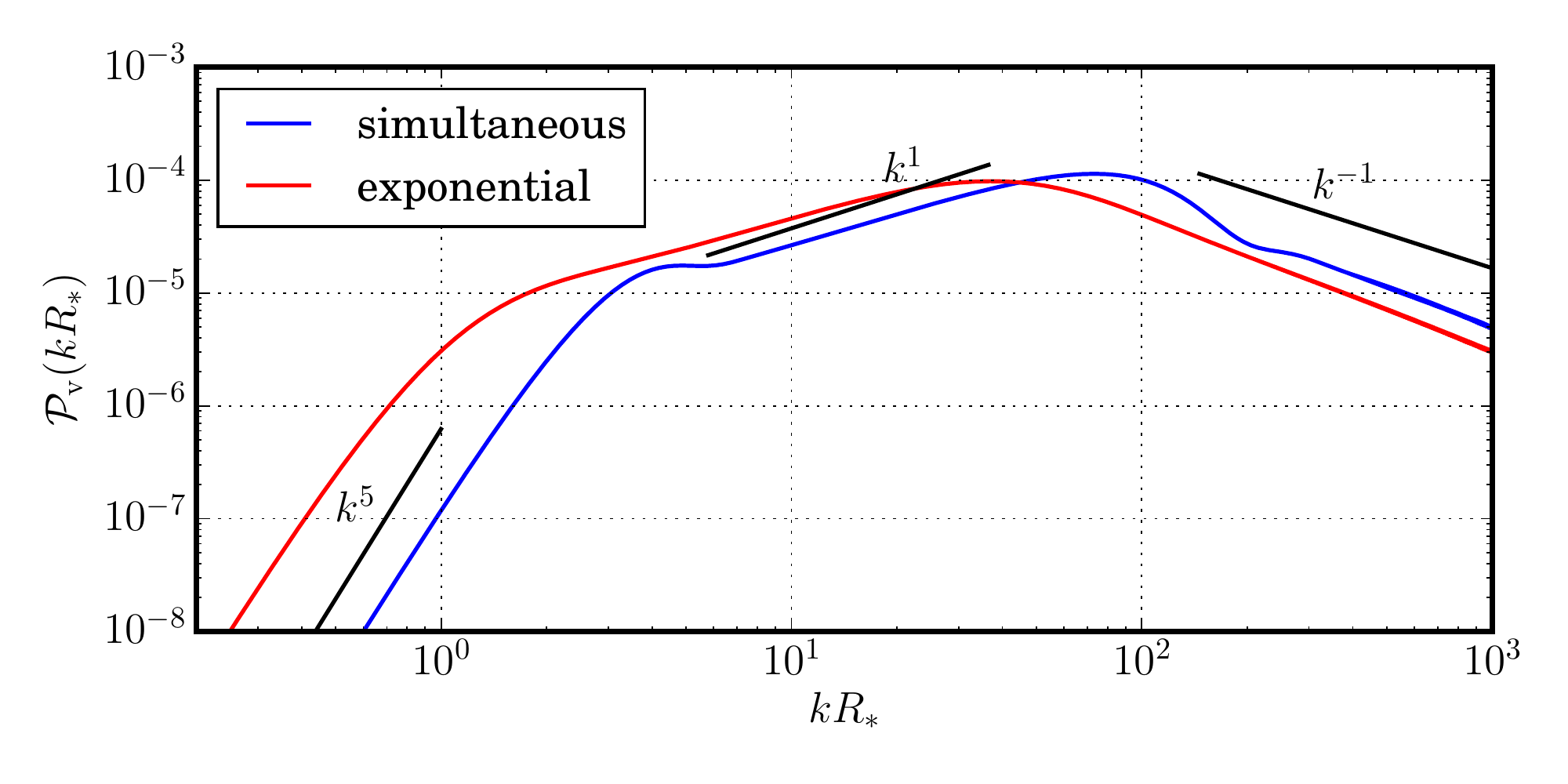}}
\hfill	
\subfigure[\ Intermediate, $\vw=0.56$]{\includegraphics[width=0.49\textwidth, clip=true]{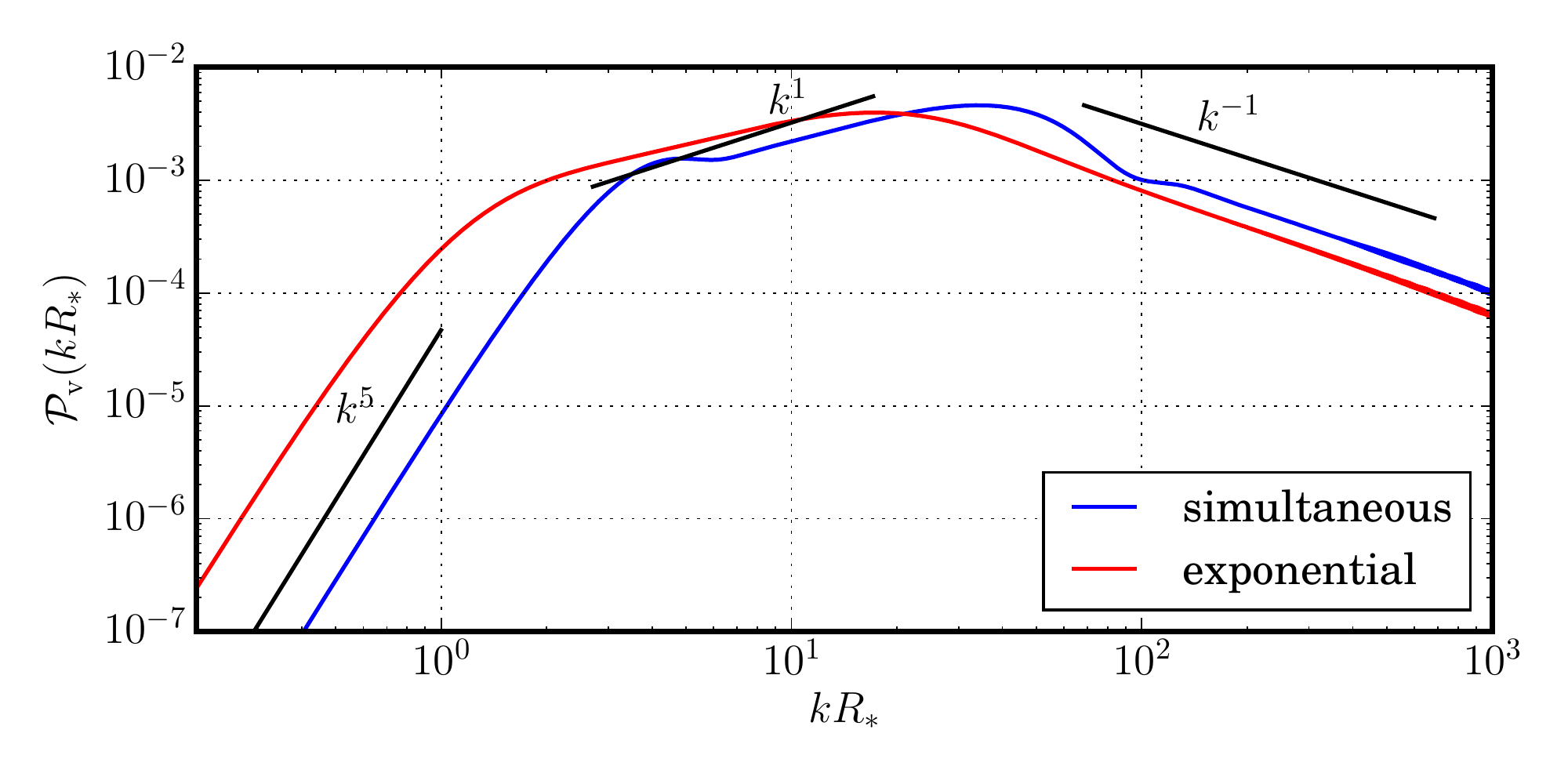}}\\
\subfigure[\ Weak, $\vw=0.44$]{\includegraphics[width=0.49\textwidth, clip=true]{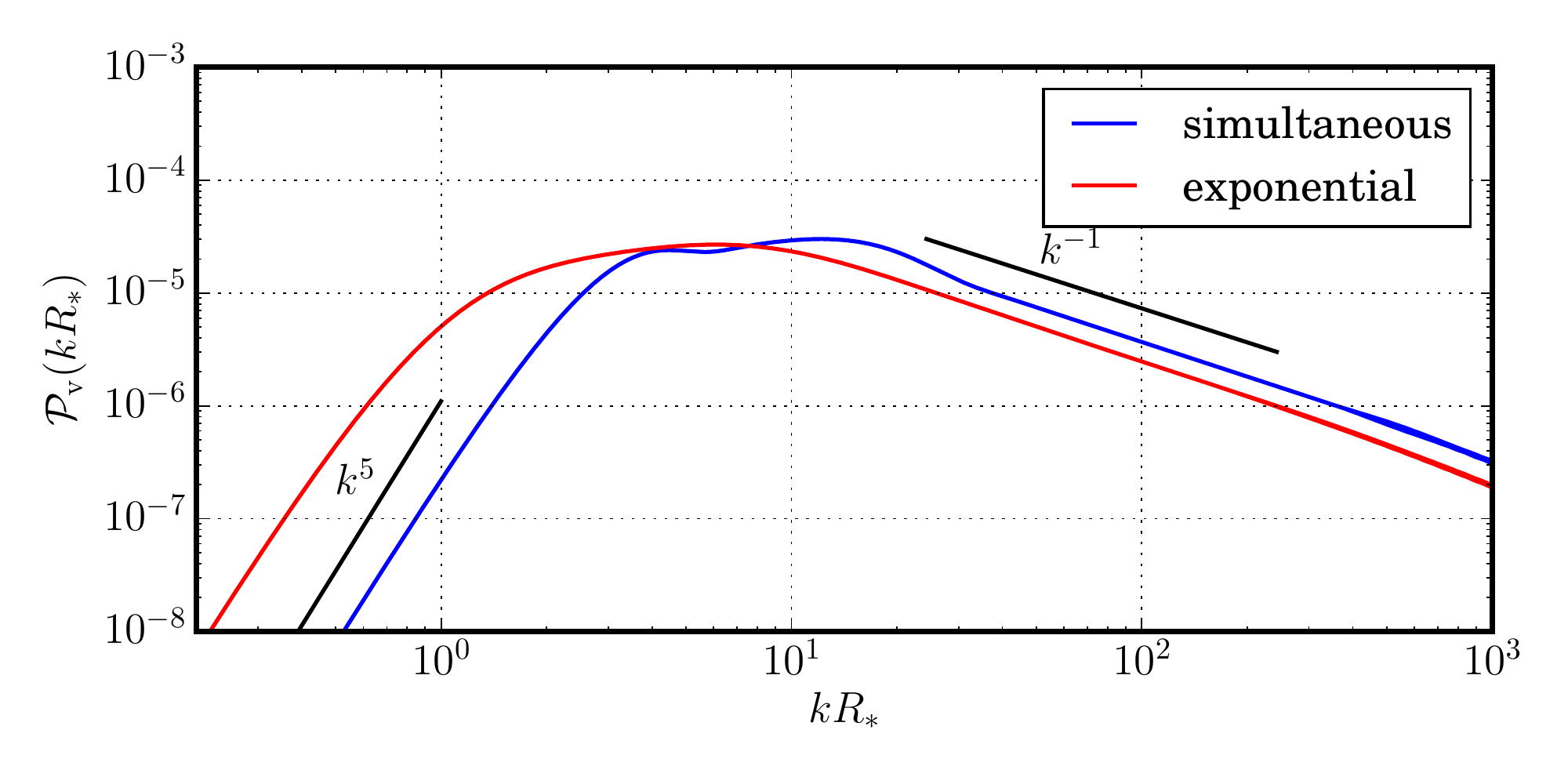}}
\hfill	
\subfigure[\ Intermediate, $\vw=0.44$]{\includegraphics[width=0.49\textwidth, clip=true]{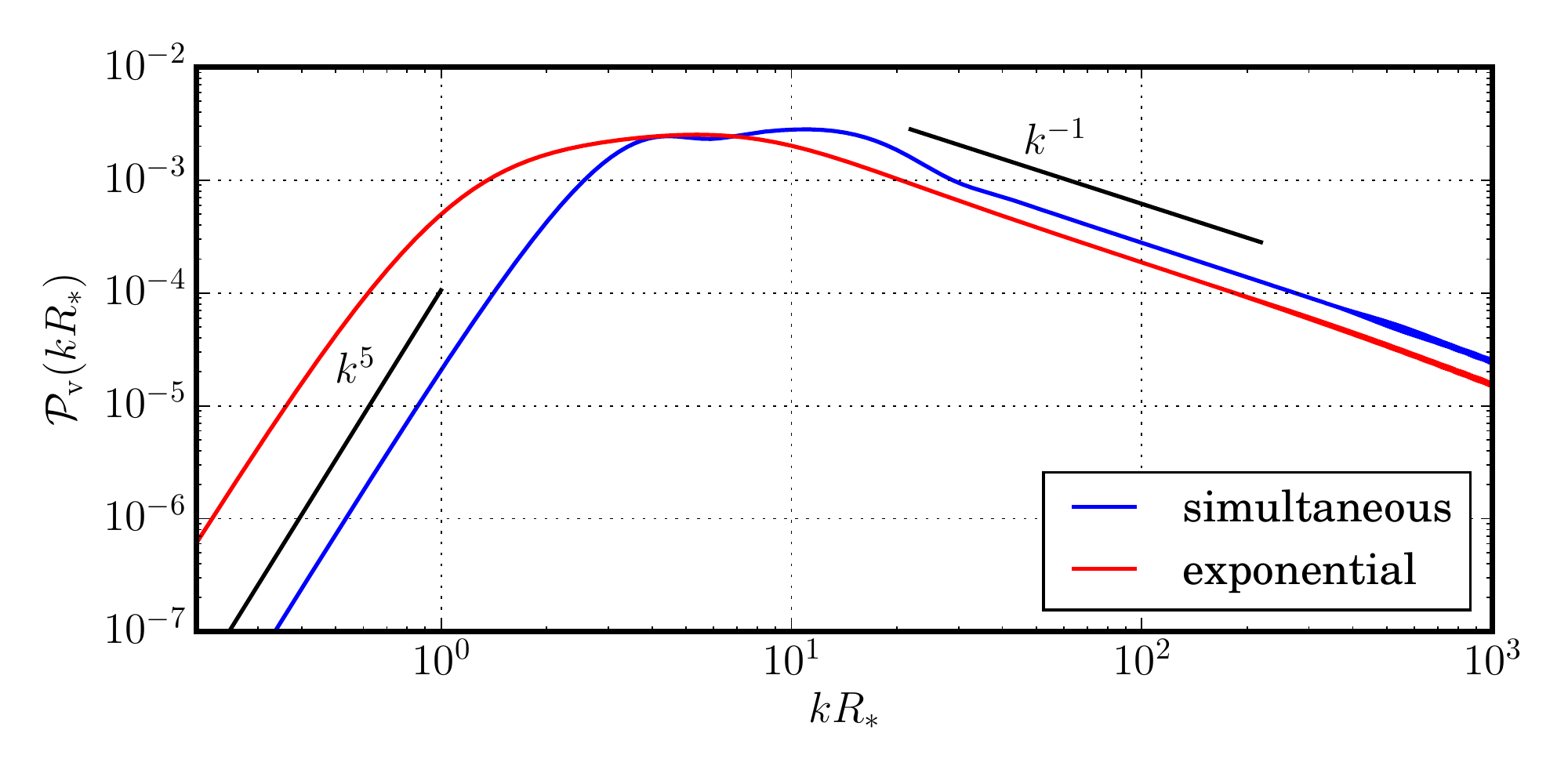}}\\
\caption{\label{f:VelPSDef}
Velocity power spectra for deflagrations. 
Predictions of the sound shell model with simultaneous nucleation are shown in blue, with exponential nucleation in red. 
Left are weak phase transitions, with $\vw = 0.56$ and $0.44$. 
Right are intermediate transitions with the same wall speeds. 
}
\end{figure}

Figs.~\ref{f:VelPSDet} and \ref{f:VelPSDef} show the predicted velocity power spectra for detonations and deflagrations, with ``weak'' transitions ($\aln = 0.0046$) on the left and ``intermediate'' strength transitions ($\aln = 0.05$) on the right.  Curves for simultaneous nucleation are shown in blue, and exponential nucleation in red. 
One can compare the blue curves with Figs.~3, 4 and 5 in Ref.~\cite{Hindmarsh:2017gnf}.
We will refer to this paper as HHRW17.

\begin{figure}[t!]

\subfigure[\ Weak, $\vw=0.92$]{\includegraphics[width=0.49\textwidth,clip]{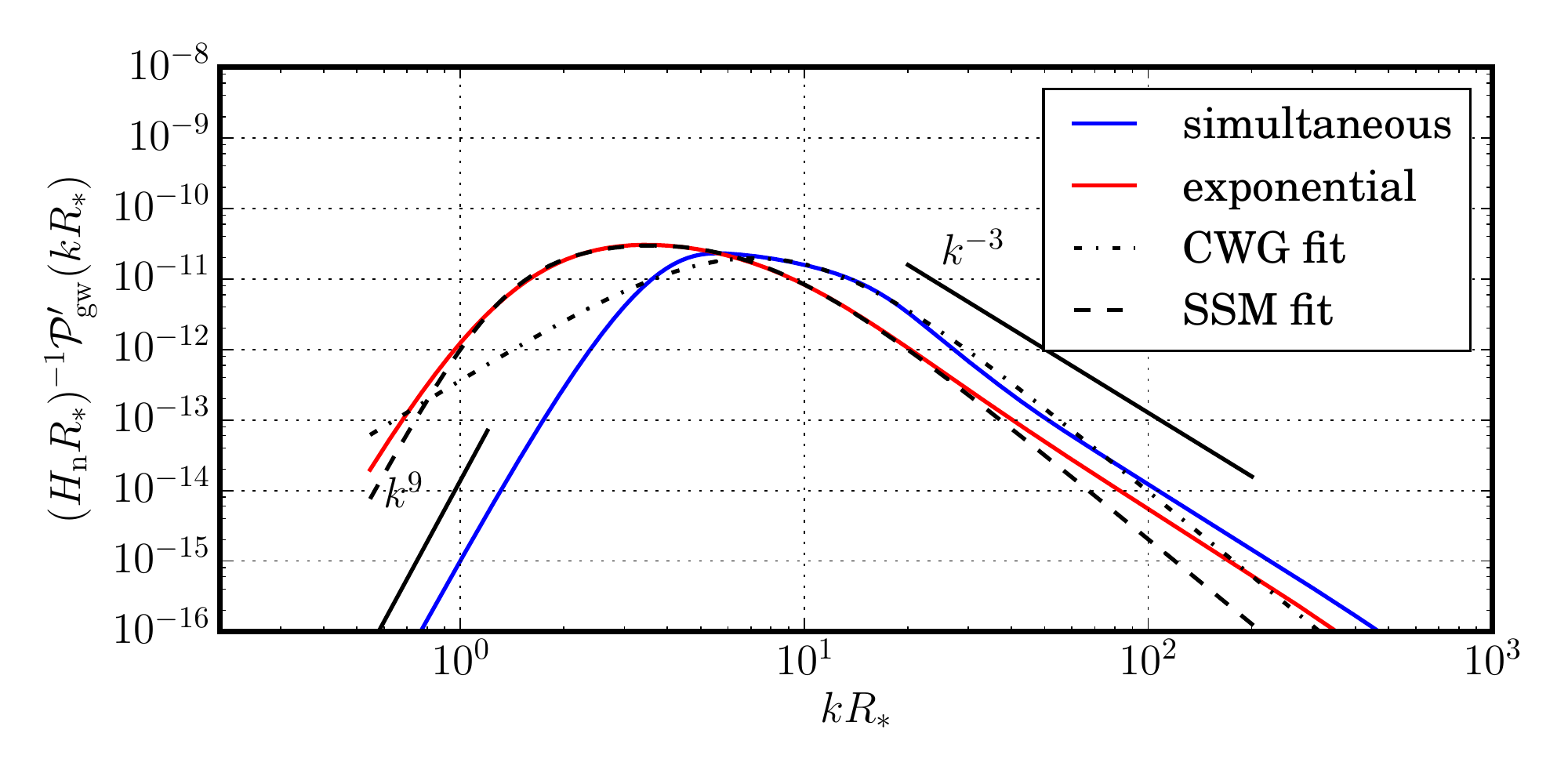}}
\hfill	
\subfigure[\ Intermediate, $\vw=0.92$]{\includegraphics[width=0.49\textwidth,clip]{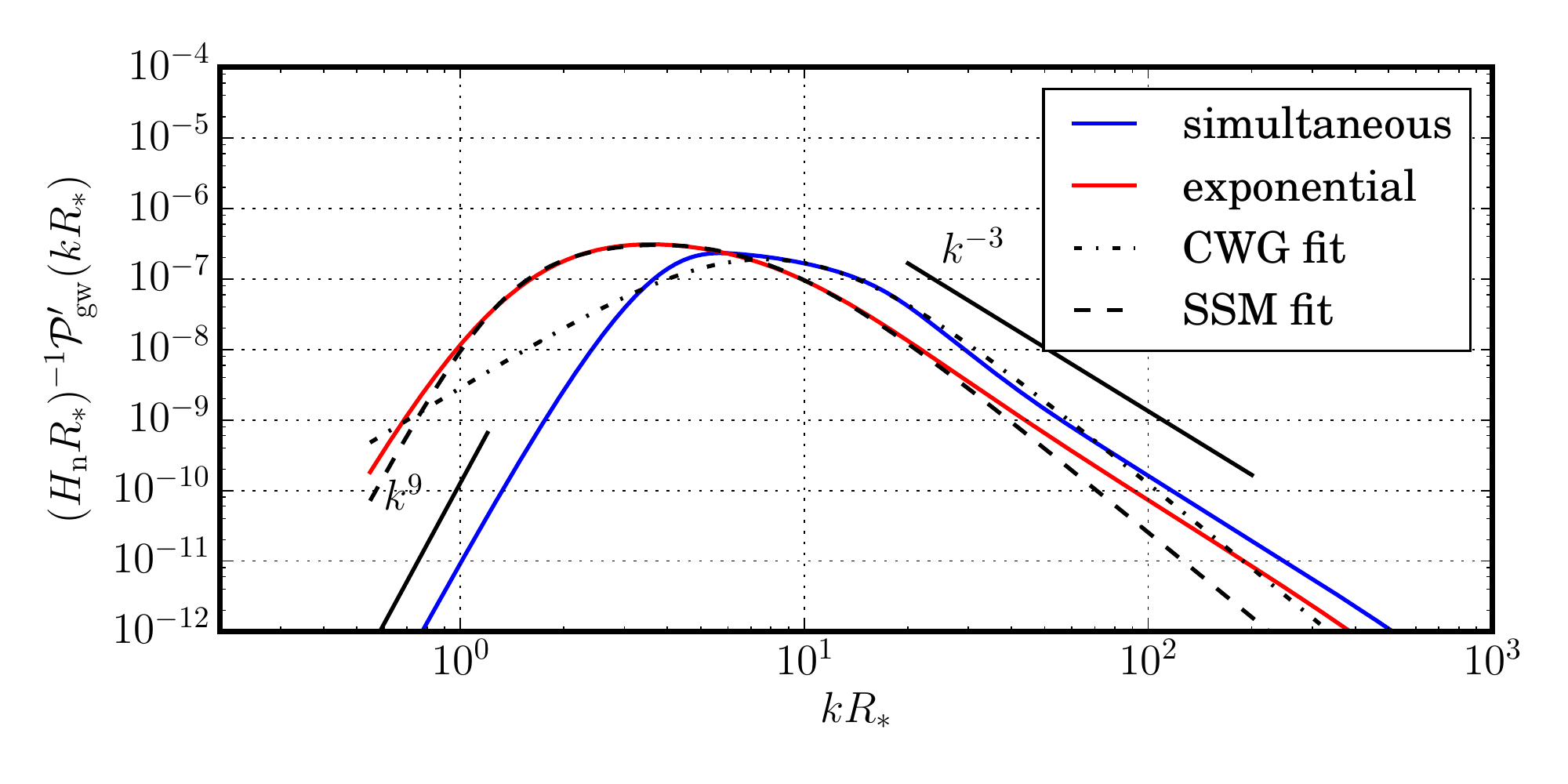}}\\
\subfigure[\ Weak, $\vw=0.80$]{\includegraphics[width=0.49\textwidth,clip]{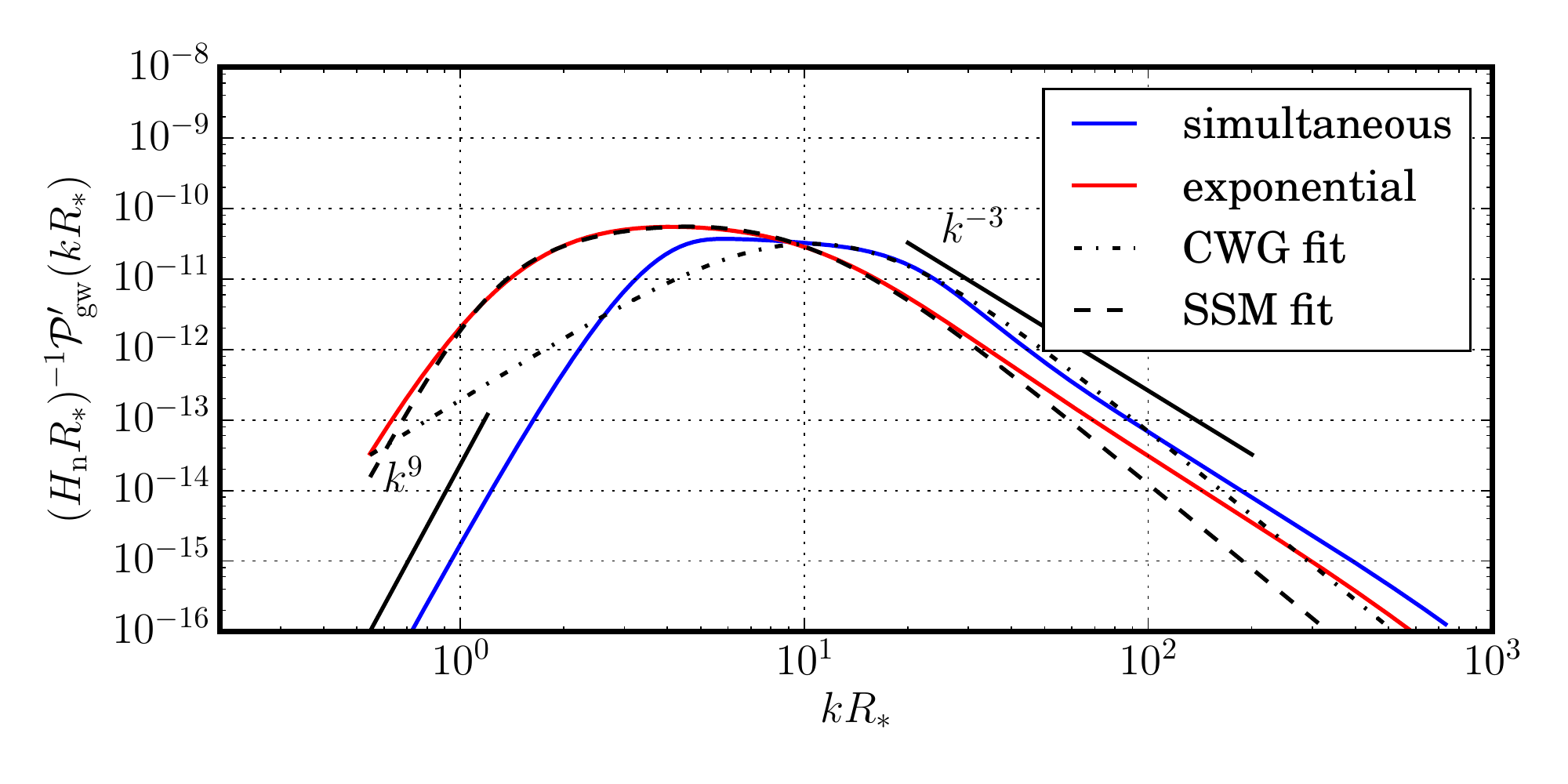}}
\hfill
\subfigure[\ Intermediate, $\vw=0.80$ ]{\includegraphics[width=0.49\textwidth,clip]{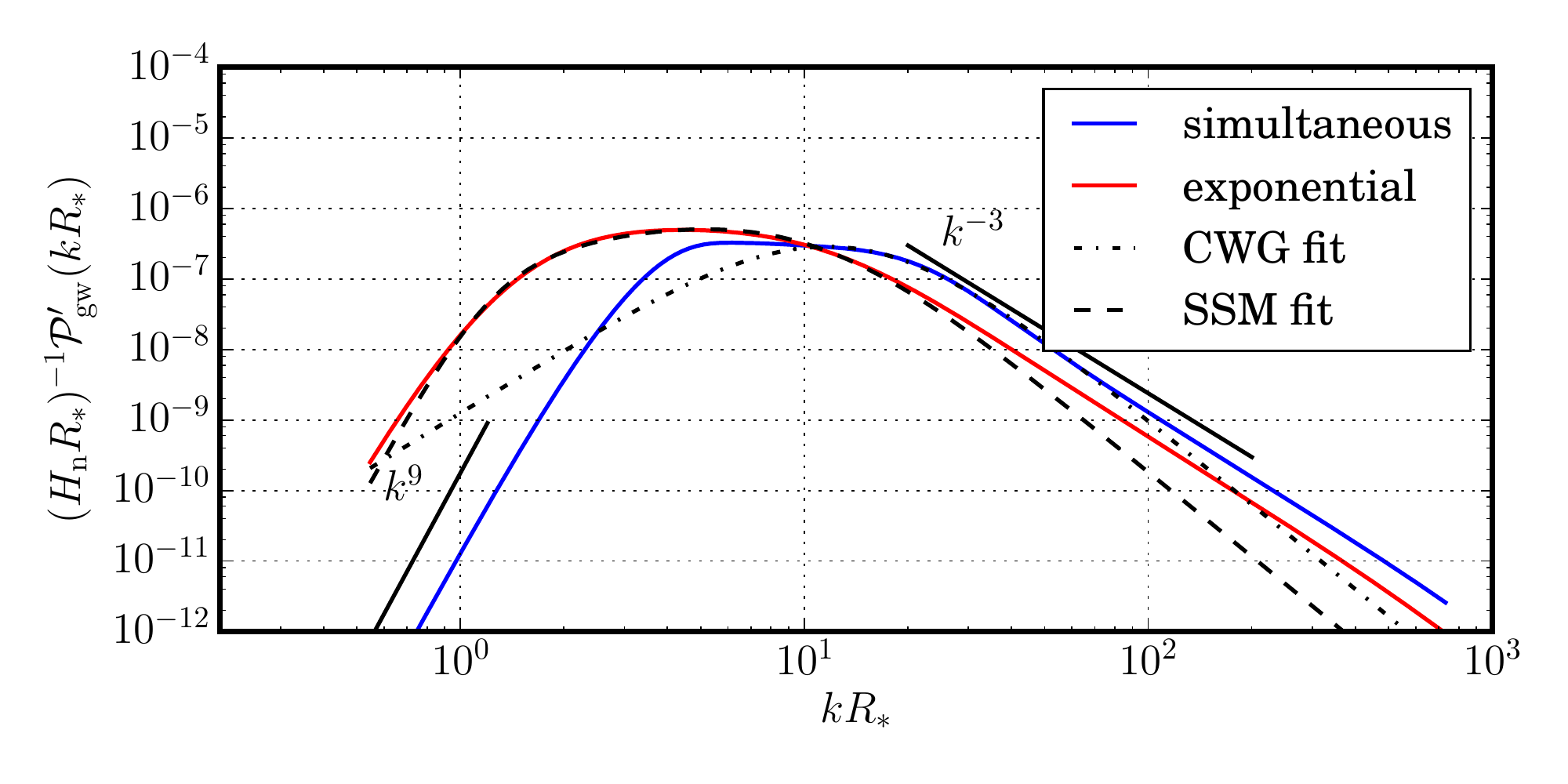}}
\\
\subfigure[\ Weak, $\vw=0.68$]{\includegraphics[width=0.49\textwidth,clip]{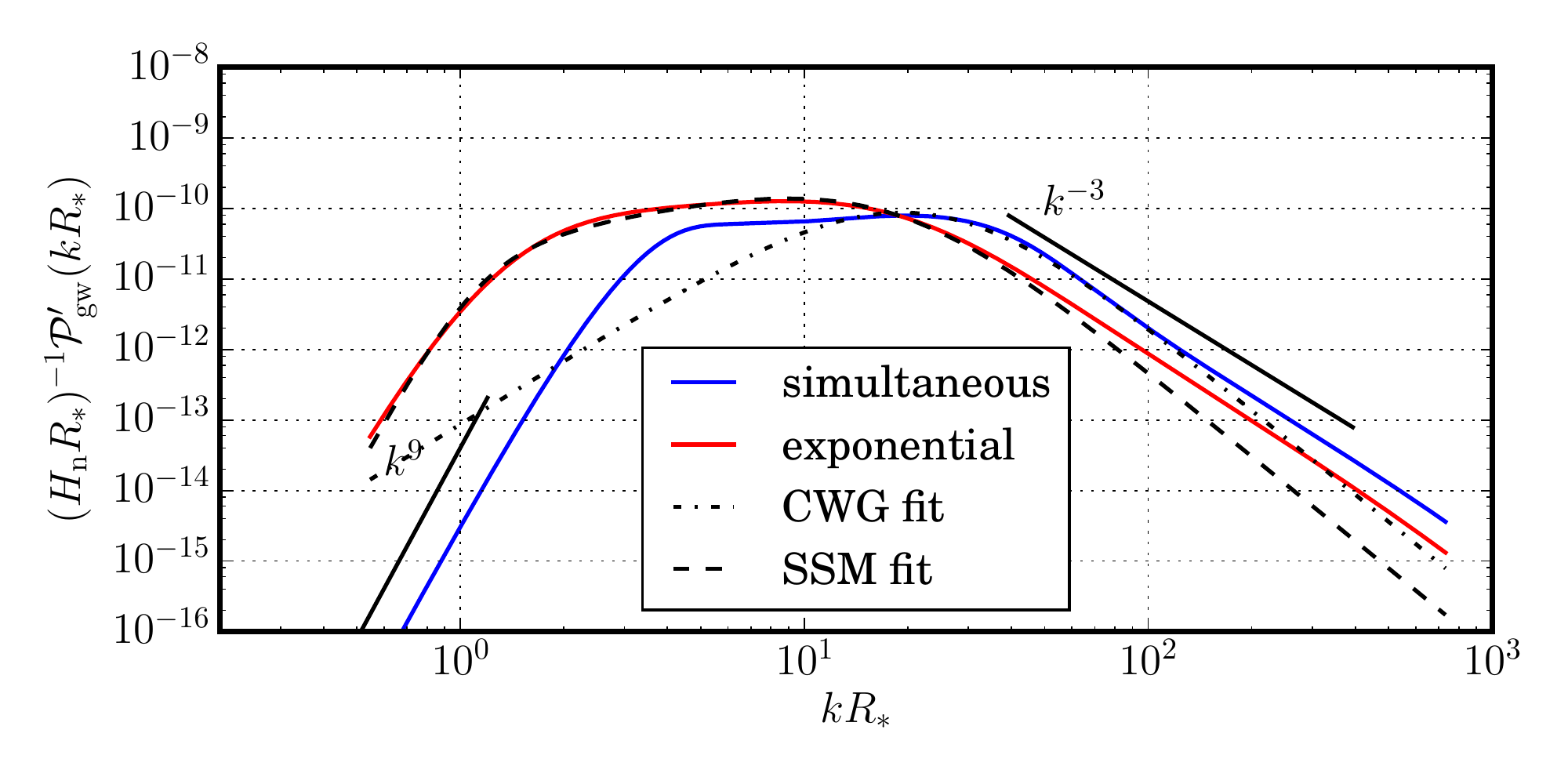}}
\hfill
\subfigure[\ Intermediate, $\vw=0.731$]{\includegraphics[width=0.49\textwidth,clip]{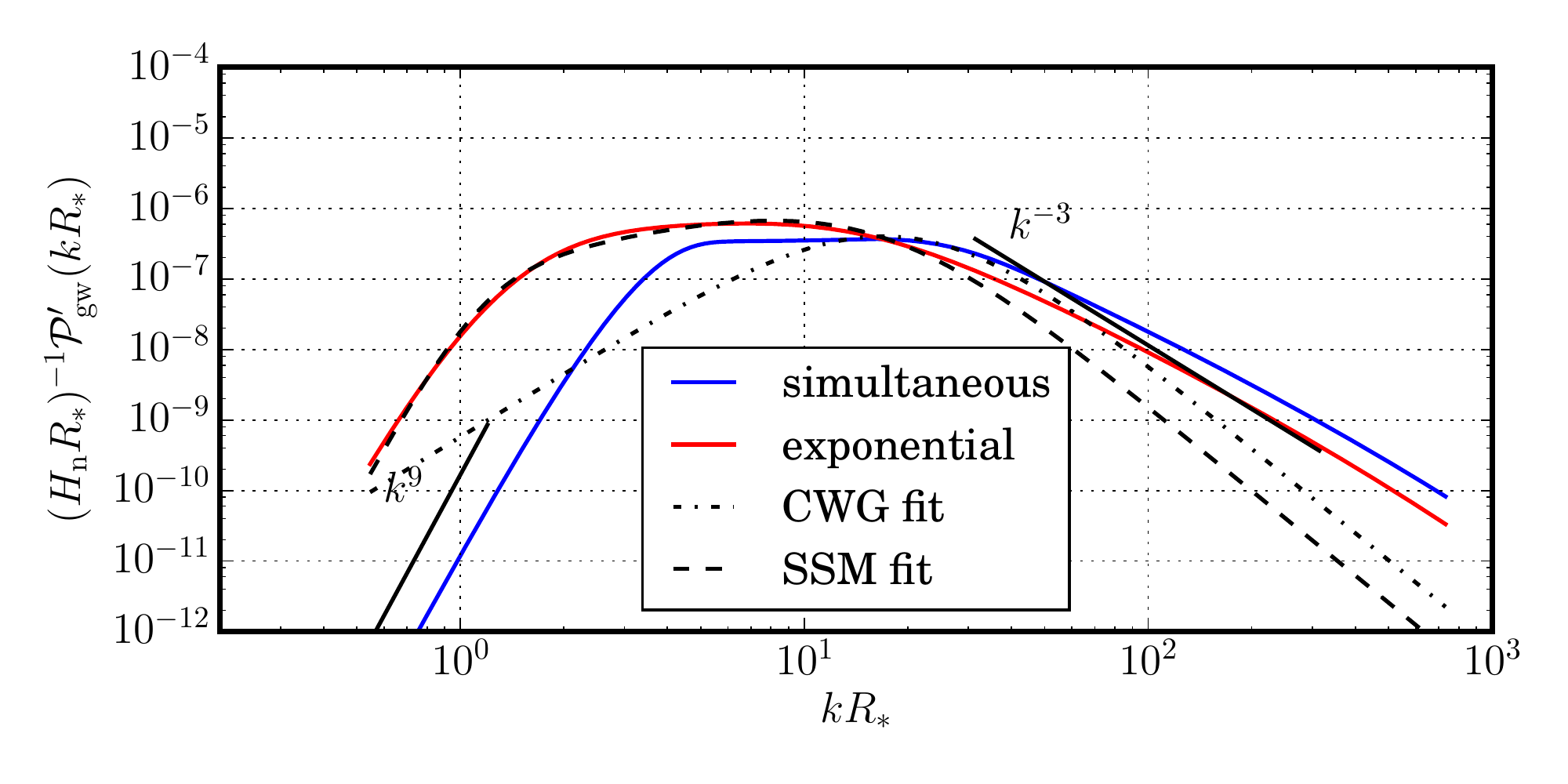}}

\caption{\label{f:GWPSDet} Scaled power spectra of fractional energy density
  in gravitational waves for detonations, defined in Eq.~(\ref{e:GWGroRat}),
  along with fits, 
  plotted against wave number scaled by the mean bubble separation.  
  Left are weak phase
  transitions, showing detonations with $\vw = 0.92$, $0.80$ and
  $0.68$ (top to bottom).  Right are intermediate phase transitions,
  with wall speeds $\vw=0.92$ and $\vw = 0.731$.  
  The dash-dot lines show fits to the simultaneous nucleation calculation,
  using the functional form (\ref{e:CWGFitFun}) \cite{Caprini:2015zlo}.
  Dashed lines show fits to the exponential nucleation calculation, 
  using the double broken power law form (\ref{e:SSMFitFun}) put forward here. 
  Best fit parameters for the peak power $A$ and the values of $k\Rbc$ at the two breaks 
  are given in Table \ref{t:ParComTabSSM}.
  }

\end{figure}

\begin{figure}

\subfigure[\ Weak, $\vw=0.56$]{\includegraphics[width=0.49\textwidth,clip=true]{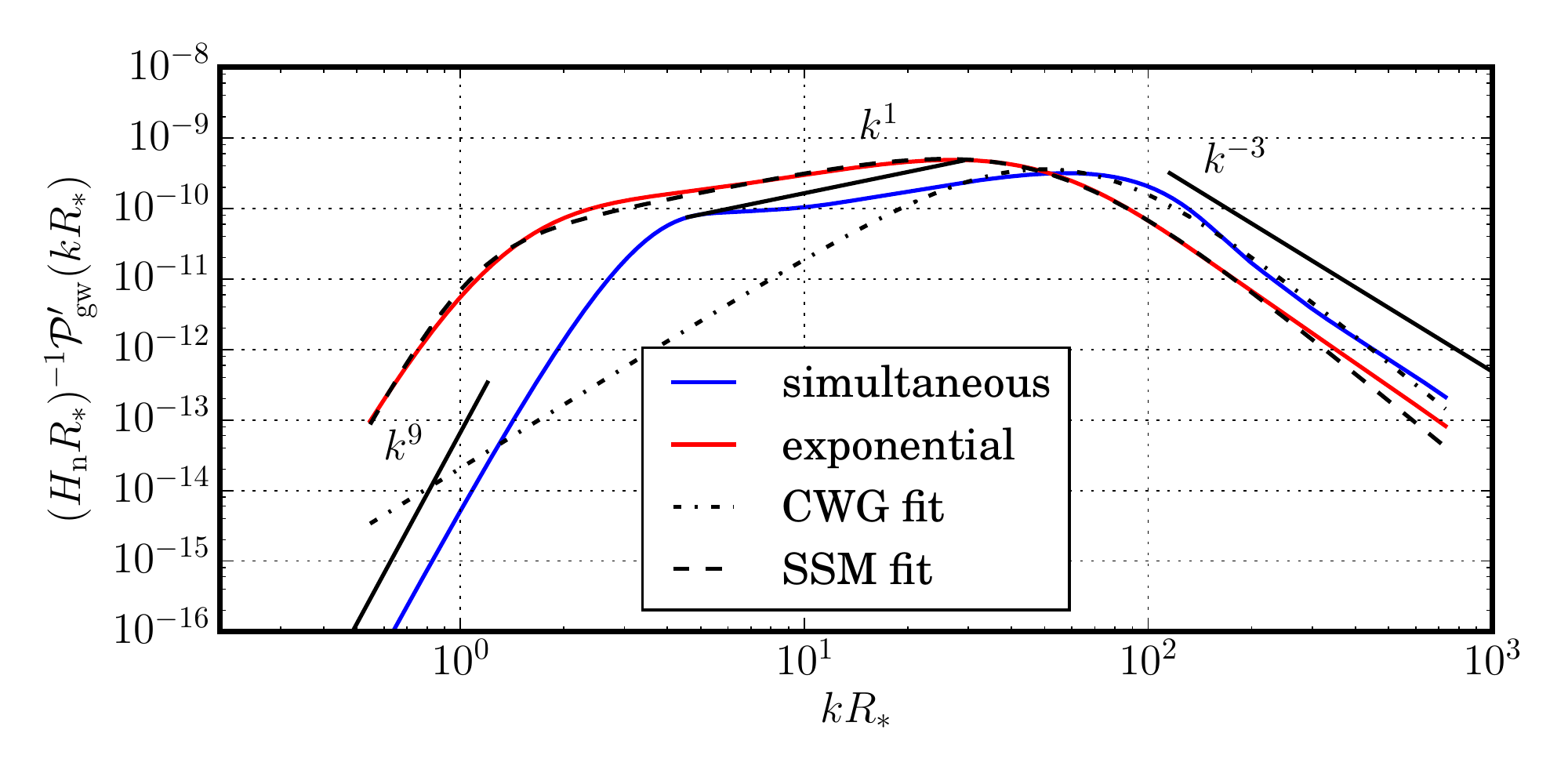}}
\hfill	
\subfigure[\ Intermediate,
  $\vw=0.56$]{\includegraphics[width=0.49\textwidth,clip=true]{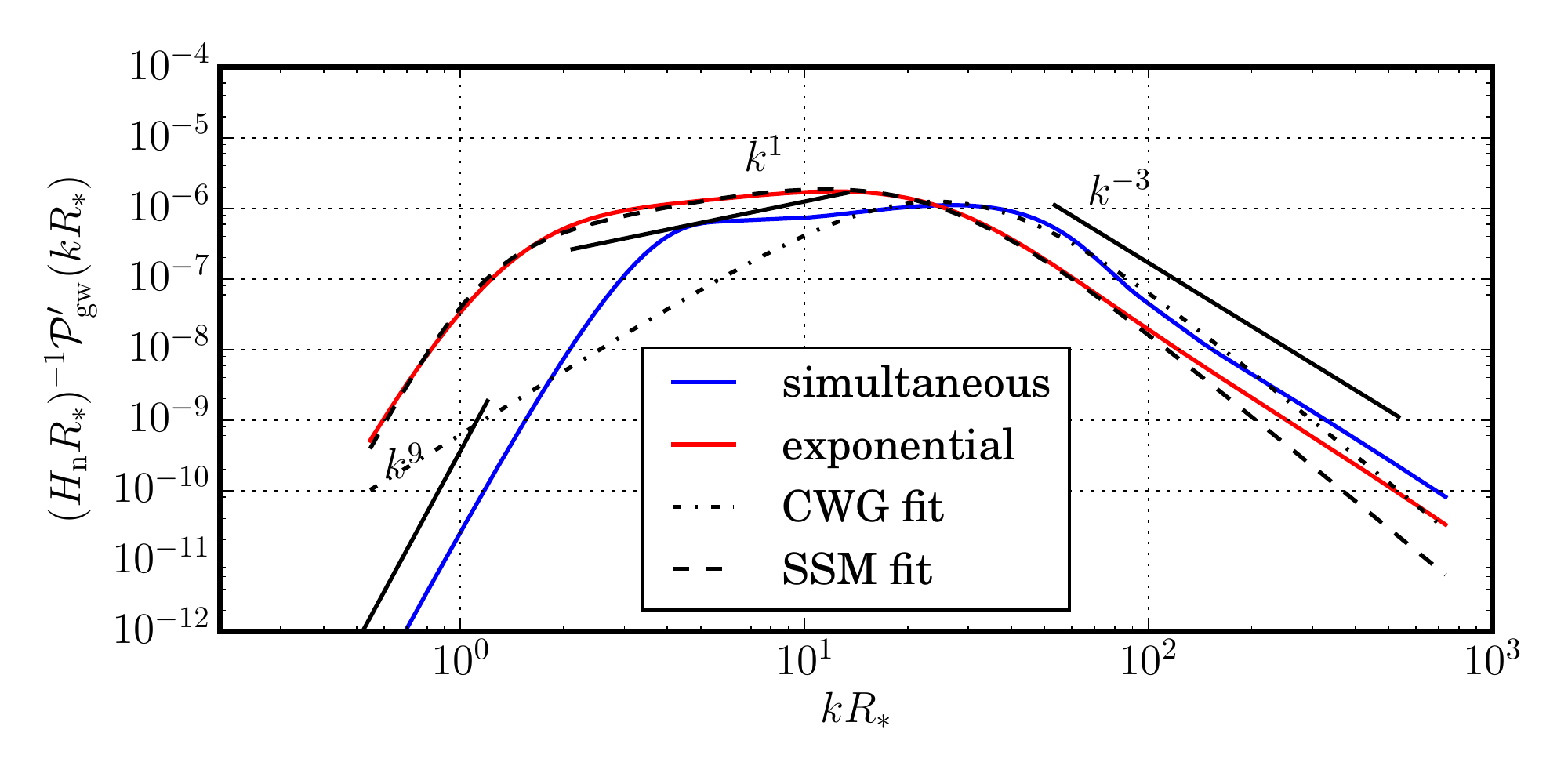}}
\\  
\subfigure[\ Weak,
  $\vw=0.44$]{\includegraphics[width=0.49\textwidth,
    clip=true]{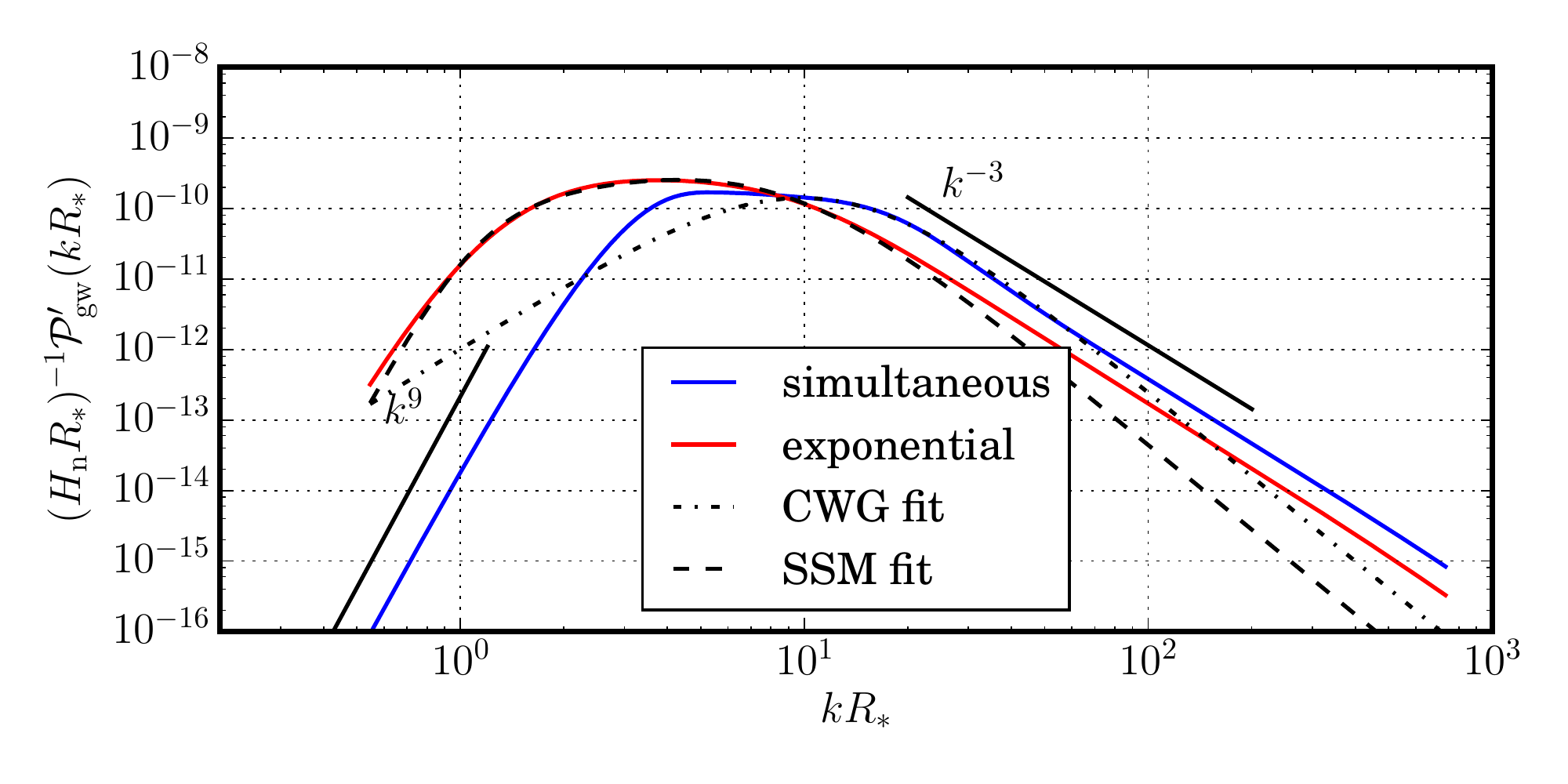}}
\hfill \subfigure[\ Intermediate,
  $\vw=0.44$]{\includegraphics[width=0.49\textwidth,clip=true]{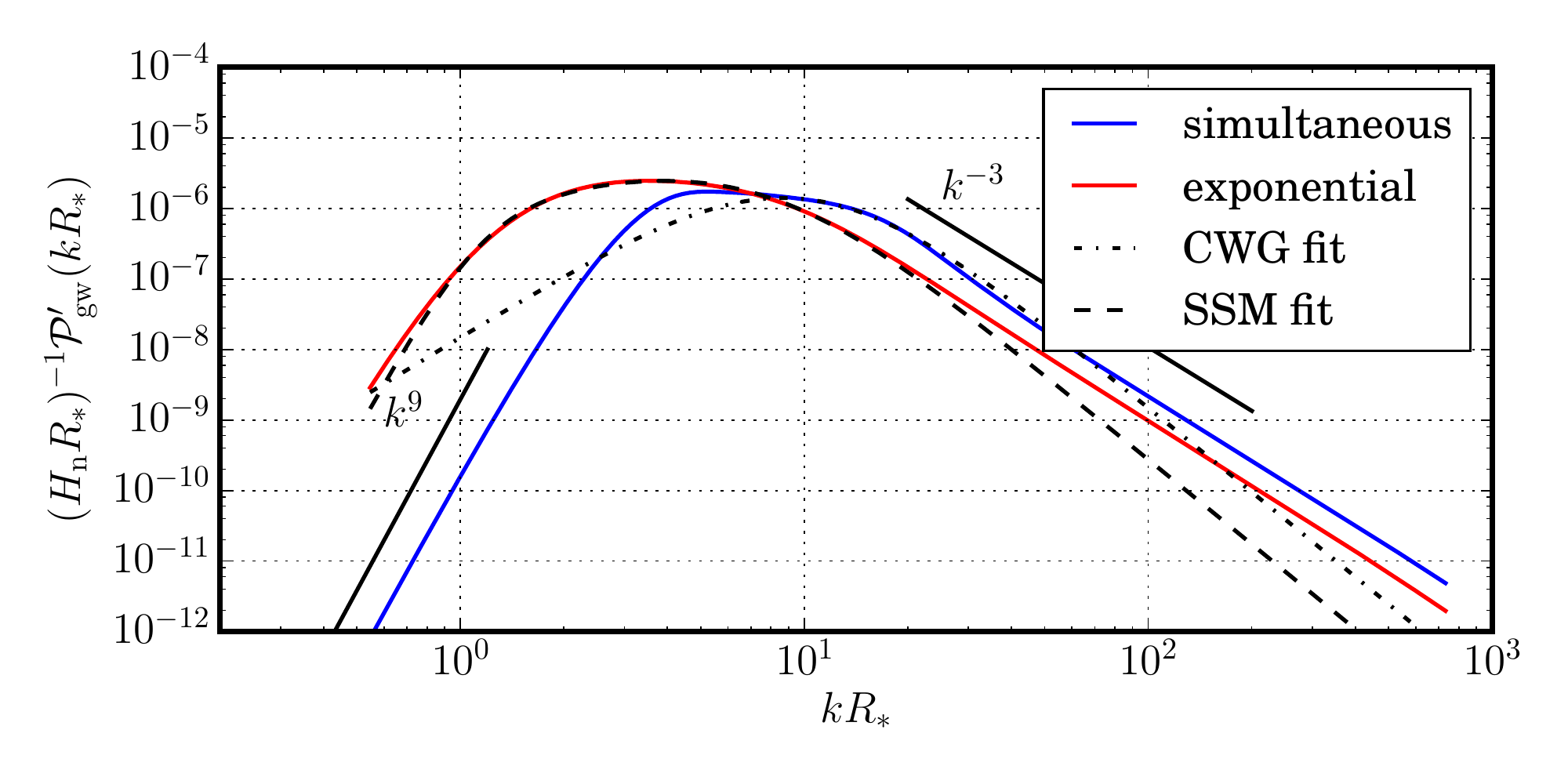}}

\caption{\label{f:GWPSDef} 
Scaled power spectra of fractional energy density
  in gravitational waves for deflagrations, defined in Eq.~(\ref{e:GWGroRat}),
  along with fits, 
  plotted against wave number scaled by the mean bubble separation.  
  Left are weak phase
  transitions, showing detonations with $\vw = 0.56$ and
  $0.44$ (top to bottom).  Right are intermediate phase transitions,
  with the same wall speeds.   
  Fit information is given in the caption to Fig.~\ref{f:GWPSDet}.
  } 

\end{figure}

Figs.~\ref{f:GWPSDet} and \ref{f:GWPSDef} show the gravitational wave power spectra for detonations and deflagrations, again with weak transitions on the left and intermediate strength transitions on the right, and the same colour code distinguishing simultaneous and exponential nucleation.
Fits are shown in black dots and dashes. 
The fit for simultaneous nucleation uses the simple form (\ref{e:CWGFitFun}) presented in \cite{Caprini:2015zlo}, fitted for $k\Rbc > 10$, 
while exponential nucleation uses the double broken power law form (\ref{e:SSMFitFun}) dictated by the sound shell model, fitted over the whole range. One can compare the blue curves with HHRW17 Figs. 6, 7 and 8 \cite{Hindmarsh:2017gnf}. 

The general shapes of both velocity and gravitational wave power spectra are in good agreement for $k\Rbc$ at and above 
the peak at $(k\Rbc)_\text{p}\simeq 10$, particularly for weak detonations.  The low-$k$ spectra have some differences. 
This partly due to a lack of dynamic range, as in most simulations there was only a factor of a few between the bubble separation and the simulation box size. 

The low-$k$ behaviour is best examined in the run with the smallest mean bubble separation, 
whose velocity power spectrum is 
shown at the bottom of HHRW17 Fig.~5, with ($\vw ,\aln) = (0.56,0.0046)$ (weak). 
There the low-$k$ power spectrum is clearly steeper than $k^3$, and consistent with the predicted $k^5$ around $k\Rbc = {\rm O}(1)$. 
The slowly-rising $k^1$ plateau towards a peak at $k\Rbc$ significantly higher than $10$ is also clearly visible,
reflecting the thinness of the fluid shell in this case. The detailed agreement of the peak position is not so good, but 
in the numerical simulations the fluid shells collide before they have reached the self-similar form used in the Sound Shell Model 
calculations, and so they are not as thin or as sharply-defined.

The Sound Shell Model velocity power spectra predictions all have a bump at  $k\Rbc \simeq 4$ which is perhaps present in the 
$(\vw,\aln) = (0.56,0.0046)$ simulation with the smallest mean bubble separation, but is not visible in others. 
The absence of the bump is possibly a finite volume effect, but may also be weakness of the model. 
Larger simulations can resolve this issue.

In the gravitational wave spectra (Figs.~\ref{f:GWPSDet} and \ref{f:GWPSDef}), the low-$k$ behaviour is quite different from that seen 
in HHRW17 Figs.~6--8. The Sound Shell Model predicts a very steep, $k^9$, rise to the peak, which is not seen in the 
numerical simulations.

However, the numerical simulations contain features generated in the bubble collision phase, not included in the Sound Shell Model. 
The power generated in the collision phase would be subdominant if the subsequent evolution had been prolonged to a realistic degree, and the gravitational wave power from the acoustic phase allowed to build up.  
Another, computationally less expensive, way to compare numerical simulations with the Sound Shell Model would be to start 
the gravitational wave evolution after the end of the bubble collision phase.

The predicted bump in the velocity power spectra at around $k\Rbc \simeq 4$ is also seen in the gravitational wave power spectra predictions.  As the bump is not clearly present in the velocity power spectra numerical simulations, it is not surprising that it is 
also absent in the gravitational wave power spectra.

Beyond the peak, the Sound Shell Model also has the ``dome'' to the high-wavenumber side of the peak noted in HHRW17. 
It is this dome which pushes the high wavenumber power-law index of the fitting function towards $k^{-4}$. 
Beyond the dome, the emergence of the $k^{-3}$ power law is clearly seen, as the general considerations 
in subsection \ref{ss:PowLaw} and Ref.~\cite{Hindmarsh:2016lnk} require.

A detailed quantitative comparison of the power spectra 
will be given elsewhere.  A summary comparison of a few global quantities and fitting parameters is given 
in Table \ref{t:ParComTabHyd}.

The first pair of results columns compare the prediction of the RMS fluid velocity  $\fluidV$ (column 3) from a simultaneously nucleated bubbles 
with that measured in the numerical simulations (column 4).  It can be seen that the agreement is excellent for detonations (better than 5\%), while the RMS fluid velocity is under-predicted for deflagrations. The agreement is worse for stronger transitions and slower bubbles, 
with the $\vw=0.44$ deflagration in the intermediate strength phase transition RMS fluid velocity about 70\% below the Sound Shell Model prediction.

The gravitational wave prediction is first compared by computing the gravitational wave production efficiency factor $\tilde\Omega_{\rm gw}$, defined in Eq.~(\ref{e:OmGWscaled}). The resulting numbers (scaled by a factor of 100) are shown in the next two columns. Here the agreement is again better for detonations than for deflagrations. 

A second comparison of the gravitational wave power spectra is derived by fitting the Sound Shell Model predictions to the same functional form as in Ref.~\cite{Hindmarsh:2017gnf} (see Eq.~\ref{e:CWGFitFun}). 
The parameters of the fit are a peak amplitude $A$ and a peak wavenumber in units of the bubble separation 
$z_\text{p} = k_\text{p} \Rbc$.  The Sound Shell Model predictions are fitted over the range $k\Rbc \ge 10$.
The agreement in the peak wavenumber supports the qualitative impression that the shapes are well-predicted by the Sound Shell Model.  There are significant differences in the amplitude for deflagrations, particularly for the intermediate strength transition, as one would 
already expect from the over-prediction of the RMS velocities $\fluidV$, and the fact that the gravitational wave power depends $\fluidV^4$.
 
The over-prediction of the velocity and the peak gravitational power in the case of deflagrations indicates that the naive modelling of the bubble collisions needs improvement.  As the transition strength increases, less of the available energy of the phase transition is going into fluid kinetic energy than the Sound Shell Model predicts.  
This kinetic energy deficit has been observed and studied in recent numerical simulations \cite{Cutting:2019zws}, but 
further numerical simulations are required to understand quantitatively the effect.

On the other hand, the peak frequencies are in better overall agreement, with the worst-performing case being again $\vw=0.44$ deflagration in the intermediate strength phase transition, where the peak frequency is reproduced to within 20\%.

We remind the reader that the case of simultaneous bubble nucleation, while convenient for numerical simulations, 
requires rather special tuning in the bubble appearance action.
Exponential nucleation is the more generic situation, and the figures show that it produces power spectra 
which are peaked at wavenumbers about a factor two lower, 
consistent with numerical simulations of gravitational wave production in the envelope approximation \cite{Weir:2016tov}. 

This can be understood as a result of the different bubble size distributions,   
coupled with the velocity power spectrum weighting the single-bubble power spectra by $\tilde{T}^6$ (\ref{e:VelPowSpe}).
The larger bubbles in the exponential nucleation then boosts the gravitational wave signal, which is proportional to the 
fluid flow length scale.

In Table \ref{t:ParComTabSSM}, we show a quantitative comparison between 
the gravitational wave power spectra produced by 
simultaneous and exponential nucleation. 
The first pair of columns shows that the gravitational wave production efficiency is about 50\% larger in the 
exponential nucleation case.
The next two pairs of columns compare fit parameters: peak amplitude and peak wavenumber.  
The simultaneous nucleation case is fitted to the simpler two-parameter broken power law 
form (\ref{e:CWGfit}) for wavenumbers $k\Rbc > 10$, 
while the exponential nucleation case is fitted to the more accurate three-parameter double broken 
power law form (\ref{e:SSMfit}) over the wavenumber range $0.1 < k\Rbc < 10^3$.

It can be seen how the amplitude is higher in the exponential nucleation case (as expected from the gravitational wave power efficiency factors) 
and the peak frequency is about a factor two lower.  
In the double broken power law the change between the long-distance power law $k^9$ and the intermediate power law 
happens at $(k\Rbc)_\text{b} \equiv z_\text{b} \simeq 1$.  The Sound Shell Model 
predicts that the ratio between $z_\text{p}$ (peak $k\Rbc$) and $z_\text{b}$ 
is related to the relative thickness of the sound shell, which 
at small $\aln$ should be given by
\ben
\frac{z_\text{b}}{z_\text{p}} \simeq \De_\text{w},
\een
where $\De_\text{w} = | \vw - \cs|/\vw$. 
The last column gives $z_\text{p} \De_\text{w}$ for comparison with $z_\text{b}$.
The estimate agrees to $20\%$ or better, except for the intermediate strength transition with $\vw = 0.56$, 
where the true sound shell is thicker than the naive estimate $| \vw - \cs|$ (see Fig.~\ref{f:BubPro}).

\begin{table}

\input{table_3dh_compare}

\caption{\label{t:ParComTabHyd}
Comparison between sound shell model predictions from simultaneously nucleated bubbles, denoted with superscript `sim'
and parameters derived from 3D hydrodynamic simulation data \cite{Hindmarsh:2017gnf}, 
denoted with the superscript `3dh'.
Where no simulation data exists, the entry is blank.
The comparison is between the enthalpy-weighted RMS fluid velocity $\bar{U}_{f,3} = 10^3\fluidV$
in the sound shell model with simultaneous bubble nucleation and that recorded in the 3D simulations (taken at the maximum);
the dimensionless gravitational wave power parameter $\tilde\Omega_{\rm gw,2} = 10^2\OmGWscaled$ (see Eq.~\ref{e:OmGWscaled});
the amplitude parameter $A$ 
and the peak value of $x = k\Rbc$ in the fitting function (\ref{e:CWGfit}).
}

\end{table}

\begin{table}

\input{table_nuc_compare}

\caption{\label{t:ParComTabSSM}
Comparison between sound shell model predictions with different bubble lifetime distributions, 
according to whether the bubble nucleation was simultaneous (sim) or at a rate exponentially growing with time (exp). 
The comparison is between 
the dimensionless gravitational wave power parameter $\tilde\Omega_{\rm gw,2} = 10^2\OmGWscaled$ (see Eq.~\ref{e:OmGWscaled});
the amplitude parameter, 
the peak value of $x = k\Rbc$, 
and the value of $x$ at the break of the fitting function (\ref{e:SSMfit}). 
The final column shows an estimate of the position of the break  valid for small fluid velocities 
$z_\text{p}\De_\text{w}$, where $\De_\text{w} = | \vw - \cs|/\vw$.
}

\end{table}

\section{Conclusions}
In this paper we have detailed the methods of 
the sound shell model of acoustic gravitational wave production, first outlined in \cite{Hindmarsh:2016lnk}. 
The Sound Shell Model predicts the gravitational wave power spectrum from a first order phase transition in the early Universe, 
under the assumption that the sound waves remain linear. 
With this assumption, and a knowledge of the time-dependence of the bubble nucleation rate, 
the Sound Shell Model gives the shape and amplitude of the spectrum as a function of the transition strength parameter $\aln$ and the wall speed $\vw$, and the peak frequency as a function of the transition temperature $\TN$ and the mean bubble separation to the Hubble length $\HN\Rbc$. 
In the standard case of an exponentially growing bubble nucleation rate, 
the shape is well approximated by the double broken power law form given in Eq.~(\ref{e:SSMFitFun}), with a peak power which can be estimated with O(1) accuracy from the average kinetic energy fraction around a single bubble $K = \Gamma \fluidV^2$ according to Eq.~(\ref{e:SSMPeaPow}). 

The method involves the following steps:
the calculation of the self-similar enthalpy and radial velocity functions around an expanding bubble of stable phase, using 
the equations of ultra-relativistic hydrodynamics;
calculating the two-point correlation function of the velocity field formed from a superposition of randomly-placed self-similar fluid shells with a size (or equivalently collision time) distribution computed from the bubble nucleation rate; 
convolving the two-point velocity correlators to obtain the shear stress correlator, 
and integrating the shear stress correlator with the gravitational wave Greens function to obtain the gravitational wave power spectrum. 
This procedure has been implemented in a Python module PTtools, which 
will be made available for public use. 

We have applied the procedure to phase transitions with a range of wall speeds and strength parameters $\aln = 0.0046$ (``weak'') and $\aln= 0.05$ (``intermediate''), and two different models of the bubble nucleation rate,  
covering the transitions explored with numerical simulations in \cite{Hindmarsh:2017gnf}. 
A preliminary comparison of quantities such as peak amplitudes and frequencies, qualitative shapes around the peak, and asymptotic 
power laws is promising, particularly for the weak transition where the linearity assumption is presumably best.

There are differences: the gravitational wave power is over-predicted for deflagrations, by an amount which grows with the transition 
strength. The over-prediction is a result of an over-estimate of the velocity power spectrum, indicating that there is a mechanism 
which suppresses the kinetic energy of the fluid at stronger transitions.  This suppression needs to be explored in further 
numerical simulations.

The other difference clear on a visual inspection is a bump on the low wavenumber side of the peak in the both the gravitational wave and 
velocity power spectra, in the simultaneous bubble nucleation model. 
It is not clear whether this is a problem of the model of the collision time distribution (which can be improved),  or one  
stemming from the finite volume of the numerical simulations.  
A more detailed comparison will be carried out elsewhere. 

The Sound Shell Model is only a beginning. Observable transitions are likely to have enthalpy-weighted RMS fluid velocities 
$\fluidV \gtrsim 0.05$ (see the signal-to-noise ratio curves in Fig.~9 of Ref.~\cite{Hindmarsh:2017gnf}), making non-linear effects important unless the mean bubble separation is larger than about 5\% of the Hubble length.
A more sophisticated understanding of how non-linear effects change the power spectrum is required; at the moment the 
best that can be done is to conservatively stop the growth of power at the non-linearity timescale $\tNL = \Rbc/\fluidV$. 
It is more realistic to expect any or all of the following: a decrease in the amplitude, an increase in the peak frequency, a change in 
the asymptotic power laws, and the increasing importance of turbulence 
\cite{Gogoberidze:2007an,Caprini:2009yp,Niksa:2018ofa}. 
With extreme supercooling \cite{Jinno:2019jhi} the true power spectra could look quite different.
However, the Sound Shell Model model should be able, with suitable modifications taking into account the recently discovered  kinetic energy suppression effect \cite{Cutting:2019zws}, 
to give the power spectrum of the compressive modes 
of the velocity field right after the phase transition completes, which will set the overall power and peak frequency scale of the 
gravitational waves.  

The importance of a detailed and accurate understanding of the gravitational wave power spectrum as a function of the phase transition parameters lies in the possibility of parameter estimation at LISA, and the interface the parameters provide to 
underlying physics beyond the Standard Model. 
Simplified modelling with two or three parameters, while useful, will produce degeneracies which will prevent 
the determination of the four main phase transition parameters $(\aln,\vw,\HN\Rbc,\TN)$. 
The goal is to develop sufficient understanding to extract as much information as possible from gravitational wave 
observations at LISA
Further large-scale numerical simulations are essential to realise this ambition, with the Sound Shell Model as a basis for further 
investigation.


\begin{acknowledgments}
Mark Hindmarsh thanks Mudhahir Al-Ajmi, Daniel Cutting, Stephan Huber, Jose-Miguel No, Kari Rummukainen, and David Weir for discussions.
He has also benefited from many useful exchanges in the LISA Cosmology Working Group,
and assistance with scripting by Jacqueline Lindsay and Michael Soughton. 
The research was supported by the Science and Technology Facilities Council (grant number ST/P000819/1), 
the Academy of Finland (grant number 286769) 
and the Nordita programme {``Gravitational waves from the early universe''}.
\end{acknowledgments}

\appendix

\section{Bubble nucleation}
\label{a:BubNuc}

Once the Universe is below the critical temperature $\Tc$, bubbles of the low-temperature phase can expand. Bubbles are nucleated at a rate per unit volume, 
\ben
p(t') = p_0 e^{-S(t')}.
\een
where $p_0 \sim \Tc^4$, and the bubble production action $S$ decreases rapidly from infinity for $T < \Tc$ and $t > \tc$, the time at which the Universe reaches the critical temperature. 
For temperatures much below the critical temperature, $S$ may either vanish altogether, or reach a minimum and increase again to a larger zero-temperature value.  The first case corresponds to a first order phase transition whose barrier is generated by thermal fluctuations; the second to a phase transition which is present even in the zero-temperature theory. 
We will see that the first case gives an exponentially rising nucleation rate, while the second gives a nucleation strongly peaked around the 
time at which $S$ reaches a minimum, so strongly that one can regard the nucleation as being simultaneous.

\subsection{Bubble nucleation: general considerations}

Let $\Vu$ be the volume in the metastable phase, and $\Vb$ the volume in the stable (Higgs or broken) phase, out of a total volume $\Vtot$, such that 
$$
\Vtot = \Vu + \Vb.
$$
Following \cite{Enqvist:1991xw} we denote the fraction in the metastable phase by $h$, so that 
\ben
h = \Vu/\Vtot.
\een
First, consider the reduction in the volume of the metastable phase between times $t$ and $t+dt$ due to the growth of bubbles nucleated between $t'$ and $t'+dt'$:
\ben
d^2\Vu(t,t') = - d\Nb(t') 4\pi {R}^2 dR \frac{V(t)}{V(t')},
\een
where $d\Nb$ is the number of bubbles nucleated in that time interval, and $R$ is the radius of those bubbles at time $t$.  
We see that
\ben
R = \vw(t - t'), \;\; dR = \vw dt 
\een
and that 
\ben
d\Nb = p(t') V(t'). 
\een
The factor $V(t)/V(t')$ takes into account the fact that only parts of the bubbles growing into the metastable phase will change the volume of that phase.

The nucleation probability is non-zero only below the critical temperature $\Tc$, which is reached at time $\tc$, so the change in the volume of the metastable phase is, in total,
\ben
d\Vu(t) = -\vw\Vu(t) dt \int_{\tc}^{t} dt' p(t') 4\pi \vw^2(t - t')^2. 
\een
Dividing by the total volume $\Vtot$, we have a differential equation for $h$, the fraction in the metastable phase:
\ben
\label{e:hDE}
\frac{dh}{dt} = -\vw h(t) \int_{\tc}^{t} dt' p(t') 4\pi \vw^2(t - t')^2. 
\een
The solution to this equation is 
\ben
h(t) = \exp\left( -  \frac{4\pi}{3}\int_{\tc}^{t} dt' p(t')  \vw^3(t - t')^3 \right).
\een
Saddle-point solutions to the integral $I(t) = -\ln h(t)$ are possible in both exponential and simultaneous nucleation.

\subsection{Exponential nucleation}
In this case, the bubble appearance action is decreasing with time at the relevant epoch and we can make a 
Taylor expansion $p(t') = \exp(-S_0 + \be(t' - t_0))$, with $\be = d \ln p /dt' |_{t_0}$, and $t_0$ a time to be chosen later. 
Therefore, the logarithm of the fraction in the metastable phase is  
\bea
I(t) &=& \frac{4\pi}{3}p_0 e^{-S_0}  \vw^3 \int_{\tc}^{t} dt' e^{\be(t' - t_0)} (t - t')^3 \nonumber\\
&=&\frac{4\pi}{3} p_0 e^{-S_0 + \be(t - t_0)}  \vw^3 \frac{6}{\be^4}.
\eea
Defining a reference time $\tf$ such that $I(\tf) = 1$, we have
\ben
\label{e:hExpNuc}
h(t) = \exp\left( - e^{\beta(t - \tf)} \right),
\een
with the time  $\tf$ determined by the equation 
\ben
8\pi p_0 e^{-S_0 + \be(\tf - t_0)}  \frac{\vw^3 }{\be^4} = 1.
\een
The number density of bubbles $\nb$ obeys the equation
\ben
\label{e:BubDenEqu}
\frac{d \nb}{dt} = p(t) h(t),
\een
as bubbles can nucleate only in the symmetric phase. 
With exponential nucleation (\ref{e:hExpNuc}), this equation can be integrated to give the asymptotic bubble density
\ben
\label{e:BubDenExp}
\nb = \frac{p_0}{\be} e^{-S_0 + \be(\tf - t_0)} = \left( 8\pi \frac{\vw^3 }{\be^3} \right)^{-1}. 
\een
Defining the mean bubble separation as $\Rbc = \nb^{-1/3}$, we have a relationship between the mean bubble density and the transition rate parameter $\be$,
\ben
\Rbc = \left( 8\pi  \right)^{\frac13} \frac{\vw }{\be}. 
\een
Note that (\ref{e:BubDenExp}) implies that the best choice of reference time is $t_0 = \tf$, so that $\be = -S'(\tf)$, otherwise higher order terms in the expansion of $S(t)$ will interfere with the accuracy.
Denoting the bubble nucleation rate at $\tf$ by $p_\text{f}$, we see from (\ref{e:BubDenExp}) that 
\ben
\label{e:NbPf}
\nb = p_\text{f}/\be .
\een

\subsection{Simultaneous nucleation}
In the case where the nucleation rate peaks at a time we denote $t_0$, we expand the action for bubble appearance around its minimum, 
so that 
\ben
p(t') = p_0 \exp(-S_0 - \half \be_2^2(t' - t_0)^2).
\een
Here, $ \be_2^2 = S''(t_0)$.  
In this case, 
\bea
I(t) &=& \frac{4\pi}{3}p_0 e^{-S_0}  \vw^3 \int_{\tc}^{t} dt' e^{-\half \be_2^2(t' - t_0)^2} (t - t')^3.
\eea
In the case that $p_0 e^{-S_0}\vw^3/\be_2^4 \ll 1 $, and also that 
$\be_2(t_0 - \tc) \gg 1$, the solution at late times  ($\be_2( t - t_0) \to \infty$) has the simple form 
\ben
\label{e:hSimNucPre}
h(t) \to \exp\left( - \frac{4\pi}{3} n_0 \vw^3( t - t_0)^3 \right),
\een
where
\ben
n_0 =  \frac{\sqrt{2\pi}}{\be_2}p_0e^{-S_0}.
\een
The bubble density is again obtained by integrating (\ref{e:BubDenEqu}), giving
\ben
\nb = f\left(\frac{n_0\vw^3}{\be_2^3}\right) n_0,
\een
where $f(0) = 1$, which is the limit we are considering. 
Hence we may write 
\ben
\label{e:hSimNuc}
h(t) \to \exp\left( - \frac{1}{6} \be_\text{eff}^3( t - t_0)^3 \right),
\een
with 
\ben
\be^3_\text{eff} = 8\pi \nb \vw^3. 
\een
Note that the asymptotic solution (\ref{e:hSimNuc}) is equivalent to a nucleation rate per unit volume 
\ben
p(t') = \nb \de(t - t_0),
\een
hence our referring to this case as simultaneous nucleation.  
Note also that if $n_0 \gg (\be_2/\vw)^3$, there is a solution to $I(t) = 1$ at a time $\tf < t_0$, and we revert to the first case, where $S(t)$ can be expanded to linear order around $\tf$.

\section{Hydrodynamics of the expanding bubble}
\label{s:HydExpBub}

\begin{figure}[h] 
   \centering
\includegraphics[scale=1.0]{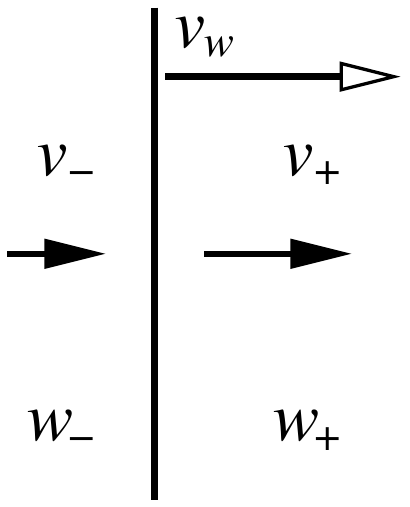}
\hspace{20mm}
\includegraphics[scale=1.0]{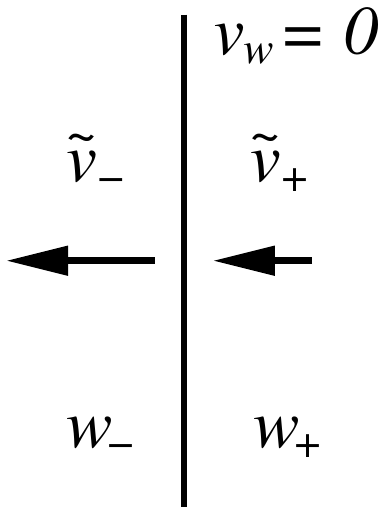} 
   \caption{Fluid velocities close the wall in a supersonic deflagration. 
   On the left, are fluid velocities in the Universe frame, where the bubble centre is at rest. The wall moves to the right with speed 
   $\vw$. The fluid velocities just ahead and just behind are also to the right, with speeds less than $\vw$, and $v_+ > v_-$. 
   On the right are the flows in the wall frame, where the wall is at rest, and the fluid velocities are to the left, with $\vWF_- > \vWF_+$.  
   In a subsonic deflagration, $v_- = 0$ and  $\vWF_- = \vw$, while in a detonation, $v_+ = 0$ and $\vWF_+ = \vw$. 
   The enthalpy $w$ also changes across the wall: see Eqs.~(\ref{e:EneCon},\ref{e:MomCon}) for the 
   relationship between the enthalpies and speeds in the wall frame. The Universe frame wall speeds are related to 
   the wall frame wall speeds through a Lorentz transform (\ref{e:VelLorTra}).}
   \label{f:WalFluVel}
\end{figure}

As the bubble expands, sharp changes in the fluid enthalpy and velocity arise across the bubble wall as the field interacts with the fluid. 
However, conservation laws enable us to match the fluid variables across the bubble wall, which can be extended to the interior and 
exterior using the differential form of energy-momentum conservation. 
This section is based on the discussion in Ref.~\cite{Espinosa:2010hh}, and is included for ease of reference.

In the generic case, the phase boundary quickly reaches a terminal speed $\vw$, and most of the available energy of the transition goes into the thermal and kinetic energy of the fluid.  
Assuming a perfect fluid, the energy-momentum tensor is given by 
\ben
T^{\mu\nu}_\text{f}=w U^{\mu}U^{\nu} + g^{\mu\nu}p, 
\een
where $w = \enDen + p$ is the enthalpy density, $p$ is the pressure and $\enDen$ is the energy density.
We assume a stationary fluid slow with spherical symmetry around the bubble. 
Conservation of energy and momentum density across the phase boundary ensure that, in the frame moving with the wall, 
\bea
w_{+}\gaWF^{2}_{+} \vWF^{2}_{+}+p_{+} &=& w_{-}\gaWF^{2}_{-} \vWF^{2}_{-}+p_{-}, \label{e:EneCon}\\
w_{+}\gaWF^{2}_{+}\vWF_{+} &=& w_{-}\gaWF^{2}_{-}\vWF_{-}, \label{e:MomCon}
\eea
where $\vWF$ is the fluid speed in the wall frame, with  $+$ and $-$ referring to points just ahead and just behind the wall, 
and $\gaWF_\pm = (1 - \vWF_\pm^2)^{-\half}$ (see Fig.~\ref{f:WalFluVel}).

These equations can be rearranged to give
\bal
\vWF_+ \vWF_- &= \frac{1 - (1 - 3\al_+)r}{3 - 3(1 + \al_+)r}, \\
\frac{\vWF_+}{\vWF_-} &= \frac{3 + (1 - 3\al_+)r}{1 + 3(1 + \al_+)r}, 
\end{align}
where
\ben
r =  \frac{w_+}{w_-}, \quad \al_+ = \frac{4}{3} \frac{\th_+ - \th_-}{w_+},
\label{e:RAlpDef}
\een
and $\th = \frac{1}{4} (\enDen - 3p)$ is the trace anomaly.
For the bag equation of state, $\th_- = 0$ and $\th_+$ is a temperature-independent constant, called $\ep$ in Ref.~\cite{Espinosa:2010hh}. 
For a physical equation of state, $\th$ is temperature-dependent and non-zero in both phases, but $\th_-  < \th_+$.

The above equations can be solved for speeds $\vWF_{+}$ and $\vWF_{-}$ to get 
\begin{eqnarray}
\vWF_{+}&=&\frac{1}{2(1+\alpha_+)} \left[ \left(\frac{1}{3\vWF_{-}} + \vWF_{-}\right)
\pm 
\sqrt{ \left(\frac{1}{3\vWF_{-}} - {\vWF_{-}} \right)^{2}+4\al_+^{2}+\frac{8}{3}\al_+ } \right],  
\label{e:vPluEqn}\\
\vWF_- &=&\frac{1}{2} \left[ \left(  ({1+\al_+})\vWF_+ + \frac{1 - 3\al_+}{3\vWF_+} \right)
\pm 
\sqrt{ \left( ( {1+\al_+})\vWF_+ + \frac{1 - 3\al_+}{3\vWF_+} \right)^{2} - \frac{4}{3} } \right],
\label{e:vMinEqn}
\end{eqnarray}
In the first equation, the upper sign is taken for $\vWF_- > 1/\sqrt{3}$, and the lower for $ \vWF_- \le 1/\sqrt{3}$,
otherwise physical solutions do not exist, as we will explain. 
In the second, the upper sign is taken for $\vWF_+ < 1/\sqrt{3}$, and the lower for $ \vWF_+ \ge 1/\sqrt{3}$, 
for the same reason.

Note that $\vWF_{+}$ is required to be positive, as the fluid must flow through the wall from the metastable phase. 
This means we must have $\al_+ < 1/3$. 

In the Universe frame, the fluid speeds either side of the wall are the appropriate Lorentz transforms of the fluid speeds in the wall frame, 
\ben
v_\pm = \frac{\vw - \vWF_\pm}{1 - \vw \vWF_\pm} .
\label{e:VelLorTra}
\een
See Fig.~\ref{f:WalFluVel} for a graphical representation of the relationship.

The enthalpy $w$ and radial fluid velocity $v$ around the rest of the expanding bubble follow from the continuity equation, 
\ben
\pa_\nu T^{\mu\nu} = 0.
\een
Two independent equations can be obtained by contracting with the fluid 4-velocity $u^\mu$ and the space-time orthonormal vector $\bar{u}^\mu = \gamma(-v, \hat{\bv})^\mu$.
Spherically symmetric similarity solutions exist, depending on $r$ and $t$ through the combination $\xi = r/t$, for which the continuity equations can be re-expressed as 
\bea
\frac{dv}{d\xi} &=& \frac{2v(1-v^2)}{\xi(1-v\xi)}\bigg[\frac{\mu^{2}}{c_{s}^{2}}-1\bigg]^{-1}, 
\label{e:VelEqn}\\
\frac{dw}{d\xi} &=& w \left( 1 + \frac{1}{\cs^2}\right) \ga^2 \mu \frac{dv}{d\xi},
\label{e:EntEqn}
\eea
where 
\ben
\mu(\xi,v)=\frac{\xi-v}{1-\xi v}
\een 
is the fluid velocity at $\xi$ in a frame moving outward with speed $\xi$, and 
\ben
\cs^2 = \frac{d p}{d \enDen}
\een
is the speed of sound.  
For expanding bubbles, the equations are defined on the region ($0 \le v < 1$, $0 \le \xi < 1$). The enthalpy density $w$ is of course
always positive. 

In the bag model, $\cs^2=1/3$. With a more physical equation of state, $\cs^2$ is temperature dependent, dropping below $1/3$ near the phase transition.
It is very convenient to study the solutions with $\cs^2=1/3$, as then the enthalpy equation can be integrated separately.

For deflagrations (whether or not supersonic), $\al_+$ is not known from the outset, as it requires knowledge of the temperature either side of the wall, which is fixed by the enthalpy profile of the solution itself.  Instead, one knows the nucleation temperature $\TN$ 
from the solution of equation (\ref{e:tfEq}), 
from which one can calculate $\strPar(\TN)$, defined in (\ref{e:StrParDef}). 
One can then implement a shooting algorithm so that the solution reaches the correct temperature, and hence the correct $\aln$, beyond the shock.
A shooting method is always required if the speed of sound depends on the temperature. 

\begin{figure}[htbp] 
   \centering
   \includegraphics[width=0.48\textwidth]{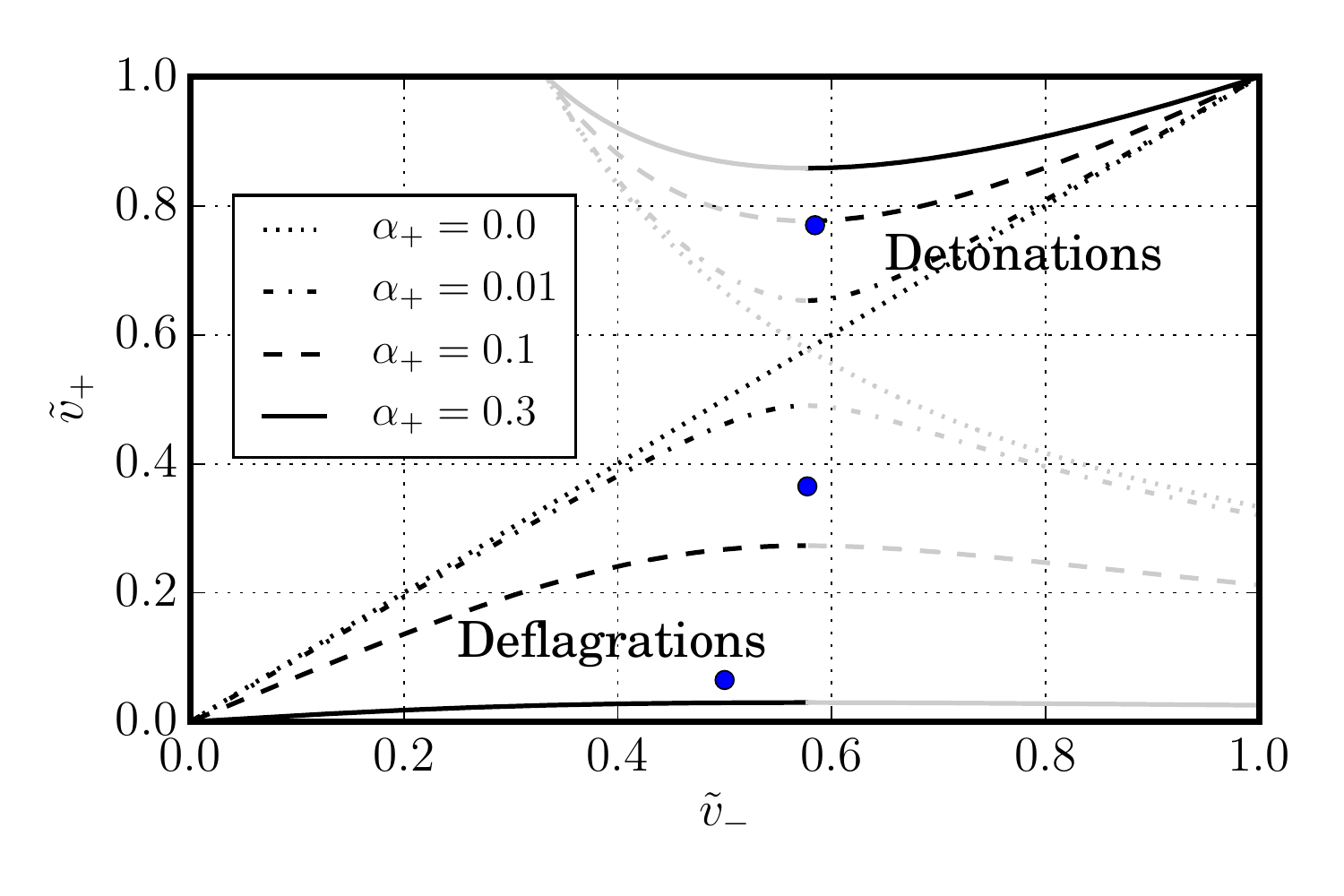} 
   \includegraphics[width=0.48\textwidth]{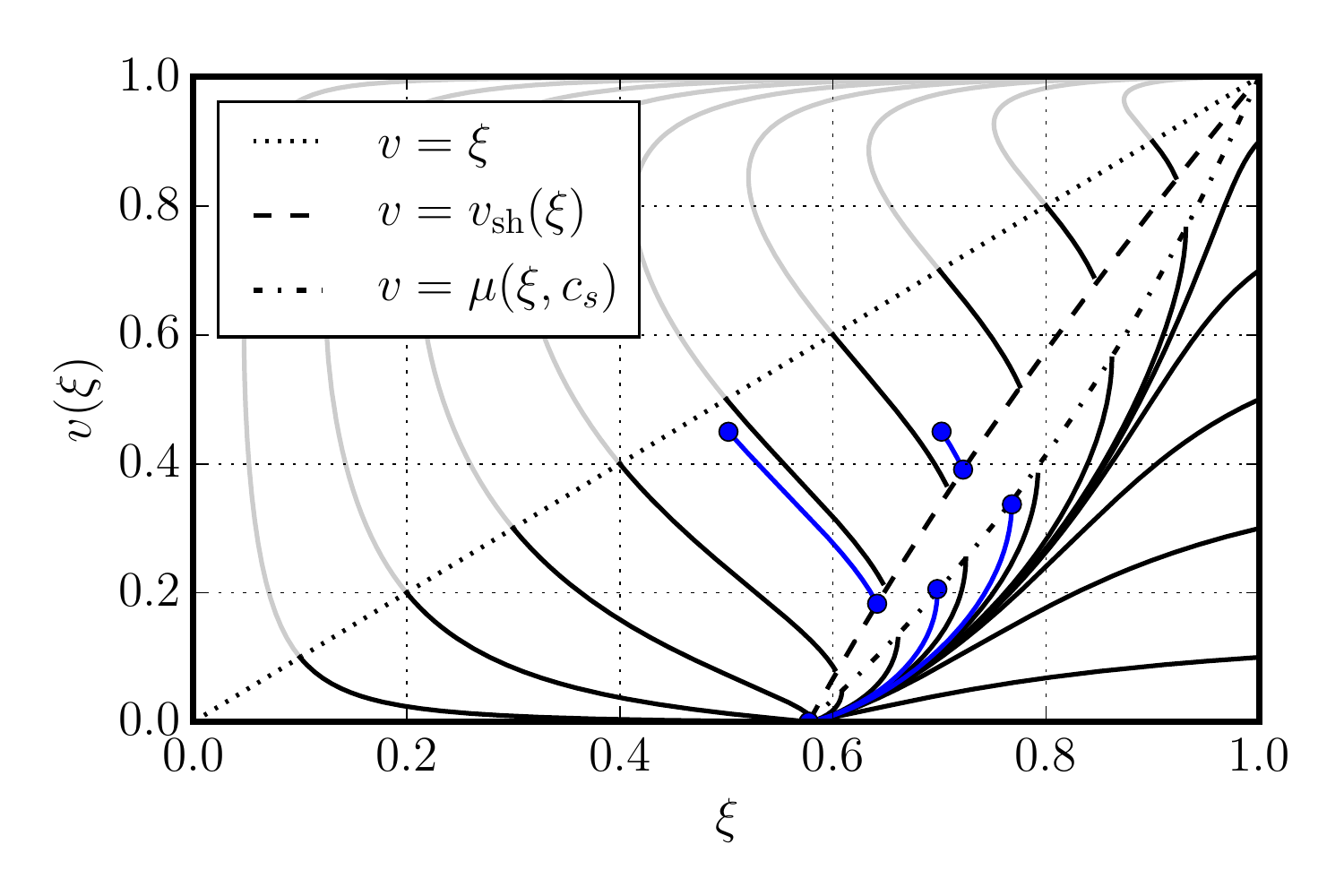} 
   \caption{Left: wall frame fluid speed 
   just ahead of the wall $\vWF_+$ as a function of the fluid speed just behind $\vWF_-$ for several values of the 
   transition strength parameter at the wall $\al_+$ (see Eq.~\ref{e:vPluEqn}).  Both $\vWF_+$ and $\vWF_-$ must be subsonic 
   (deflagrations) or supersonic (detonations) for physical solutions to exist. Other parts of the curves are drawn in grey.  
   Marked as blue dots are the wall frame fluid speeds either side of the wall for the solutions plotted in Fig.~\ref{f:DefHybDet}.
   Right: solutions to the differential equations (\ref{e:EntEqn},\ref{e:ParEquXi}) with sound speed $\cs = 1/\sqrt{3}$.  
   Unphysical parts of the curves, which are never realised in a fluid shell, are drawn in grey. 
   The solutions used in constructing the velocity curves in Fig.~\ref{f:DefHybDet} are shown in blue, with the 
   end points marked with blue dots. 
   }
   \label{f:ParSol}
\end{figure}

In practice it is easier to integrate the equations in parametric form, with 
\bea
\label{e:ParEquXi}
\frac{d\xi}{d\tau} &=& \xi \left[ (\xi - v)^2 - \cs^2 ( 1 -\xi v)^2 \right] , \\
\frac{dv}{d\tau} &=& 2v\cs^2(1-v^2)( 1 -\xi v) , \\
\frac{dw}{d\tau} &=& w \left( 1 + \frac{1}{\cs^2}\right) \ga^2 \mu \frac{dv}{d\tau} .
\eea
The parametric equations have fixed points at $(\xi,v)=(0,0)$, $(\xi,v)=(\cs,0)$ and $(\xi,v)=(1,1)$.  
All non-trivial solutions originate at $(\xi,v)=(1,1)$ and asymptote to $(\xi,v)=(\cs,0)$ (see Fig.~\ref{f:ParSol}, right).

\subsection{Solutions}

Solutions to the hydrodynamic equations (\ref{e:VelEqn}, \ref{e:EntEqn}) 
are obtained by integrating away from $\xiw$, the position of the wall, which we 
can regard as infinitesimally thin. 

We denote the initial conditions $(v_\pm,w_\pm,\xi_\pm)$, 
with $ \xi_\pm = \vw \pm \delta$, and $\delta$ infinitesimal and positive.
Behind the wall, 
the fluid speed $v$ must vanish at $\xi = 0$, as the fluid is forced to be at rest at the centre of the bubble by the radial symmetry. 
In front of the wall, 
$v$ must also vanish as $\xi \to 1$, by causality: the fluid is assumed to be undisturbed 
until a signal from the expanding bubble arrives.

There are only two ways to continuously approach $v=0$: 
(a) start at $v=0$, or (b) start in the region $\xi > \cs$, $\mu(\xi,v) > \cs$ (so that $dv/d\xi > 0$) and integrate backwards in $\xi$.
The only other way to reach $v=0$ is by removing the continuity condition, that is through a shock.

Hence, considering the fluid inside the bubble, we can divide solutions into two classes according to whether the fluid is at rest or not.

\subsubsection{Subsonic deflagrations: $\vw < {1}/{\sqrt{3}}$}

In a subsonic deflagration, the fluid is at rest everywhere inside the bubble wall, 
so that $v_- = 0$, and hence $\vWF_{-}=\vw$.
From (\ref{e:vPluEqn}), 
the wall frame fluid speed just in front of the wall is $\vWF_+(\al_+,\vw)$. 
In the Universe frame the fluid speed just ahead of the wall is 
$v_+ = \mu(\vWF_{+},\xi_\text{w})$, from which one can see that 
the negative sign must be chosen in the square root in 
Eq.~(\ref{e:vPluEqn}) in order that $v_+$ be positive.

\begin{figure}[!h]
\begin{center}
\includegraphics[width=\textwidth]{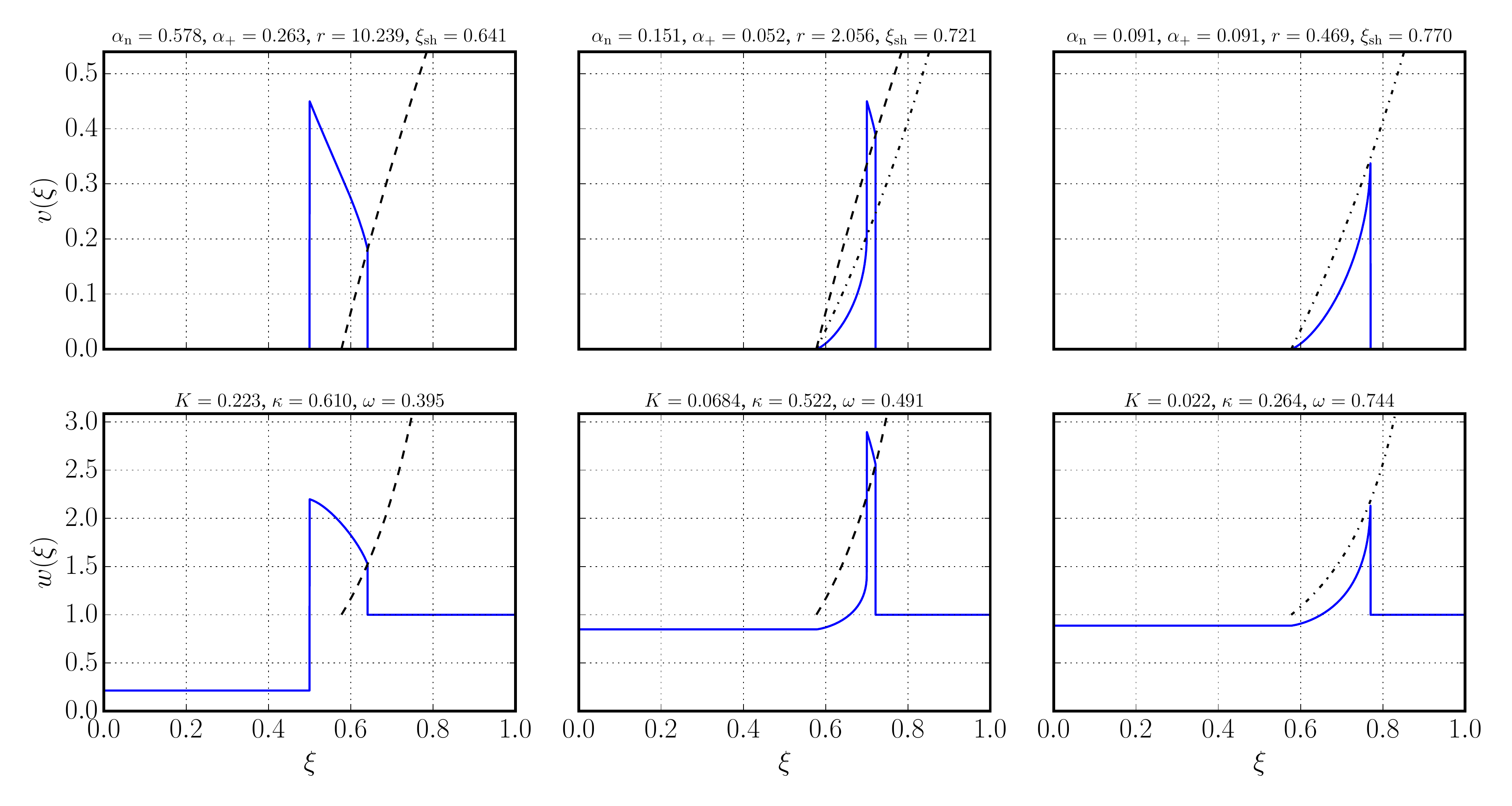}
\caption{Self-similar fluid profiles for a deflagration ($\vw = 0.5$), a hybrid ($\vw = 0.7$), and a detonation ($\vw = 0.77$). 
The dashed line indicates the curves in the $(\xi,v)$ and $(\xi,w)$ planes on which the shock must lie, $v_\text{sh}(\xi)$ (\ref{e:vShoCur}) 
and $w_\text{sh}(\xi)$ (\ref{e:wShoCur}).
The dash-dot line indicates the maximum possible fluid velocity behind a wall in all cases, 
and the maximum possible enthalpy behind a detonation. 
The values of $(\vw,\alpha_+)$ are chosen to equal those of Fig.~4 in Ref.~\cite{Espinosa:2010hh} for comparison purposes. 
The titles show 
$\aln$, the transition strength parameter (\ref{e:StrParDef});
$\al_+$, the transition strength parameter at the wall and 
$r = w_+/w_-$, the ratio of enthalpy densities either side of the wall (\ref{e:RAlpDef});
$\xish$, the shock speed (\ref{e:vShoCur});
$K$, the kinetic energy fraction (\ref{e:KinEneFra});
$\ka$ the kinetic efficiency parameter and
$\om$ the thermal efficiency parameter (\ref{e:EffParDef}).
}
\label{f:DefHybDet}
\end{center}
\end{figure}

Integrating Eqs.~(\ref{e:VelEqn}, \ref{e:EntEqn}), outwards from the wall, 
the fluid speed decreases, until a shock is encountered 
at $\xish$, outside of which the fluid velocity drops to zero. 
In the frame of the shock, $\vSF_{+}\vSF_{-}=1/3$; outside the shock, the fluid is at rest, so that $\vSF_{+} = \xish$.
In the Universe frame, this  translates to the condition that $\xish \mu(\xish,v(\xish))=1/3$;
the shock is therefore encountered when the curve $v(\xi)$ crossed the curve
\ben
\label{e:vShoCur}
\vsh(\xi) = \frac{3\xi^2 - 1}{2\xi} .
\een
Using $\vSF_{-}=1/3\xish$ it follows from (\ref{e:MomCon}) that the analogous equation for the enthalpy is 
\ben
\label{e:wShoCur}
w_\text{sh}(\xi) = w_\text{n} \frac{9\xi^2 - 1}{3(1 - \xi^2)}.
\een
Velocity and enthalpy for a deflagration with $\vw=0.5$ and $\alpha_+=0.263$, along with $\vsh(\xi)$,  are shown in Figure \ref{f:DefHybDet} (left). The shock curves (\ref{e:vShoCur}) and (\ref{e:wShoCur}) are shown as dashed lines.
The speed of sound has been taken as $\cs = 1/\sqrt{3}$.

\subsubsection{Detonations}

In a detonation, the condition $v \to 0$ as $\xi \to 1$ is met by $v=0$  everywhere 
outside the bubble wall, so that 
$\vWF_{+}=\vw$. 
From (\ref{e:vMinEqn}) 
the wall frame fluid speed just behind the wall is then $\vWF_-(\al_+,\vw)$, 
leading to a 
Universe frame fluid speed just behind the wall 
$v_- = \mu(\vWF_{-},\xi_\text{w})$. 
The positive sign must be chosen in the square root in (\ref{e:vMinEqn}) to 
ensure that $\vWF_- = \mu(\vw,v_-) > \cs$, so that $dv/d\xi>0$. 

The condition that the argument of the square root in (\ref{e:vMinEqn}) is non-negative
gives a minimum value of $\vWF_{+}$ (and hence $\vw$),  
called the Chapman-Jouguet speed $\vCJ$. 
The wall speed in a detonation is then larger than 
\ben
\label{e:CJvel}
v_\text{w}^\text{det} > \vCJ = \frac{1}{\sqrt{3}(1 + \al_+)} \left( 1 + \sqrt{\al_+ + 3\al_+^2} \right) .
\een
At the Chapman-Jouguet speed, one can see that 
\ben
\vWF_-(\vCJ,\al_+) = \frac{1}{\sqrt{3}}, 
\een
i.e.~the wall frame fluid exit velocity is equal to ${1}/{\sqrt{3}} $.

Having found the initial condition $v_-$ for the backwards integration of 
Eqs.~(\ref{e:VelEqn}, \ref{e:EntEqn}), 
the fluid velocity $v(\xi)$ decreases gradually the further inside, 
until it comes smoothly to rest at $\xi=c_{s}$. 
Figure \ref{f:DefHybDet} (right) shows a detonation with 
$\vw=0.77$, and $\alpha_+ = 0.091$. 

Note that this detonation has close to the maximum possible $v_-$, which is set by the wall frame exit speed being 
the speed of sound, here $\cs=1/\sqrt{3}$. The curve $\mu(\xi,\cs)$ for $\xi>\cs$ is shown with a dash-dot line in 
Fig.~\ref{f:DefHybDet}.

\subsubsection{Supersonic deflagrations (hybrids): $\vw > {1}/{\sqrt{3}}$}

In the case that the wall frame exit velocity is $\vWF_- = 1/\sqrt{3}$, another solution of the equations in front of the wall is possible. 
Examining (\ref{e:vPluEqn}), we see that a physical solution $\vWF_+(1/\sqrt{3},\alpha_+)$ is possible provided 
it exceeds the wall speed, so that $v_+$ is positive.  Hence $\alpha_+$ must satisfy 
\ben
\frac{1}{\sqrt{3}(1 + \al_+)} \left( 1 + \sqrt{\al_+ + 3\al_+^2} \right) > v_\text{w}^\text{def}  .
\een
This is a complementary condition to that obtained for the detonation (\ref{e:CJvel}).

With a non-zero $v_+$, the hydrodynamic solution in front of the wall behaves in exactly the same way as a deflagration, decreasing 
from $(\xi_w,v_+)$ until it reaches a shock. These solutions, with compression waves both ahead and behind, 
existing only if $\vw > 1/\sqrt{3}$, are called supersonic deflagrations or hybrids \cite{KurkiSuonio:1995pp}.
As $\vw \to \vCJ$ from above, the compression wave in front of the bubble wall tends to zero thickness, while the 
trailing part evolves smoothly into the curve for a detonation at  $\vw = \vCJ$.

Figure \ref{f:DefHybDet} (centre) shows a supersonic deflagration with $\vw=0.7$ and $\alpha_+ = 0.052$.

\subsection{Energy redistribution}
In the sound shell model, the kinetic energy fraction of a single bubble is approximately equal to the kinetic energy fraction of 
the resulting fluid flow as a whole (see Section \ref{ss:RMSFluVel}). 
Hence, the kinetic energy fraction of a single bubble helps fix 
the peak power of the gravitational wave signal, along with an O($10^{-2}$) gravitational wave 
efficiency constant $\tilde{\Om}_\text{gw}$ 
\cite{Hindmarsh:2015qta,Hindmarsh:2017gnf}.
Roughly speaking, the expanding bubble converts the potential energy of the field into kinetic energy and heat.  
Here, we show how this statement can be made quantitative.

Conservation of energy around a single bubble means that 
\ben
E = 4\pi \int_0^R dr r^2 T^{00}.
\een
is a constant for large enough $R$. 
Note that 
\ben
T^{00} = w\ga^2 - p = w\ga^2 v^2 + \enDen =  w\ga^2 v^2 + \frac{3}{4}w + \th,
\een
Hence we can write
\ben
\label{e:BubEneBud}
e_K + \Delta e_Q = - \Delta e_\theta
\een
where 
\begin{align}
e_K &= 4\pi\int_0^{\xi_\text{max}} d\xi \xi^2w \ga^2 v^2, \\
\De e_Q &= 4\pi\int_0^{\xi_\text{max}} d\xi \xi^2 \frac{3}{4}(w - w_\text{n}), \\
\De e_\theta &= 4\pi\int_0^{\xi_\text{max}} d\xi \xi^2 (\th - \theta_\text{n}).
\end{align}
and $\xi_\text{max} = \max(\vw,\xish)$.
We can interpret these three contributions as volume-averaged 
kinetic energy density, thermal energy density and the trace anomaly, which can be thought of as 
the potential energy available for transformation.
The trace anomaly is not quite equal to the thermal potential energy density $V_T(\phi)$, as 
\ben
\th = V_T - \frac{1}{4} T \frac{\pa V_T}{\pa T},
\een
so that not all of the potential energy difference is available to be turned into kinetic and thermal energy.

We can quantify the distribution of the available potential energy by 
defining a bubble volume averaged trace anomaly $\ep$, so that 
\ben
\De e_\theta  = \frac{4\pi}{3}\vw^3 \ep. 
\een
In the bag model, $\ep$ is just the bag constant. 

Kinetic and thermal efficiency factors, quantifying the fraction of the available energy in the scalar field is converted to 
kinetic and thermal energy, can be defined as 
\ben
\kappa = \frac{e_K}{|\Delta e_\th|}, \quad
\omega = \frac{\Delta e_Q}{|\Delta e_\th|},
\label{e:EffParDef}
\een
with $\ka + \om = 1$. 
They can be expressed in terms of the enthalpy and velocity fields and the average trace anomaly $\ep$ as
\ben
\kappa = \frac{3}{\ep\vw^3}\int_0^{\xi_\text{max}} d\xi \xi^2w \ga^2 v^2, \quad
\omega = \frac{3}{\ep\vw^3}\int_0^{\xi_\text{max}} d\xi \xi^2\frac{3}{4}(w - w_\text{n}).
\label{e:EffParExp}
\een
The captions to Fig.~\ref{f:DefHybDet} give $\ka$ and $\om$ for selected solutions, 
and their sum differs from unity by less than a percent, giving an estimate of the numerical errors 
associated with the integration of the fluid equations with $5000$ points along the $\xi$ axis.

Another useful quantity is the enthalpy-weighted mean square fluid 4-velocity around the bubble,
\ben
\label{e:OneBubUbar}
\fluidV^2 = \frac{3}{4\pi\bar{w}\vw^3} e_K.
\een
As shown in Table \ref{t:UbarCom}, this is a good predictor of the mean  square fluid velocity of the fluid after the collision of 
a population of randomly nucleated bubbles, for weak and intermediate strength transitions. 

For the gravitational wave power, it is also useful to define a kinetic energy fraction, the kinetic energy density as a 
fraction of the total average energy density   $\bar{\enDen}$, 
\ben
\label{e:KDef}
K = \frac{e_K}{\bar{\enDen}}.
\een
It is useful to introduce an enthalpy-weighted mean square fluid 4-velocity $\fluidV$ through the relation  
\ben
\label{e:KinEneFra}
K = \Ga \fluidV^2,
\een
where 
\ben
\Ga = \frac{\bar{w}}{\bar{\enDen}}
\een
is the mean adiabatic index of the fluid in the broken (stable) phase. For non-relativistic fluids, $\fluidV$ approaches the RMS 3-velocity.

Note that for the rapid phase transitions we are considering, 
$\bar{\enDen} = \bar{\enDen}_\text{s}(\TN)$, 
the average energy density in the metastable phase at the nucleation temperature.\footnote{For a slower transition, the universe can expand significantly between the nucleation temperature and the completion of the phase transition.} 

The kinetic energy fraction is related to the kinetic efficiency parameter, as 
\ben
K = \kappa \frac{|\Delta e_\th|}{\bar{\enDen}}
\een
Using $\enDen = (3/4)w + \theta$, and recalling the definition of the strength parameter (\ref{e:StrParDef}) we have
\ben
K = \frac{\ka \aln}{1 + \aln + \de_\text{n}},
\een
where
\ben
\de_\text{n} = \frac{4 \th_\text{b}(\TN)}{3 w_\text{s}(\tN)}.
\een
In the bag model, the trace anomaly in the broken phase $\th_\text{b}$ vanishes,  
and we can write 
\ben
\label{e:KbagAlp}
K_\text{bag} = \frac{\ka \aln}{1 + \aln }. 
\een
Note that the bag approximation to relationship between the kinetic energy fraction $K$, the efficiency parameter $\kappa$, and the strength parameter $\aln$ 
(\ref{e:KbagAlp}) is incorrect in general, and there is no particular reason to expect $\de_\text{n} \ll \al_\text{n}$.

\section{Initial conditions, energy-momentum conservation and causality}
\label{s:AppIniCon}

In the work introducing the sound shell model \cite{Hindmarsh:2016lnk}, the plane wave amplitudes were derived from 
\ben
\label{e:OldIniCon}
\vel{\bq}^i = \half \left( \tilde{v}^i_\bq(\tInit) + \frac{i}{\om} \dot{\tilde{v}}^i_\bq(\tInit) \right) e^{i \om \tInit}.
\een
At first sight, this looks identical to (\ref{e:PlaWavCoeCor}), as 
in a sound wave, $\dot\vfft_\bq^i(\tInit) = -i\cs^2 q^i \lafft_\bq(\tInit)$ through the equation of motion, the velocity field is 
longitudinal, and $\om = \cs q$.  We recall that $\lafft_\bq = \enDen_\bq/\bar{w}$, where $\enDen$ is the fluid rest frame 
energy density, and $\bar{w}$ is the mean enthalpy.

However, when one bases the initial condition for the sound waves on the self-similar fluid profile around an expanding bubble, 
one must bear in mind that this is a forced solution. Gradients in the the scalar field potential are accelerating the fluid, and so 
the full equation for the acceleration is 
\ben
\label{e:ForEquMot}
w \dot\vfft_\bq^i + iq^i p_\bq =  \tilde{F}^i_\bq(t),
\een
where the right hand side is the Fourier transform of the scalar forcing term
\ben
F^i = - \left(\frac{\pa V}{\pa \phi} + \eta \dot\phi \right) \pa^i\phi.
\een
Here, $V$ is the thermal effective potential of the scalar field $\phi$, and $\eta$ is the damping coefficient \cite{Hindmarsh:2015qta}. 
The first term is present because $p$ includes the bulk pressure of the scalar field.

When the bubbles collide, the forcing term is removed, and the acceleration of the fluid is given by the pressure gradient of the fluid alone. 
Hence the initial condition for the free evolution is set by the pressure fluctuation established around the expanding bubble.

The result of using (\ref{e:OldIniCon}) is to give an incorrect  $q$-dependence, which violates the energy-momentum constraint 
on the velocity field of the sound waves. 
From (\ref{e:ForEquMot}), one can establish that around a single bubble, the forcing term, and hence the acceleration, goes as O($q$) as $q \to 0$, due to the derivative.  
Substituting into the initial condition (\ref{e:OldIniCon}), we see that the division by $\omega = \cs q$ in the second term results in a plane wave coefficient with low-$q$ behaviour $\vel{\bq} \sim q^0$, and hence a spectral density $P_v(q) \sim q^0$.
This behaviour violates the causality constraints established \cite{Veeraraghavan:1990yd,Durrer:2003ja}.
For completeness, we repeat the argument here.

Once the scalar field has relaxed to the vacuum, 
the fluid obeys the energy-momentum conservation equations
which in linearised form are 
\bea
\dot{\enDen}_\bq + iq_j \bar{w}\vfft_\bq^j &=& 0,  \label{e:LinEneCon}\\
\bar{w}\dot \vfft_\bq^i +  iq^i p_\bq &=& 0 . \label{e:LinMomCon}
\eea
These equations show that if pressure fluctuations with $p_\bq \sim q^m$ are established in a stationary fluid, the resulting velocity fluctuations are longitudinal with $\vfft_\bq \sim q^{m+1}$, and the energy fluctuations are $\enDen_\bq \sim q^{m+1}$.  
Randomly placed bubbles sets up white noise pressure fluctuations ($p_\bq \sim q^0$), and momentum conservation (\ref{e:LinMomCon}) implies that the resulting velocity field must have $\vfft_\bq \sim q$, and energy density fluctuations (and hence $\la_\bq$) must go as $\enDen_\bq \sim q^{2}$. 

Hence, the plane wave amplitudes (\ref{e:PlaWavCoeCor}) are determined at low $q$ by the velocity fluctuations, whose 
spectral density goes as $P_v(q) \sim q^2$.

\bibliographystyle{JHEP}
\bibliography{GWs}

\end{document}

%% file: table_1dh_compare.tex
\begin{tabular}{cc | rrr }
\hline\hline
$10^2\alpha$ & $v_{\rm w}$ & $\bar{U}_{f,3}^{\rm sim}$ & $\bar{U}_{f,3}^{\rm exp}$ & $\bar{U}_{f,3}^{\rm 1d}$ \\ 
\hline
$ 0.46 $ & 0.92 &  4.5 &  4.5 &  5.2 \\ 
$ 0.46 $ & 0.80 &  5.8 &  5.8 &  6.5 \\ 
$ 0.46 $ & 0.68 &  9.0 &  9.0 &  9.7 \\ 
$ 0.46 $ & 0.56 & 16.1 & 16.2 & 16.2 \\ 
$ 0.46 $ & 0.44 &  8.5 &  8.5 &  7.6 \\ 
$ 5.0 $ & 0.92 & 46.1 & 46.1 & 54.7 \\ 
$ 5.0 $ & 0.80 & 59.7 & 59.8 & 68.4 \\ 
$ 5.0 $ & 0.73 & 77.5 & 77.7 & 86.4 \\ 
$ 5.0 $ & 0.56 & 102.8 & 103.0 & 100.3 \\ 
$ 5.0 $ & 0.44 & 80.8 & 80.9 & 71.5 \\ 
\hline\hline
\end{tabular}

%% file: table_3dh_compare.tex
\begin{tabular}{cc | rr | rr | ll | rr}
\hline\hline
$10^2\alpha$ & $v_{\rm w}$ & $\bar{U}_{f,3}^{\rm sim}$ & $\bar{U}_{f,3}^{\rm 3dh}$ & $\tilde\Omega_{\rm gw,2}^{\rm sim}$ & $\tilde\Omega_{\rm gw,2}^{\rm 3dh}$ & \hspace{1em} $A^{\rm sim}$ & \hspace{1em} $A^{\rm 3dh}$ & $x_{\rm p}^{\rm sim}$ & $x_{\rm p}^{\rm 3dh}$ \\ 
\hline 
$ 0.46 $ & 0.92 &  4.5 &  4.6 & 1.3 & 1.2 & $ 2.0\cdot 10^{-11} $ & $ 1.4\cdot 10^{-11} $ &  7.1 &  8.6 \\ 
$ 0.46 $ & 0.80 &  5.8 &  5.8 & 1.0 & 1.4 & $ 3.1\cdot 10^{-11} $ & $ 3.1\cdot 10^{-11} $ & 10.4 & 10.4 \\ 
$ 0.46 $ & 0.68 &  9.0 &  8.7 & 0.5 & 0.6 & $ 8.7\cdot 10^{-11} $ & $ 8.1\cdot 10^{-11} $ & 19.2 & 18.3 \\ 
$ 0.46 $ & 0.56 & 16.1 & 13.8 & 0.2 & 0.3 & $ 3.6\cdot 10^{-10} $ &          & 49.6 &      \\ 
$ 0.46 $ & 0.44 &  8.5 &  7.5 & 1.0 & 1.1 & $ 1.4\cdot 10^{-10} $ & $ 8.2\cdot 10^{-11} $ &  9.8 &  9.9 \\ 
$ 5.0 $ & 0.92 & 46.1 & 43.7 & 1.4 & 2.0 & $ 1.9\cdot 10^{-7} $ & $ 1.6\cdot 10^{-7} $ &  7.7 &  8.5 \\ 
$ 5.0 $ & 0.80 & 59.7 &      & 0.9 &     & $ 2.9\cdot 10^{-7} $ &          & 11.7 &      \\ 
$ 5.0 $ & 0.73 & 77.5 & 65.0 & 0.5 & 1.8 & $ 4.0\cdot 10^{-7} $ & $ 3.7\cdot 10^{-7} $ & 17.0 & 16.1 \\ 
$ 5.0 $ & 0.56 & 102.8 &      & 0.4 &     & $ 1.2\cdot 10^{-6} $ &          & 24.2 &      \\ 
$ 5.0 $ & 0.44 & 80.8 & 54.5 & 1.2 & 1.7 & $ 1.4\cdot 10^{-6} $ & $ 4.3\cdot 10^{-7} $ &  8.7 &  6.9 \\ 
\hline\hline 
\end{tabular}

%% file: table_nuc_compare.tex
\begin{tabular}{cc | rr | ll | rr | rr}
\hline\hline
$10^2\alpha$ & $v_{\rm w}$ & $\tilde\Omega_{\rm gw,2}^{\rm sim}$ & $\tilde\Omega_{\rm gw,2}^{\rm exp}$ & \hspace{1em} $A^{\rm sim}$ & \hspace{1em} $A^{\rm exp}$ & $x_{\rm p}^{\rm sim}$ & $x_{\rm p}^{\rm exp}$ & $x_{\rm b}^{\rm exp}$ & $x_{\rm p}^{\rm exp}\De_\text{w}$ \\ 
\hline
$ 0.46 $ & 0.92 & 1.3 & 2.1 & $ 2.0\cdot 10^{-11} $ & $ 3.0\cdot 10^{-11} $ &  7.1 &  3.6 & 1.3 & 1.3 \\ 
$ 0.46 $ & 0.80 & 1.0 & 1.6 & $ 3.1\cdot 10^{-11} $ & $ 5.5\cdot 10^{-11} $ & 10.4 &  4.6 & 1.2 & 1.3 \\ 
$ 0.46 $ & 0.68 & 0.5 & 0.8 & $ 8.7\cdot 10^{-11} $ & $ 1.4\cdot 10^{-10} $ & 19.2 &  9.0 & 1.1 & 1.4 \\ 
$ 0.46 $ & 0.56 & 0.2 & 0.3 & $ 3.6\cdot 10^{-10} $ & $ 5.0\cdot 10^{-10} $ & 49.6 & 25.6 & 1.1 & 0.8 \\ 
$ 0.46 $ & 0.44 & 1.0 & 1.6 & $ 1.4\cdot 10^{-10} $ & $ 2.5\cdot 10^{-10} $ &  9.8 &  4.3 & 1.1 & 1.3 \\ 
$ 5.0 $ & 0.92 & 1.4 & 2.1 & $ 1.9\cdot 10^{-7} $ & $ 3.0\cdot 10^{-7} $ &  7.7 &  3.7 & 1.3 & 1.4 \\ 
$ 5.0 $ & 0.80 & 0.9 & 1.5 & $ 2.9\cdot 10^{-7} $ & $ 5.0\cdot 10^{-7} $ & 11.7 &  5.2 & 1.2 & 1.4 \\ 
$ 5.0 $ & 0.73 & 0.5 & 0.8 & $ 4.0\cdot 10^{-7} $ & $ 6.7\cdot 10^{-7} $ & 17.0 &  8.1 & 1.2 & 1.7 \\ 
$ 5.0 $ & 0.56 & 0.4 & 0.7 & $ 1.2\cdot 10^{-6} $ & $ 1.9\cdot 10^{-6} $ & 24.2 & 11.5 & 1.1 & 0.4 \\ 
$ 5.0 $ & 0.44 & 1.2 & 2.0 & $ 1.4\cdot 10^{-6} $ & $ 2.5\cdot 10^{-6} $ &  8.7 &  3.9 & 1.2 & 1.2 \\ 
\hline\hline
\end{tabular}